\documentclass[aps,prb,twocolumn,superscriptaddress,longbibliography,floatfix]{revtex4-2}
\usepackage{amsmath,amssymb,graphicx,bm,xcolor,dsfont}
\usepackage[colorlinks=true,linkcolor=blue,citecolor=blue,urlcolor=blue]{hyperref}
\newcommand{\ii}{\mathrm{i}}
\newcommand{\Jpp}{J_{\pm\pm}}
\newcommand{\Jzp}{J_{z\pm}}

\begin{document}

\title{Spin--orbit-locked symmetry and the stability of the triangular Dirac spin
liquid in the ytterbium delafossites}

\author{Yasir Iqbal}
\email{yiqbal@physics.iitm.ac.in}
\affiliation{Department of Physics, Indian Institute of Technology Madras, Chennai 600036, India}

\begin{abstract}
The Yb delafossites NaYbSe$_2$, KYbSe$_2$, CsYbSe$_2$ and NaYbO$_2$ are candidates
for the triangular U(1) Dirac spin liquid, QED$_3$ with $N_f=4$, but their exchange
is spin--orbit coupled. Under their spin--orbit-locked space group we determine
which operators the XXZ anisotropy and the bond-dependent couplings $\Jpp$, $\Jzp$
induce, and at what order. Symmetry forbids every Dirac mass, density and current at any order and,
assuming adiabatic continuity, any new monopole term through charge three. At
linear order the couplings reach only spin-2 members of the quartic multiplets
$\mathbf{20}'$ and $\mathbf{84}$, which share their dimensions with Heisenberg-allowed
operators. At second order $\Jpp$ and $\Jzp$ renormalize the anisotropy through
$\Jzp^2-2\Jpp^2$, and nine spin-1 quartics
without Heisenberg partners appear, irrelevant at $O(1/N)$. A single layer of a
stable Heisenberg spin liquid thus acquires no relevant Lorentz-scalar operator
from spin--orbit coupling. The Heisenberg liquid itself allows a Lorentz-vector
$\mathbf{20}'$ quartic of large-$N$ dimension 2.92, with two-loop $\epsilon$-expansion
estimates 2.9--3.5. Rhombohedral stacking adds triplet-monopole tunneling between
layers at first order. At the transition to 120$^\circ$ order the anisotropy is
relevant, with crossover exponent $y_T\simeq0.74$--1, and exact diagonalization
shows that easy-plane anisotropy selects the chiral-Ising class. The out-of-plane
spectrum at the K point then stays gapless, and the spin-flip to non-spin-flip
ratio grows as $\omega^{-0.56}$ at leading order. Existing KYbSe$_2$
neutron data below $J$ do not fix the K-point exponent. In-plane fields induce a chiral mass
and an out-of-plane moment proportional to $\sin3\varphi$, zero along the bonds and
canceling between CsYbSe$_2$ layers, which the planar thermal Hall effect can test.
\end{abstract}

\maketitle

\section{Introduction}
\label{sec:intro}

The triangular-lattice Heisenberg antiferromagnet with nearest- and
second-neighbor exchange has an intermediate phase between 120$^\circ$ and
stripe order~\cite{Manuel1999,Kaneko2014,ZhuWhite2015,Hu2015,Li2015CCM,Bishop2015,Iqbal2016,Drescher2023,Kovalska2026,Bader2026}, at $0.08\lesssim J_2/J\lesssim0.16$ in variational Monte
Carlo~\cite{Iqbal2016}, with the 120$^\circ$ phase ending at $J_2/J\simeq0.06$--$0.09$ in
other studies~\cite{Scheie2023,Hu2019,Wietek2024}. Variational, density-matrix and
exact-diagonalization studies identify it with the U(1) Dirac spin liquid
(DSL)~\cite{Iqbal2016,Hu2019,Ferrari2019,Wietek2024}, described at low
energy by QED$_3$ with $N_f=4$ two-component Dirac fermions coupled to an
emergent U(1) gauge field. The theory has an emergent SU(4) flavor
symmetry, and its fermion bilinears and monopoles carry the order
parameters of the competing phases~\cite{Hermele2005,Song2019,Song2020}. On
the triangular lattice the monopole quantum numbers forbid every monopole
term below charge three~\cite{Song2019,Song2020}, and the stability of the
phase is decided by four-fermion operators in the SU(4) representations
$\mathbf{20}'$ and $\mathbf{84}$~\cite{He2022}, as for other Dirac spin liquids~\cite{Hermele2004,Xu2008}.

The ytterbium delafossites $A$YbX$_2$ are among the closest material
realizations of this model~\cite{Liu2018,Baenitz2018,Bordelon2019,Ding2019,Ranjith2019,Ranjith2019b,Xing2019,Xing2019b,Schmidt2021,Xing2021}.
NaYbO$_2$ and NaYbSe$_2$ do not order down to 50\,mK~\cite{Bordelon2019,Ranjith2019,Dai2021}, CsYbSe$_2$ develops short-range
order below 0.4\,K~\cite{Xie2023}, and KYbSe$_2$ orders at
$T_N=0.29$\,K~\cite{Scheie2023,Lee2024}. Neutron scattering finds broad continua~\cite{Dai2021,Scheie2023,Xie2023}
and power laws in the spectrum at the K point~\cite{Scheie2023,Scheie2024arXiv}. Fitted exchange parameters
place the selenides near the Heisenberg $J_1$--$J_2$ model with small XXZ
and bond-dependent anisotropies~\cite{Scheie2024,Scheie2023,Villanova2023}. These
compounds are where the stability of the DSL against realistic exchange can be tested.

The Yb$^{3+}$ ion is a Kramers doublet with strong spin--orbit coupling,
and its exchange is not SU(2) symmetric. The allowed nearest-neighbor
exchange contains, besides $J$ and the XXZ ratio $\Delta$, the
bond-dependent couplings $\Jpp$ and $\Jzp$~\cite{Li2016,Rau2018,Maksimov2019}.
With these couplings the lattice operations are symmetries only when
combined with spin rotations. In existing variational Monte Carlo studies the projected Dirac
and spinon-Fermi-surface wave functions are SU(2) invariant, so the $\Jpp$ and $\Jzp$
terms average to zero in them and leave their variational energies
unchanged~\cite{Iaconis2018}. The model has also been studied with projective
symmetry groups~\cite{Li2017}, by density-matrix renormalization
group~\cite{Zhu2018} and in spin-wave theory~\cite{Maksimov2019}. The XXZ anisotropy alone has been treated by
self-consistent mean-field theory combined with the random-phase
approximation~\cite{WillsherKnolle2025}.
The symmetry analysis of the QED$_3$ theory assumes exact SO(3) spin
symmetry~\cite{Song2019,Song2020}, and the operator content of the
spin--orbit-coupled exchange has not been determined.

In this paper we classify all operators of the continuum theory under the
spin--orbit-locked symmetry group of the delafossites and determine the
order in the couplings at which each is induced. We find that a single layer of a
stable Heisenberg DSL acquires no relevant Lorentz-scalar operator from
spin--orbit coupling, and that the anisotropy matters at the transition to
120$^\circ$ order and through monopole tunneling between layers. The lattice symmetries,
each with its locked spin rotation, are projected onto the Dirac zero
modes. Equivariant maps, the linear maps from products of couplings to
operators that commute with the symmetries of the Heisenberg model, then give the
order in the couplings at which each operator appears. The main results are listed
in Table~\ref{tab:summary}, and a short account of them is given in Ref.~\cite{Letter}.

The paper is organized as follows.
Section~\ref{sec:conventions} fixes the structures, conventions and
parameters, and Sec.~\ref{sec:locked} the symmetry group.
Section~\ref{sec:psg} constructs the projective symmetry group of the Dirac
theory, Sec.~\ref{sec:bilinears} classifies the bilinears and the gradient operators, and
Sec.~\ref{sec:spurion} gives the order in the couplings at which each operator appears.
Sections~\ref{sec:quartic} and \ref{sec:monopoles} treat the quartic operators and the monopoles.
Section~\ref{sec:critical} treats the transition to 120$^\circ$ order and
Sec.~\ref{sec:order} the selection of the ordered state.
Sections~\ref{sec:field} and \ref{sec:hall} derive the field selection rules and
the planar thermal Hall response, Sec.~\ref{sec:disorder} classifies quenched disorder, and
Sec.~\ref{sec:data} compares the results with the data and lists the
predictions for each compound. The appendices give the methods, the large-$N$
graphs, the NMR form factors, the one-loop disorder flow, the large-$N$ dimensions of the
quartic operators and the details of the KYbSe$_2$ refit.

\begin{table*}[t]
\caption{Summary of results. ``Order'' is the lowest order in the couplings
at which the operator is induced (nearest neighbors unless stated). The
dimensions are large-$N$ values at $N_f=4$ unless stated. All rows except the
interlayer row refer to a single layer. The tilt $\mathcal O_R$ is defined in Eq.~\eqref{eq:OR}, the
order parameter $\phi_a$ and its crossover exponent $y_T$ in Sec.~\ref{sec:critical}.}
\label{tab:summary}
\begin{ruledtabular}
\begin{tabular}{p{0.2\textwidth}p{0.22\textwidth}p{0.08\textwidth}p{0.38\textwidth}}
perturbation & induced operator & order & consequence\\
\hline
$\Delta$, $\Jpp$, $\Jzp$, $J_2$ & Dirac mass, density, current & none at any order & no explicit gap, no node displacement\\
same & monopole below charge 3 & none & only $\Phi_1\Phi_2\Phi_3+\mathrm{h.c.}$ at charge 3 (adiabatic continuity)\\
$\Delta-1$ & spin-2 parts of $\mathbf{20}'$, $\mathbf{84}$ & linear & relevant only if the Heisenberg DSL is unstable\\
$\Jpp$, $\Jzp$ & spin-2 parts of $\mathbf{84}$ and of the $\mathbf{20}'$ vector and tensor & linear & irrelevant ($\Delta_{\mathbf{84}}\simeq4.54$); relevant only if the Heisenberg $\mathbf{20}'$ vector is\\
$J(\Delta-1)\Jzp$ & tilt $\mathcal O_R$ & second & irrelevant ($\Delta_{\mathcal O_R}>3$)\\
$\Jzp^2-2\Jpp^2$ & spin-2 part of $\mathbf{20}'$ & second & renormalizes the XXZ anisotropy\\
$\Jzp^2$, $\Jpp^2$, $(\Delta-1)\Jzp$ & nine spin-1 non-scalar quartics & second & irrelevant: dimension $4$ or $4+128/(3\pi^2N_f)$, or null\\
anisotropic $J_\perp$ (R$\bar3$m) & triplet monopole tunneling between layers & linear in $J_\perp$ & dimension $2\Delta_\Phi<3$, relevant\\
$\Delta-1$ at the DSL--120$^\circ$ point & spin-2 bilinear of $\phi_a$ & linear & relevant, $y_T\simeq0.74$--$1$\\
$h_z(\Jzp^2-2\Jpp^2)$ & chiral mass & $h\lambda^2$ & $c$-axis Zeeman field induces scalar chirality; orbital term of similar size\\
$\Jpp\Jzp|h_\parallel|^3\sin3\varphi$ & chiral mass, $M_z$ & $h^3\lambda^2$ & zero along bonds; layer-alternating in CsYbSe$_2$\\
non-magnetic defects & singlet monopoles & --- & random valence-bond field, eigenvalue $\simeq2$\\
random Dzyaloshinskii--Moriya & quantum-spin-Hall masses, $\phi_a$ & --- & spin--orbit-specific random channel\\
\end{tabular}
\end{ruledtabular}
\end{table*}

\section{Crystal structures, conventions and parameters}
\label{sec:conventions}

\subsection{Structures and site symmetry}
NaYbSe$_2$, KYbSe$_2$ and NaYbO$_2$ crystallize in R$\bar3$m with one Yb
per primitive cell, and CsYbSe$_2$ in P6$_3$/mmc with two Yb layers per
cell~\cite{Ranjith2019,Ding2019,Xing2019b,Xing2021,Hashimoto2003}. We determined the
site and bond stabilizers from the reported structures by an exhaustive
search over the operations of the hexagonal lattice. In all four
compounds the Yb site has $D_{3d}$ symmetry with its twofold axes along
the nearest-neighbor bonds. The nearest-neighbor bond center has $2/m$
symmetry, with the twofold axis along the bond, and the second-neighbor
bond center has $2/m$ symmetry with the twofold axis perpendicular to the
bond. Both bond centers are inversion centers, which excludes
Dzyaloshinskii--Moriya exchange on first-neighbor bonds~\cite{Rau2018,Schmidt2021} and on second-neighbor bonds.
This antisymmetric part of the exchange transforms as a spin-1 vector under spin
rotations, so the remaining anisotropy is pure spin 2.

\subsection{Frame, ligand setting and pseudospin gauge}
\begin{figure*}[t]
\includegraphics[width=0.96\textwidth]{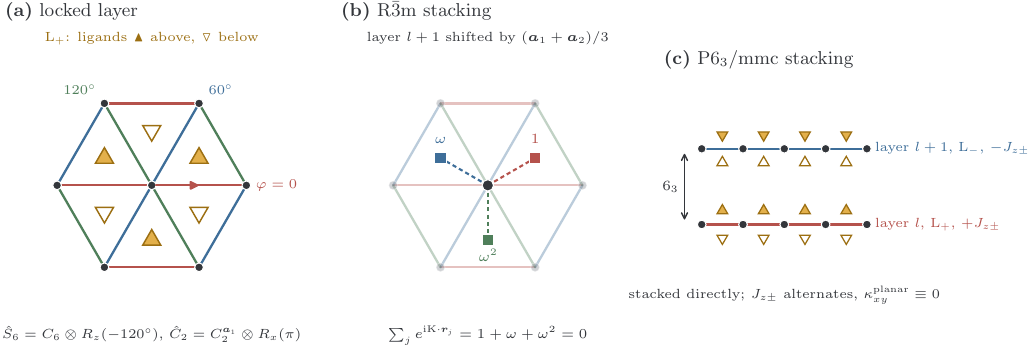}
\caption{Structure and stacking. (a) One Yb layer with the three bond orientations
$\varphi=0$, $60^\circ$ and $120^\circ$, which fix the phases $e^{-2\ii\varphi_{ij}}$ of $\Jpp$ and $\Jzp$ in
Eq.~\eqref{eq:Hmaster}, the ligand setting L$_+$ (ligands above the layer at azimuths $30^\circ$, $150^\circ$,
$270^\circ$, below at $90^\circ$, $210^\circ$, $330^\circ$), and the generators of the locked group,
$\hat S_6$ and $\hat C_2$. (b) R$\bar3$m stacking, in which the three nearest neighbors of a Yb ion in the next
layer, shifted by $(\bm a_1+\bm a_2)/3$, occupy the three sublattices of the 120$^\circ$ pattern, so
their K-point phases $1$, $\omega$, $\omega^2$ cancel and the isotropic interlayer exchange has no Fourier
component at the K point. (c) Side view of the P6$_3$/mmc stacking of CsYbSe$_2$, in which the layers are stacked directly,
related by the $6_3$ screw, with opposite ligand settings and hence opposite signs of $\Jzp$.}
\label{fig:structure}
\end{figure*}

We take $x$ along a nearest-neighbor Yb--Yb bond and $z\parallel c$, with
bond vectors $\bm a_1=a(1,0)$, $\bm a_2=a(\tfrac12,\tfrac{\sqrt3}2)$ and
$\bm a_3=\bm a_2-\bm a_1$. The sense of $x$ along the bond is part of the
convention. Each Yb ion has three ligands above the layer and three below. In
the setting we call L$_+$ the upper ligands lie at in-plane azimuths
$30^\circ$, $150^\circ$ and $270^\circ$ [Fig.~\ref{fig:structure}(a)], and in L$_-$ they lie at $90^\circ$,
$210^\circ$ and $330^\circ$. The two settings are related by a rotation by
$60^\circ$ about $z$, or by reversing the sense of $x$ (a rotation of the
frame by $\pi$ about $z$), which leave $J$, $\Delta$ and $\Jpp$
unchanged and reverse $\Jzp$. With $x$ along $+\bm a$ in the obverse
hexagonal setting, every layer of the R$\bar3$m compounds is L$_+$. In
CsYbSe$_2$ the two layers of the cell are related by the $6_3$ screw and
have opposite settings, so in one global frame $\Jzp$ alternates in sign
from layer to layer. For the structure in the Supporting Information of
Ref.~\cite{Xing2019b} (Yb at $2a$, Se at $4f$ with $z=0.5924$) the layer at $z=0$ is L$_+$.

A rotation of the Kramers pseudospin by $\pi$ about $z$ preserves the form of the
Hamiltonian below and maps $(\Jzp,g_{ab},g_c)$ to $(-\Jzp,-g_{ab},g_c)$, so among these only the
sign of $\Jzp g_{ab}$ is a matter of convention. The sign of $g_c$, that of $\det g$, is physical. No
unitary relabeling reverses it, and reversing it maps the system to one in which every
time-reversal-odd response, such as the field-induced chirality and $\kappa_{xy}$, changes sign.
We fix $g_{ab},g_c>0$, and a source written in a basis with
$g_{ab}<0$ requires $\Jzp\to-\Jzp$.

\subsection{Hamiltonian}
With effective spin-$\tfrac12$ operators and each bond counted once,
\begin{align}
H=\sum_{\langle ij\rangle}\Big[&J(S_i^xS_j^x+S_i^yS_j^y+\Delta S_i^zS_j^z)\nonumber\\
&+\Jpp(\gamma_{ij}S_i^+S_j^++\gamma_{ij}^*S_i^-S_j^-)\nonumber\\
&+\ii\Jzp\big(\gamma_{ij}^*S_i^zS_j^+-\gamma_{ij}S_i^zS_j^-\nonumber\\
&\qquad+i\leftrightarrow j\big)\Big]\nonumber\\
&+(\text{second neighbors: }J_2,\Delta_2,J_{\pm\pm,2},J_{z\pm,2})\nonumber\\
&-\sum_i\bm h\cdot\bm S_i,
\label{eq:Hmaster}
\end{align}
with $\gamma_{ij}=e^{-2\ii\varphi_{ij}}$, $\varphi_{ij}$ the bond angle, and the Zeeman energy
$\bm h=\mu_B(g_{ab}B_x,g_{ab}B_y,g_cB_z)$ for a magnetic field $\bm B$.
Equation~\eqref{eq:Hmaster} is Eq.~(1) of Ref.~\cite{Iaconis2018} with $J=2J_\pm$ and
$J\Delta=J_z$ in their notation. The exchange tensor of a bond at angle
$\varphi$, $S_i^aJ^{ab}S_j^b$, is
\begin{multline}
J^{ab}=J\,\mathrm{diag}(1,1,\Delta)+2\Jpp\!\begin{pmatrix}c&s&0\\s&-c&0\\0&0&0\end{pmatrix}\\
-2\Jzp\!\begin{pmatrix}0&0&s\\0&0&c\\s&c&0\end{pmatrix},
\label{eq:tensor}
\end{multline}
with $c=\cos2\varphi$ and $s=\sin2\varphi$. For the bond along $x$,
$J^{xx}-J^{yy}=4\Jpp$, $J^{yz}=-2\Jzp$ and $J^{xz}=0$, as the twofold axis
along the bond requires. On second-neighbor bonds the same formula gives
a tensor invariant under the twofold axis perpendicular to the bond.

\subsection{Conversion of existing parameter sets}
Existing parameter sets differ in frame, bond phases and normalization, so
they can be compared only through the full exchange tensor. Matching that tensor
to Eq.~\eqref{eq:tensor} on all three bond orientations, with the ligand setting
taken from the crystal structure of each source, brings every set into the form of
Eq.~\eqref{eq:Hmaster}.\footnote{The $\Jzp$ of Refs.~\cite{Li2016,Li2017,Zhu2018,Maksimov2019} must be multiplied
by $-\tfrac12$, and the frames of Refs.~\cite{Maksimov2019,Rau2018} are
L$_-$. The tabulated $\Jzp$ of Ref.~\cite{Villanova2023} follows the
uncorrected Eq.~(45) of Ref.~\cite{Maksimov2019}, corrected in
Ref.~\cite{Maksimov2022}. The bond phases given in Refs.~\cite{Schmidt2021,Ranjith2019}
are not compatible with the $D_{3d}$ site symmetry and require
$\varphi\to-\varphi$, and the parameter dictionaries tabulated in these references give the entries of
Ref.~\cite{Zhu2018} a factor of 2 too large.} The resulting parameters are collected in
Table~\ref{tab:params}.
\begin{table*}[t]
\caption{Exchange parameters in the convention of Eq.~\eqref{eq:Hmaster}
(energies in meV). Entries held fixed in the fit are marked ``(fixed)''. $\eta$ is the undetermined sign of $g_{ab}$ in
the pseudospin basis of Ref.~\cite{Villanova2023}. ESR denotes electron spin resonance.}
\label{tab:params}
\begin{ruledtabular}
\begin{tabular}{lp{0.2\textwidth}llllll}
compound & source (method) & $J$ & $\Delta$ & $\Jpp/J$ & $\Jzp/J$ & $J_2/J$ & $g_{ab}/g_c$\\
\hline
KYbSe$_2$ & \cite{Scheie2024} (spin waves, heat capacity, diffuse scattering) & 0.438(8) & 1 (fixed) & 0 (fixed) & 0 (fixed) & 0.044(5) & 3.0(2)/1.8(6)~\cite{Scheie2023}\\
 & \cite{Scheie2023} (diffuse scattering; $J=0.56(3)$ from Schwinger bosons) & ratios only & 0.98(6) & $+0.004(2)$ & $\lesssim0.015$ (derived) & 0.047(7) & \\
 & \cite{Villanova2023} (density functional) & 0.474 & 0.956 & $+0.0012$ & $0.013\,\eta$ & 0.004 & \\
NaYbSe$_2$ & \cite{Scheie2024} (spin waves) & 0.551(10) & 1 (fixed) & 0 (fixed) & 0 (fixed) & 0.071(15) & 3.13(4)/1.01(1)~\cite{Ranjith2019}\\
 & \cite{Ranjith2019,Schmidt2021} (susceptibility, magnetization) & 0.39--0.46 & 0.50--0.54 & & & & \\
 & \cite{Villanova2023} (density functional) & 0.551 & 0.970 & $-0.0016$ & $0.011\,\eta$ & $-0.0009$ & \\
CsYbSe$_2$ & \cite{Xie2023} (field-polarized spin waves) & 0.395(8) & $\simeq1$ (fitted) & 0 (fixed) & 0 (fixed) & 0.028(10) & 3.20(6); 3.25/0.3 (ESR)\\
 & \cite{Pocs2021} (crystal field, magnetotropic) & 0.44 & 0.16 & & & & \\
NaYbO$_2$ & \cite{Bordelon2019} (representative point) & 0.51 & 0.88 & & & & 3.294(8)/1.726(9)\\
 & \cite{Ranjith2019b,Ding2019} (Curie--Weiss) & 0.33--0.34 & & & & & \\
\end{tabular}
\end{ruledtabular}
\end{table*}
\section{Spin--orbit-locked symmetry group}
\label{sec:locked}

The Heisenberg group $G_H$ consists of the layer group of the Yb lattice
acting on positions, global SO(3) spin rotations and time reversal
$\mathcal T$. The exchange of Eq.~\eqref{eq:Hmaster} with $\Jpp,\Jzp\neq0$ is
invariant only under a subgroup $G_L\subset G_H$, the spin--orbit-locked
group. It is generated by
\begin{itemize}
\item the translations $T_1$, $T_2$ by $\bm a_1$, $\bm a_2$, which do not act on spin;
\item $\hat S_6$: the site-centered rotation $C_6$ by $+60^\circ$ combined with
an active spin rotation by $-120^\circ$ about $z$;
\item $\hat C_2$: the twofold axis along the $\bm a_1$ bond, which acts in
the plane as the mirror $M_x:y\to-y$ and on spin as a rotation by $\pi$
about $x$;
\item time reversal $\mathcal T$, $\bm S\to-\bm S$.
\end{itemize}
These are equivalent to the generators of Ref.~\cite{Iaconis2018}, whose $S_6$
spin matrix is the rotation by $-120^\circ$ and whose twofold axis lies along
another bond. Equation~\eqref{eq:tensor} is invariant under each generator, $C_6$
combined with a $+60^\circ$ spin rotation is not a symmetry, and the
generators reproduce the space-group operations computed from the
reported structures. The spin part of $\hat S_6$ is a rotation by
$-120^\circ$ because the in-plane rotation of the
Yb positions by $60^\circ$ is realized in the crystal by the
roto-reflection $S_6$, which equals the inversion through the Yb site
times a rotation by $-120^\circ$ about $c$. Inversion leaves the spin, an
axial vector, unchanged, so the spin rotates by $-120^\circ$.

Every element of $G_L$ is an element of $G_H$. The action of $G_L$ on any
operator follows from its action under $G_H$ by restriction,
with each lattice generator accompanied by its spin rotation. The invariants of $G_L$ in the space $V$ of
symmetric nearest-neighbor exchange tensors on the three bond
orientations ($\dim V=18$) form a four-dimensional subspace, spanned by
$J$, $J\Delta$, $\Jpp$ and $\Jzp$. With second neighbors the locked space
is eight-dimensional and coincides with the span of the parametrization
in Eq.~\eqref{eq:Hmaster}. Under $G_H$, $V$ decomposes into spin 0 and
spin 2 (a symmetric tensor has no spin-1 part), each tensored with the
permutation representation $A_1\oplus E_2$ of the three bond
orientations. The isotropic $J$ is spin 0 and lattice $A_1$, the XXZ
anisotropy contains a spin-2 lattice-$A_1$ part, and $\Jpp$ and $\Jzp$ are
two components of one spin-2 $\otimes E_2$ multiplet.

A rotation by $60^\circ$ about $z$ maps the ligand setting L$_+$ onto
L$_-$ and acts on the couplings as $\Jzp\to-\Jzp$, leaving $J$, $\Delta$
and $\Jpp$ unchanged. Quantities even under this rotation are
therefore even in $\Jzp$, and quantities odd under it are odd in $\Jzp$. In CsYbSe$_2$ the $6_3$ screw relates the two layers,
so every layer-resolved quantity odd in $\Jzp$ alternates along $c$.

\section{Projective symmetry group and Dirac theory}
\label{sec:psg}

\subsection{Mean-field state and zero modes}
The DSL is the staggered $\pi$-flux state of Abrikosov fermions $f_{i\alpha}$,
with $\pi$ flux through one set of triangles and zero through the
other~\cite{Lu2016,Iqbal2016,Song2020}. We use the real gauge
\begin{align}
t(\bm r,\bm r+\bm a_1)&=1,\nonumber\\
t(\bm r,\bm r+\bm a_2)&=t(\bm r,\bm r+\bm a_2-\bm a_1)=(-1)^{n_1},
\end{align}
with $\bm r=n_1\bm a_1+n_2\bm a_2$. On $L\times L$ tori with
$L\equiv0\bmod4$ the spectrum has four zero modes per spin species, at two
valleys. Writing the spinful fermions as Majorana operators gives 16 real
zero modes $\xi$. The continuum theory is
\begin{equation}
\mathcal L=\bar\psi\,\ii\gamma^\mu(\partial_\mu-\ii a_\mu)\psi+\dots,
\end{equation}
with $\psi$ an eight-component field carrying spin ($\sigma^a$), valley
($\tau^b$) and Dirac indices, $N_f=4$ two-component flavors and an
emergent SU(4) symmetry generated by $\sigma^a$, $\tau^b$ and
$\sigma^a\tau^b$. The velocities along $x$ and $y$ are equal, as required
by $C_6$.

\subsection{Projection of the symmetry generators}
Each generator $g$ acts on the lattice as $f_i\to\eta_i\,U_g\,f_{g(i)}$, with
$U_g$ its spin rotation and $\eta_i=\pm1$ a sign gauge, or, for $C_6$ and the
mirrors, as $f_i\to\eta_i\,U_g(\ii\sigma^y)f^\dagger_{g(i)}$, a spin-singlet
particle--hole transformation. The gauge $\eta_i$ is obtained by
propagating $t_{g(i)g(j)}=\pm\eta_i\eta_jt_{ij}$ over the bonds, and each result is
an exact symmetry of the Majorana Hamiltonian. Projected
onto the zero modes, $g$ acts as $g\,\xi(\bm r)\,g^{-1}=O\xi(R\bm r)$ with an
orthogonal $16\times16$ matrix $O$ and the spatial matrix $R$. Bilinears
$(\ii/4)\xi^T\mathsf M\xi$ with $\mathsf M$ antisymmetric and commuting with the U(1)
generator transform as
\begin{equation}
\mathsf M\to\sigma_g\,O^T\mathsf MO,
\label{eq:Mtrans}
\end{equation}
and gradient bilinears $(\ii/4)\xi^T\mathsf M_j\partial_j\xi$ with $\mathsf M_j$ symmetric as
\begin{equation}
\mathsf M'_k=\sigma_g\sum_jR_{kj}O^T\mathsf M_jO,
\label{eq:Mgrad}
\end{equation}
with $\sigma_g=-1$ for antiunitary $g$ and $+1$ otherwise. The Dirac kinetic term is invariant under
every generator and is a spin singlet. The relations $C_6^6=M_x^2=\mathcal T^2=1$
and $[T_1,T_2]=0$ hold on operator space. The locked generators are
obtained by inserting the spin rotations of Sec.~\ref{sec:locked} into $U_g$.

\subsection{Masses}
Table~\ref{tab:masses} lists the 16 Dirac masses and their transformation
under the Heisenberg generators. It reproduces the $T_1$, $T_2$, $C_6$ and
$\mathcal T$ columns of Table~VII of Ref.~\cite{Song2020}. The reflection column of
Table~VII of Ref.~\cite{Song2020} includes a particle--hole transformation, and its block characters could
not be matched to a mirror or glide of the lattice in our conventions, so our
classification rests on the lattice mirror $M_x$ instead. Its entries for $\bar\psi\psi$ and $\bar\psi\sigma^a\psi$ agree with our $M_x$ column. The quantum-spin-Hall masses $\bar\psi\sigma^a\psi$ are $\mathcal T$
even, $C_6$ odd and carry spin 1. The valence-bond masses $\bar\psi\tau^b\psi$
are $\mathcal T$ even spin singlets at the three M points. The chiral mass
$\bar\psi\psi$ carries the scalar spin chirality and is $\mathcal T$ odd, as are
the nine spin--valley masses.

\begin{table}[t]
\caption{The 16 Dirac masses. The column $(T_1,T_2)$ gives the translation
eigenvalues. The valence-bond and spin--valley masses occupy the three M points
and are permuted by $C_6$. The last column gives the traces of the locked
$\hat S_6$ and $\hat C_2$ on each multiplet.}
\label{tab:masses}
\begin{ruledtabular}
\begin{tabular}{lccccc}
mass & $\mathcal T$ & spin & $(T_1,T_2)$ & $C_6$, $M_x$ & $\hat S_6$, $\hat C_2$\\
\hline
$\bar\psi\psi$ & $-$ & 0 & $(+,+)$ & $+$, $-$ & $1$, $-1$\\
$\bar\psi\sigma^a\psi$ & $+$ & 1 & $(+,+)$ & $-$, $+$ & $0$, $-1$\\
$\bar\psi\tau^b\psi$ & $+$ & 0 & M points & permuted & $0$, $-1$\\
$\bar\psi\sigma^a\tau^b\psi$ & $-$ & 1 & M points & permuted & $0$, $-1$\\
\end{tabular}
\end{ruledtabular}
\end{table}

\section{Fermion bilinears at all orders}
\label{sec:bilinears}

\subsection{No mass, density or current}
The gauge-invariant zeroth-order bilinears form a 64-dimensional space:
16 masses, 16 densities and 32 spatial currents. We computed the
invariant subspace under $G_H$ and under $G_L$ as the null space of
$\{g-1\}$ over the generators, on $L=8$ and $L=12$. Both groups leave no bilinear invariant.

For the masses the result can be read off Table~\ref{tab:masses}. The
chiral and spin--valley masses are odd under $\mathcal T$, which is not
locked. The valence-bond masses carry nonzero momentum, and translations
are not locked either. The quantum-spin-Hall masses are $C_6$ odd. Under
$\hat S_6$ the component $a=z$ is unchanged by the spin rotation and stays
odd, while the in-plane pair is rotated by
$120^\circ$ and has no fixed vector. No combination of masses is invariant.

Because only the symmetry group of the spin model enters, this holds at
any order in $\Delta-1$, $\Jpp$, $\Jzp$ and the second-neighbor couplings. The
spin--orbit-coupled exchange of the delafossites cannot gap the
DSL through an explicit mass. Since no density or current is invariant,
it cannot generate a constant flavor gauge field, and the Dirac nodes are
not displaced. Neither statement holds for the kagome DSL, where the
out-of-plane Dzyaloshinskii--Moriya coupling is a spin-1, $\mathcal T$-even
perturbation that couples to a quantum-spin-Hall mass, at linear order in
the out-of-plane and second order in the in-plane components~\cite{Hermele2008}. The mass gaps
the spinons~\cite{Dodds2013,Hermele2008}, and monopole proliferation then produces magnetic
order~\cite{Hermele2008,Dodds2013}. In the delafossites inversion at every bond
center removes the spin-1 part of the exchange. The statement concerns
explicit local operators in the symmetric phase. It does not exclude
spontaneous symmetry breaking or confinement, which are controlled by the
quartic and monopole operators of
Secs.~\ref{sec:quartic}--\ref{sec:monopoles}.

\subsection{Linear order}
At linear order no bilinear is induced, because the
couplings are spin 0 or spin 2, and fermion bilinears carry spin 0 or 1
(the spin index enters through $\sigma^a$ only). A spin-2 coupling can
therefore induce a bilinear only at second or higher order. A mean-field
decoupling makes this explicit. The hopping induced by an exchange tensor
$J^{ab}$ is proportional to
$\sum_{ab}J^{ab}\sigma^a\sigma^b=\mathrm{tr}J+\ii\epsilon_{abc}J^{ab}\sigma^c$,
which vanishes for a symmetric traceless $J$.

\subsection{Gradient operators}
With one spatial gradient there are 128 structures. The Heisenberg group
leaves only the kinetic term invariant. The locked group allows exactly one
further operator,
\begin{equation}
\mathcal O_R=\psi^\dagger\big[\sigma^x(-\ii\partial_y)-\sigma^y(-\ii\partial_x)\big]\psi,
\label{eq:OR}
\end{equation}
a valley singlet carrying spin 1 that tilts the cones of opposite spin in
a spin--momentum-locked pattern. With one time derivative (64
structures) the locked invariants reduce to $\mathcal O_R$, total
derivatives and terms that vanish by the equations of motion. The Lorentz
content of $\mathcal O_R$ is the $(0j)$ component of the traceless
flavored tensor $\bar\psi\sigma^a\gamma^{(\mu}\ii\partial^{\nu)}\psi$, up to a total
derivative of the quantum-spin-Hall masses. This operator is a
non-conserved primary of Lorentz spin 2. In three dimensions unitarity
requires $\Delta\ge j+1$ for a primary of spin $j\ge1$, with equality only for
conserved currents~\cite{Minwalla1998}, so $\Delta_{\mathcal O_R}>3$.
At large $N$ its dimension and that of the adjoint masses are
(Appendix~\ref{app:largeN})
\begin{equation}
\Delta_{\rm adj}=2-\frac{64}{3\pi^2N_f},\qquad
\Delta_{\mathcal O_R}=3+\frac{64}{5\pi^2N_f},
\end{equation}
that is, $1.46$ and $3.32$ at $N_f=4$. The tilt is irrelevant.

\section{Order in the couplings}
\label{sec:spurion}

\subsection{Equivariant maps}
Symmetry fixes which operators are allowed, and equivariant maps fix the
lowest order in the couplings at which each appears. An operator $O$ can be induced at order $n$ in the couplings only if there
is a $G_H$-equivariant map from $\mathrm{Sym}^n(V)$ to the representation
spanned by $O$ under $G_H$. Evaluated on the physical, $G_L$-invariant
couplings, such a map gives the allowed coefficients. We
compute these maps as linear null spaces with exact spin generators, or
by sampling invariant polynomials of the couplings and the operator
coordinates at random points.

\subsection{The tilt}
For nearest-neighbor couplings there is a single quadratic map onto
$\mathcal O_R$. On the physical couplings it gives
\begin{equation}
c_R\propto J(\Delta-1)\Jzp ,
\end{equation}
with no contribution from $\Jpp^2$, $\Jzp^2$, $\Jpp\Jzp$ or $\Jpp(\Delta-1)$. Spin 1 appears in $\mathrm{Sym}^2$ of the spin-2
couplings only through the antisymmetric part of their lattice factor, and
$\Lambda^2(A_1\oplus E_2)$ contains $E_2$, the lattice representation of
$\mathcal O_R$, only through the cross term between the $A_1$ (XXZ) and $E_2$
($\Jzp$) couplings. At third order the maps give, besides $J$ times the second-order term,
$\Jzp^3-2\Jpp^2\Jzp$, $J(\Delta-1)\Jpp\Jzp$ and $[J(\Delta-1)]^2\Jzp$. With second neighbors five
second-order terms appear,
\begin{equation}
\begin{gathered}
J(\Delta-1)\Jzp,\ J_2(\Delta_2-1)J_{z\pm,2},\ J(\Delta-1)J_{z\pm,2},\\
J_2(\Delta_2-1)\Jzp,\ \Jzp J_{\pm\pm,2}-\Jpp J_{z\pm,2},
\end{gathered}
\end{equation}
the last of which needs no XXZ anisotropy.

If every $\Jzp$-type coupling vanishes, the model is invariant under a
spin rotation by $\pi$ about $z$ alone, under which $\mathcal O_R$ is odd.
The coefficient of $\mathcal O_R$ then vanishes at all orders. Equivalently,
$c_R$ is odd in $\Jzp$ and reverses between the ligand settings L$_\pm$.

\section{Quartic operators}
\label{sec:quartic}

\subsection{Classification}
The four-fermion operators built from the 15 adjoint masses decompose
under SU(4) as $\mathrm{Sym}^2(\mathbf{15})=\mathbf1\oplus\mathbf{15}\oplus\mathbf{20}'\oplus\mathbf{84}$.
We projected onto the irreducible components with the SU(4) Casimir and
counted invariants (Table~\ref{tab:quartic}). The Heisenberg-allowed members of
$\mathbf{20}'$ and $\mathbf{84}$ are the operators $(T,S)$ and $(84,S)$ of
Ref.~\cite{He2022}, whose irrelevance is the condition for a stable
triangular DSL. The locked group adds one spin-2 invariant to the
$\mathbf{20}'$ and three to the $\mathbf{84}$. Their $G_H$ content is
spin~2~$\otimes$~(lattice-trivial) in the $\mathbf{20}'$, and spin
2~$\otimes A_1$ plus spin 2~$\otimes E_2$ in the $\mathbf{84}$.

\begin{table}[t]
\caption{Invariants among the quartic operators
$\mathrm{Sym}^2(\mathbf{15})$ of the adjoint masses.}
\label{tab:quartic}
\begin{ruledtabular}
\begin{tabular}{lccc}
irrep & $G_H$ & lattice and $\mathcal T$ only & $G_L$ (spin content)\\
\hline
$\mathbf1$ & 1 & 1 & 1 (0)\\
$\mathbf{15}$ & 0 & 0 & 0\\
$\mathbf{20}'$ & 1 & 6 & 2 (0, 2)\\
$\mathbf{84}$ & 1 & 6 & 4 (0, 2, 2, 2)\\
\end{tabular}
\end{ruledtabular}
\end{table}

\subsection{Linear order}
The equivariant maps from $V$ to the quartic multiplets show that the
$\mathbf{20}'$ receives only $J$ and the XXZ anisotropy, while $\Jpp$ and $\Jzp$
do not contribute. The $\mathbf{84}$ receives all couplings, with $\Jpp$
and $\Jzp$ entering through one shared map.

All members of an SU(4) multiplet have the same scaling dimension at the
QED$_3$ fixed point. The spin-2 members reached by spin--orbit coupling are
therefore relevant if and only if the Heisenberg-allowed members are. A
stable Heisenberg DSL requires $\Delta_{\mathbf{20}'}>3$ and
$\Delta_{\mathbf{84}}>3$, and the conformal bootstrap then bounds the lowest
monopole dimension, $\Delta_\Phi>1.046$, from the first
condition~\cite{He2022}. If these conditions hold, the spin--orbit coupling
is irrelevant at linear order, and the spin-rotation-breaking parts of all
correlation functions decay as $(T/J)^{\Delta_{\mathbf{20}'}-3}$ (XXZ) and
$(T/J)^{\Delta_{\mathbf{84}}-3}$ ($\Jpp$, $\Jzp$), so spin-rotation symmetry
is restored at low energy. If instead the Heisenberg DSL is a multicritical
point with $\Delta_{\mathbf{20}'}<3$, the XXZ anisotropy adds one relevant
direction with crossover scale $T_{\rm SO}\sim J|\delta_{\rm eff}|^{1/(3-\Delta_{\mathbf{20}'})}$, with $\delta_{\rm eff}$
the effective anisotropy of Eq.~\eqref{eq:deff} below,
while $\Jpp$ and $\Jzp$ remain irrelevant at linear order. At large $N$,
$\Delta_{\mathbf{20}'}=4-64/(\pi^2N_f)\simeq2.38$ and
$\Delta_{\mathbf{84}}=4+64/(3\pi^2N_f)\simeq4.54$~\cite{Xu2008,Chester2016,He2022}, whereas an $\epsilon$-expansion
estimate quoted in Ref.~\cite{Song2019} gives 3.17 for the Heisenberg $\mathbf{20}'$ quartic. The tuning quartic of the
120$^\circ$--valence-bond transition proposed in Ref.~\cite{Jian2018} has the same large-$N$ dimension 2.38. With the eigenvalue $0.62$ and $|\delta_{\rm eff}|=|c_1(\Delta-1)|$,
$T_{\rm SO}$ is 10--40\,mK for the anisotropy of KYbSe$_2$ and $|c_1|=1$, and 2--80\,mK for
$|c_1|=0.3$--1.6 (Sec.~\ref{sec:scales}). KYbSe$_2$ itself lies on the 120$^\circ$ side of the
transition, where the detuning in $J_2$ sets a larger scale. The estimate states that, even at a
multicritical Heisenberg DSL, an anisotropy of this size acts only below tens of mK. At large $N$
the Heisenberg-allowed $\mathbf{20}'$ vector discussed below is a further relevant direction of that
point.

The multiplet argument covers the Lorentz-scalar quartics, the symmetric
products of the adjoint masses in Table~\ref{tab:quartic}. Classifying all
derivative-free, gauge-neutral quartics of the 16 Majorana zero modes (784
operators) by their SU(4) and Lorentz content, we find that the locked
group also allows nine spin-1 operators with no Heisenberg-invariant
partner in their multiplet: three Lorentz-vector components and one
component of a rank-2 tensor in the $\mathbf{15}$, and five Lorentz-vector
components in the $\mathbf{45}\oplus\overline{\mathbf{45}}$. The classification holds for either sense of the locked rotation. The couplings carry spin 0 and spin 2, so they
cannot reach spin-1 operators at linear order. The equivariant maps from
symmetric powers of the coupling space to these blocks show that they
first arise at second order, through $\Jzp^2$, $\Jpp^2$ and $(\Delta-1)\Jzp$ (and the
corresponding products with second-neighbor couplings), but not through
$(\Delta-1)^2$, $(\Delta-1)\Jpp$ or $\Jpp\Jzp$. These orders follow from a character count, which compares, for each spin, how the lattice
group acts on the $n$-th powers of the coupling vector $\lambda$ and on the operator
block. Their coefficients are of the
same order as the other second-order terms, below $10^{-3}$ in KYbSe$_2$.

To decide whether they are relevant we computed the $O(1/N)$ dimensions of all
784 derivative-free quartics (Appendix~\ref{app:quartic}). Table~\ref{tab:quarticdims} collects
the invariants and the dimensions by SU(4) irrep and Lorentz spin $j$. Operators
containing the U(1) gauge current $\bar\psi\gamma^\mu\psi$ as a factor vanish by the equation
of motion of the gauge field, as for the singlet $(\bar\psi\gamma^\mu\psi)^2$ of
Ref.~\cite{Chester2016}. They span 189 of the 784 dimensions and lie only in the singlet and
the $\mathbf{15}$. The five Lorentz vectors in the $\mathbf{45}\oplus\overline{\mathbf{45}}$ have
$\Delta=4+O(1/N^2)$. The rank-2 tensor in the $\mathbf{15}$ is the product of the gauge current
with an adjoint current and vanishes. Modulo such operators, the three Lorentz vectors
in the $\mathbf{15}$ are the product of the singlet mass $\bar\psi\psi$ with an adjoint current, of
dimension $\Delta_{\bar\psi\psi}+2=4+128/(3\pi^2N_f)\simeq5.08$ at this order. All nine are
irrelevant at large $N$. Every other derivative-free quartic allowed by the locked group
shares an SU(4) multiplet with a Heisenberg-allowed one, so the multiplet argument
covers all derivative-free quartics.

The same table shows that the Heisenberg group itself allows non-scalar
quartics, which the list of lattice-singlet quartics in Ref.~\cite{He2022}, restricted to
Lorentz scalars, does not include: a spatial component of a Lorentz vector and two components
of a rank-2 tensor in the $\mathbf{20}'$. The vector has
$\Delta=4-128/(3\pi^2N_f)\simeq2.92$ at large $N$, below 3, and the tensor $4+O(1/N^2)$. A stable
Heisenberg DSL therefore also requires the $\mathbf{20}'$ vector to be irrelevant at $N_f=4$, a
condition that, as for the scalar $\mathbf{20}'$, the large-$N$ value does not meet. The locked
group adds two spin-2 components to this vector multiplet, which $\Jpp$ and $\Jzp$ reach at
linear order through one shared map, and three to the tensor, one reached by $\Delta-1$ and two by
$\Jpp$, $\Jzp$. These components are relevant only if the Heisenberg-allowed members are.

\begin{table}[t]
\caption{Derivative-free quartics by SU(4) irrep and Lorentz spin $j$: number of invariants of
the Heisenberg group $G_H$ and of the locked group $G_L$, and the
dimension at $O(1/N)$, $\Delta=4+8\lambda/(\pi^2N_f)$, with values at $N_f=4$. ``Null'' marks operators
that contain the U(1) gauge current and vanish by its equation of motion, and the singlet
scalars mix with $F_{\mu\nu}F^{\mu\nu}$~\cite{Chester2016}.}
\label{tab:quarticdims}
\begin{ruledtabular}
\begin{tabular}{lcccl}
irrep & $j$ & $G_H$ & $G_L$ & $\lambda$, $\Delta$\\
\hline
$\mathbf1$ & 0 & 2 & 2 & mix with $F^2$~\cite{Chester2016}\\
$\mathbf1$ & 2 & 1 & 1 & null\\
$\mathbf{15}$ & 1 & 0 & 3 & $16/3$, 5.08\\
$\mathbf{15}$ & 2 & 0 & 1 & null\\
$\mathbf{20}'$ & 0 & 1 & 2 & $-8$, 2.38\\
$\mathbf{20}'$ & 1 & 1 & 3 & $-16/3$, 2.92\\
$\mathbf{20}'$ & 2 & 2 & 5 & 0, 4\\
$\mathbf{45}\oplus\overline{\mathbf{45}}$ & 1 & 0 & 5 & 0, 4\\
$\mathbf{84}$ & 0 & 1 & 4 & $8/3$, 4.54\\
\end{tabular}
\end{ruledtabular}
\end{table}

The large-$N$ value 2.92 lies close to 3, so we computed the dimension of the
$\mathbf{20}'$ Lorentz vector also to two loops in the $\epsilon$ expansion of QED in $d=4-2\epsilon$, following Refs.~\cite{DiPietro2017,DiPietro2018}.
The operator continues to $d=4-2\epsilon$ as
$\bar\Psi_1\Gamma^{(k+1)}_{\mu\nu_1\ldots\nu_k}\Psi_2\,\bar\Psi_3\Gamma^{(k)\,\nu_1\ldots\nu_k}\Psi_4$
with four distinct flavors and $k=0,2$ (or $k=1,3$), and mixes with an infinite tower of
evanescent operators with $k\geq4$, which we treat in the scheme of Ref.~\cite{DiPietro2018}.
With $N_f$ two-component flavors ($N_f/2$ four-component Dirac fermions in $d=4-2\epsilon$),
\begin{equation}
\Delta_{\mathbf{20}'}^{\rm vec}=6-4\epsilon-\frac{15}{N_f}\,\epsilon
+\Big(\frac{5}{2N_f}+\frac{315}{4N_f^2}\Big)\epsilon^2+O(\epsilon^3).
\label{eq:vec_eps}
\end{equation}
The calculation reproduces the
complete two-loop mixing matrix of the nonsinglet Lorentz-scalar operators of
Ref.~\cite{DiPietro2018}. Its $\epsilon^2$ coefficient does not depend on the gauge or on the choice of evanescent
operators, and its $1/N_f$ part, $5/(2N_f)$, agrees with the large-$N$ dimension evaluated
directly in $d=4-2\epsilon$, evanescent tower included. Closed fermion loops do not change the
leading order at $N_f=4$, and at two loops we use the four-flavor continuation, as
Ref.~\cite{DiPietro2018} does for the scalar.

Table~\ref{tab:vec} lists the resulting estimates at $d=3$ and $N_f=4$, together with the same
estimates for the $\mathbf{20}'$ Lorentz scalar. The two-loop $\epsilon$-expansion estimates
range from 2.92 to 3.51 for the vector and from 2.84 to 3.88 for the scalar, so at this order
the $\epsilon$ expansion does not decide whether either operator is relevant. The last two rows,
which add the $1/N_f^2$ part of the $\epsilon^2$ coefficient to the large-$N$ result, lie higher,
between 3.33 and 4.70. Splitting the fixed-order value in powers of $1/N_f$ locates the
spread. For the vector it is $4-1.72+1.23$, and every estimate above 3
depends on the $1/N_f^2$ term, $+1.23$. For the adjoint mass, whose $1/N^2$ term at $d=3$ is
known~\cite{Albayrak2022}, the corresponding $\epsilon^2$ term gives $+0.53$ at $N_f=4$, against an
exact $-0.03$, and adding the $\epsilon^3$ term of Ref.~\cite{DiPietro2017} turns it into $-2.10$.
We therefore take the large-$N$ value 2.92 as the best available estimate for the vector, with an
unknown $1/N^2$ correction, and treat the vector, like the $\mathbf{20}'$ scalar, as marginal, with its
relevance at $N_f=4$ unsettled.

\begin{table}[t]
\caption{Dimensions of the $\mathbf{20}'$ Lorentz vector and Lorentz scalar at $d=3$ and $N_f=4$.
The $\epsilon$-expansion values use $d=4-2\epsilon$ at $\epsilon=1/2$. Pad\'e $[m,n]$ refers to
$\Delta(\epsilon)$, and ``anomalous part'' is a $[0,1]$ Pad\'e of $\Delta-2(d-1)$. The last two rows
add the $1/N_f^2$ part of the $\epsilon^2$ coefficient to the large-$N$ result, directly or as a
$[1/1]$ Pad\'e in $1/N_f$. Scalar values from the two-loop results of Ref.~\cite{DiPietro2018}.}
\label{tab:vec}
\begin{ruledtabular}
\begin{tabular}{lcc}
estimate & vector & scalar\\
\hline
large $N$, $O(1/N)$ & 2.92 & 2.38\\
$\epsilon$, one loop & 2.13 & 1.75\\
$\epsilon$, two loops, fixed order & 3.51 & 3.88\\
$\epsilon$, two loops, Pad\'e $[1,1]$ & 3.15 & 3.17\\
$\epsilon$, two loops, Pad\'e $[0,2]$ & 3.28 & 3.24\\
$\epsilon$, two loops, anomalous part & 2.92 & 2.84\\
large $N$ plus $O(1/N_f^2)$ part of $\epsilon^2$ & 4.15 & 4.70\\
same, $[1/1]$ Pad\'e in $1/N_f$ & 3.50 & 3.33\\
\end{tabular}
\end{ruledtabular}
\end{table}

The operator analysis assumes that the DSL persists, and the couplings also move
its boundaries. $\Jpp$ lowers the classical energy of stripe order at linear
order, and density-matrix studies find that it
removes the spin liquid near $|\Jpp|\simeq0.1J$~\cite{Zhu2017,Zhu2018}, far above the measured
$|\Jpp|/J\approx0.004$.

\subsection{Second order}
\label{sec:seeds}
Second order is where $\Jpp$ and $\Jzp$ first reach the Lorentz-scalar $\mathbf{20}'$, the
channel that can be relevant. We computed the equivariant maps from $\mathrm{Sym}^2(V)$
to the $G_H$ span of the locked invariants of each multiplet (dimension
6 for the $\mathbf{20}'$, 16 for the $\mathbf{84}$) and evaluated them on products
of the physical couplings (Table~\ref{tab:seeds}). The spin-2 component
of the $\mathbf{20}'$ receives exactly three independent combinations,
$J(\Delta-1)$, $(\Delta-1)^2$ and $\Jzp^2-2\Jpp^2$. In this component $\Jpp^2$ and
$\Jzp^2$ enter only through the combination $\Jzp^2-2\Jpp^2$, while $J\Jpp$, $J\Jzp$, $(\Delta-1)\Jpp$,
$(\Delta-1)\Jzp$ and $\Jpp\Jzp$ do not contribute. The $\mathbf{84}$ receives every product except $J^2$.

\begin{table}[t]
\caption{Second-order seeds of the spin--orbit (spin-2) parts of the
quartic multiplets. ``yes'' marks a product of couplings that induces the spin-2 part of the
multiplet at second order, and 0 one that does not.}
\label{tab:seeds}
\begin{ruledtabular}
\begin{tabular}{lcc}
product & $\mathbf{20}'$ & $\mathbf{84}$\\
\hline
$J^2$ & 0 & 0\\
$J(\Delta-1)$, $(\Delta-1)^2$ & yes & yes\\
$J\Jpp$, $J\Jzp$, $(\Delta-1)\Jpp$, $(\Delta-1)\Jzp$ & 0 & yes\\
$\Jpp^2$, $\Jzp^2$ & yes, ratio $1:-\tfrac12$ & yes\\
$\Jpp\Jzp$ & 0 & yes\\
\end{tabular}
\end{ruledtabular}
\end{table}

With $A_\varphi$ the
bond-dependent part of Eq.~\eqref{eq:tensor}, the traceless part of
$\sum_\varphi A_\varphi^2$ is a lattice-invariant spin-2 tensor with $zz$
component proportional to $\Jzp^2-2\Jpp^2$. Character theory shows that
$\mathrm{Sym}^2$ of the spin-2~$\otimes E_2$ multiplet contains the
lattice-trivial spin-2 representation once, so this is the only such
combination. The $zz$ component of a lattice-trivial spin-2 quantity is
invariant under the joint $60^\circ$ rotation of positions and spins about
$z$, which belongs to $G_H$ and maps L$_+$ onto L$_-$, reversing $\Jzp$. Such a
quantity is even in $\Jzp$, which excludes $\Jpp\Jzp$.

The only spin--orbit operator that can be relevant through second order
is therefore the lattice-trivial spin-2 component, fed by the effective
anisotropy
\begin{equation}
\delta_{\rm eff}=c_1(\Delta-1)+c_2(\Delta-1)^2+c_3\frac{\Jzp^2-2\Jpp^2}{J^2},
\label{eq:deff}
\end{equation}
with nonuniversal $c_i$. The bond-dependent couplings act at second order
as a renormalization of the XXZ anisotropy. For the parameters of
Table~\ref{tab:params}, $(\Jzp^2-2\Jpp^2)/J^2\simeq1.7\times10^{-4}$ (KYbSe$_2$) and
$1.2\times10^{-4}$ (NaYbSe$_2$), two orders of magnitude below the best estimates of the anisotropy,
$|\Delta-1|\ge0.02$. The diffuse-scattering value $\Delta=0.98(6)$ of KYbSe$_2$ is also consistent with $\Delta=1$. These
density-functional values omit the magnetic dipole
interaction~\cite{Villanova2023}. For point moments $\mu_B g\bm S$ it adds to every
nearest-neighbor bond $\delta J=-\tfrac12g_{ab}^2D$, $\delta(J\Delta)=g_c^2D$, $\delta\Jpp=-\tfrac34g_{ab}^2D$ and
$\delta\Jzp=0$, with $D=\mu_0\mu_B^2/(4\pi a^3)$ and $a$ the in-plane lattice constant. The point-moment sum gives
$\delta\Jpp/J=-0.011$ in KYbSe$_2$ and $-0.011$ in NaYbSe$_2$, and $\delta\Delta=+0.012$ and
$+0.009$, with $a$ from Ref.~\cite{Xing2021}, the $g$ factors of Table~\ref{tab:params} and
the density-functional $J$. Only the nearest-neighbor dipolar terms are included. The
in-plane sum of $(a/R)^3$ is 1.84 times the nearest-neighbor term, and the longer-range
parts act as further-neighbor couplings. With the dipolar term, $(\Jzp^2-2\Jpp^2)/J^2\simeq-2\times10^{-5}$ in
KYbSe$_2$, of undetermined sign within the error of $g_{ab}$, and $-1.9\times10^{-4}$ in
NaYbSe$_2$, so $|\Jzp^2-2\Jpp^2|/J^2\lesssim2\times10^{-4}$ in both, still two orders below
$|\Delta-1|$. The fits of
Ref.~\cite{Scheie2023} have no separate dipolar term, so their $\Jpp=+0.004(2)J$ is a total
effective value, which differs in sign from the density-functional value with the dipolar
term, $-0.010J$, and the two agree only if the superexchange part is about $+0.016J$, an order
of magnitude above the density-functional estimate. The dipolar shift of $\Delta$ lowers the density-functional
crossover scales of Sec.~\ref{sec:scales} by a factor of 1.4--1.6.

We estimated the ratio $c_3/c_1$ from the spinons. A source $s_a$ for the
quantum-spin-Hall mass, a spin-dependent imaginary hopping $\ii s_a\sigma^a$ on the
bonds oriented counterclockwise around the up triangles, projects purely
on $\beta\sigma^a$, with $\beta$ the Dirac mass matrix,
$\bar\psi\sigma^a\psi=\psi^\dagger\beta\sigma^a\psi$. We define $\delta_{\rm eff}$ as the anisotropy
$(\Gamma_{zz}-\Gamma_{xx})/\Gamma_{\rm iso}$ of the response $\Gamma_{ab}=-\partial^2\Omega/\partial s_a\partial s_b$, with
$\Omega$ the grand potential to second order in the anisotropic exchange
(Fock term, ring, crossed and insertion graphs). The first order gives
$c_1=\Gamma_{\rm iso}J/24>0$ exactly ($0.90$--$0.96$), and $\Jpp$, $\Jzp$ give nothing at first order. At
second order $\Jpp^2$ enters with exactly $-2$ times the coefficient of $\Jzp^2$
and no cross term appears, as required by symmetry, with
$c_3/c_1=3.18$, $3.23$ and $3.28$ at $T/J=0.02$, $0.01$ and $T\to0$.
The ratio does not depend on how
the source is normalized but depends on the saddle amplitude $\chi_0$, defined
by the mean-field hopping $-\chi_0t_{ij}$ (Sec.~\ref{sec:spinondiag}). The coefficient $c_2$
has no scheme-independent value at this level, because it changes at order one
with the scheme, mostly through the renormalization of $\chi_0$ by the XXZ trace. The bond-dependent term is
therefore $3.3(\Jzp^2-2\Jpp^2)/J^2\simeq6\times10^{-4}$ compared with $|\Delta-1|=0.02$--$0.04$ in
KYbSe$_2$ in this estimate.

The spinon estimate is not controlled, and its first-order term has the
wrong sign (below), so we computed $\delta_{\rm eff}$ in the spin model itself. The
staggered vector chirality $K^a$ of Sec.~\ref{sec:critical} carries the quantum
numbers of $\phi_a$ and its Fock decoupling is the quantum-spin-Hall mass. We
take $\delta_{\rm eff}=(\chi_{zz}-\chi_\perp)/\bar\chi$, with $\chi_{ab}$ the static susceptibility of $K^a$
per site, $\chi_\perp=(\chi_{xx}+\chi_{yy})/2$ and $\bar\chi$ the mean of the three components, and
also the same anisotropy of the equal-time correlations $\langle K^aK^a\rangle$.
At $\Delta=1$ the Hamiltonian without $\Jpp$ and $\Jzp$ has only Heisenberg
couplings and is SU(2) invariant. Each bond perturbation is $\bm S_i\cdot Q_b\bm S_j$ with $Q_b$
symmetric and traceless, and the traceless part of $\chi_{ab}$ is a rank-2
tensor, so at second order it is
$\sum_{bb'}G_{bb'}[\{Q_b,Q_{b'}\}/2-\tfrac13\operatorname{tr}(Q_bQ_{b'})]$ with a kernel $G_{bb'}$ of the
SU(2)-invariant Hamiltonian. Its $zz$ component is
$\tfrac43(\Jzp^2-2\Jpp^2)\sum_{bb'}G_{bb'}\cos2(\varphi_b-\varphi_{b'})$, and the cross term
vanishes identically, so the combination of Eq.~\eqref{eq:deff} holds for any
spin-2 quantity, independently of the lattice, whereas $c_3$ itself, and its
ratio to $c_1$, are not fixed by symmetry. We diagonalized the $J_1$--$J_2$ model
with $\Delta=0.98$, $1.02$ and $(\Jpp,\Jzp)=(0,0.06)J$, $(0,0.03)J$, $(0.04,0)J$,
$(0.04,0.06)J$ on 9, 12 and 21 sites (Table~\ref{tab:deffED}). On 21 sites we
used a matrix-free Lanczos method in the full $2^{21}$-dimensional space.
All four anisotropic points give the same $c_3$, including
$(0.04,0.06)J$, where $\Jzp^2-2\Jpp^2$ nearly cancels, and on 21 sites
$(0,0.06)J$ and $(0.04,0)J$ give the same value. Extrapolated to zero coupling, the
ratio of the $\Jpp^2$ to the $\Jzp^2$ coefficient is $-2$, as the spin-2 structure requires, and
the small deviations at the couplings used come from fourth-order terms.

\begin{table}[t]
\caption{Coefficients of Eq.~\eqref{eq:deff} from exact diagonalization:
$c_1=\partial\delta_{\rm eff}/\partial\Delta$ at $\Delta=1$ and $c_3=\delta_{\rm eff}J^2/(\Jzp^2-2\Jpp^2)$, from the
static susceptibility $\chi$ of $K^a$ in the ground state and from the
equal-time correlations $\langle KK\rangle$. On 9 and 21 sites the ground multiplet is
degenerate, $\chi$ at low temperature is dominated by its Curie part, and only
$\langle KK\rangle$ is used. On 9 sites $J_2$ commutes with
the XXZ term. Free spinons give $c_1=+0.96$, $c_3=+3.16$, $c_3/c_1=+3.28$.}
\label{tab:deffED}
\begin{ruledtabular}
\begin{tabular}{llcccc}
$N_s$ & measure & $J_2/J$ & $c_1$ & $c_3$ & $c_3/c_1$\\
\hline
9 & $\langle KK\rangle$ & 0, 0.1 & $-0.62$ & 0.08 & $-0.13$\\
12 & $\chi$ & 0 & $-1.25$ & 2.43 & $-1.9$\\
12 & $\chi$ & 0.1 & $-1.03$ & 3.12 & $-3.0$\\
12 & $\langle KK\rangle$ & 0 & $-0.42$ & 1.06 & $-2.5$\\
12 & $\langle KK\rangle$ & 0.1 & $-0.33$ & 1.56 & $-4.7$\\
21 & $\langle KK\rangle$ & 0 & $-1.60$ & 2.20 & $-1.4$\\
21 & $\langle KK\rangle$ & 0.1 & $-1.06$ & 3.08 & $-2.9$\\
\end{tabular}
\end{ruledtabular}
\end{table}

For every entry of Table~\ref{tab:deffED} $c_1<0$ and $c_3>0$. On 9 sites the susceptibility
at $T=0.02J$ is dominated by matrix elements within the fourfold degenerate ground
multiplet, and its Van Vleck part alone gives $c_1=+0.375$ and $c_3/c_1=+1.0$, so
there the equal-time correlations are the reliable measure. The sign of $c_1$
agrees with the vector-chirality correlations of Sec.~\ref{sec:critical}
(easy-plane anisotropy enhances the $z$ component), whereas the spinon sign is
opposite because their first-order Fock term is repulsive in this channel. The
spinon $c_3$ has the right sign and a similar size. The ratio $c_3/c_1$ ranges
from $-0.13$ on 9 sites to $-4.7$, without a trend in cluster size, so
$|c_3/c_1|\lesssim5$ is an empirical range. With it the bond-dependent
term is at most about 5\% of $\delta_{\rm eff}$ in KYbSe$_2$ for $|\Delta-1|\ge0.02$, also for the bound
$|\Jzp|/J\lesssim0.015$ from diffuse scattering~\cite{Scheie2023,Paddison2020}, and it changes $T^*$ by a few
percent. A positive $\Jzp^2-2\Jpp^2$ acts like a reduction of $\Delta$, toward the
easy-plane side.

\section{Monopoles}
\label{sec:monopoles}

The six $2\pi$ monopoles of the triangular DSL consist of three spin
singlets $\Phi_{1,2,3}$ and a spin triplet $\Phi_{4,5,6}$. Under $\mathcal T$ the
singlets are even and the triplets odd~\cite{Song2019,Song2020}. Their
transformation under $G_H$ is given in Table~III of
Ref.~\cite{Song2020}. We recomputed the translation Berry phases by flux
insertion on tori up to $L=10$, including the background gauge charge of $-1$ per
site. The gauge transformation that accompanies each translation is fixed by angular
synchronization, which takes the site phases from the leading eigenvector of the
matrix of bond-phase mismatches between the translated and the original Hamiltonian
(Appendix~\ref{app:methods}). The triplet momenta are $(k_1,k_2)=(-2\pi/3,+2\pi/3)$ in our
convention. Combined with
the valley translation signs of the projected group they reproduce the
translation columns of Table~III of Ref.~\cite{Song2020}. The overall sign of the momenta
depends on the sign of the flux.

Translations do not act on spin, so their Berry phases survive spin--orbit
coupling unchanged. Every charge-0 product of monopoles and bilinears
carries $T_1$ phase $0$ or $\pi$, so a charge-$q$ operator has $T_1$ phase
$q\pi/3$ modulo $\pi$. Terms with $q$ not divisible by 3 are
excluded with any dressing. At charge 3 there are 56 cubic monomials,
twenty of which are translation invariant. At the Heisenberg point only
$\Phi_1\Phi_2\Phi_3+\mathrm{h.c.}$ survives. Restricting Table~III of
Ref.~\cite{Song2020} to $G_L$ leaves the same single term, for either sense
of the locked rotation relative to the table. This restriction assumes
that the anisotropic DSL is adiabatically connected to the Heisenberg DSL,
since the time-reversal and reflection phases of the monopoles were fixed
using exact SO(3) symmetry~\cite{Song2020}. Without time reversal the locked
group would admit two further charge-3 combinations, $(\Phi_4^3-3\Phi_4\Phi_5^2)-\mathrm{h.c.}$
and one built from $\Phi_i^2$ and the in-plane triplet components $\Phi_{4,5}$, but their
Hermitian forms are odd under $\mathcal T$. Spin--orbit coupling
adds no monopole term through charge 3.

\subsection{Interlayer coupling}
\label{sec:interlayer}
Stacking adds monopole tunneling between layers, which is relevant already for
Heisenberg exchange. In the three-dimensional crystals each layer carries its own emergent gauge
field, and the interlayer exchange $J_\perp$ generates the gauge-invariant
monopole tunneling $\Phi_a^{(l)}\Phi_b^{(l+1)\dagger}+\mathrm{h.c.}$ between adjacent layers. We
classified these couplings with the monopole table of Ref.~\cite{Song2020},
defining the monopoles of layer $l+1$ through a three-dimensional symmetry
that maps layer $l$ onto it, the rhombohedral centering in R$\bar3$m and the
$6_3$ screw in P6$_3$/mmc, so that every operation acts on a layer as an
element of its layer group. In R$\bar3$m the three nearest neighbors of a Yb
ion in the next layer occupy the three sublattices of the 120$^\circ$
pattern [Fig.~\ref{fig:structure}(b)], so isotropic exchange has no Fourier component at the K point. The
Heisenberg group then allows a single coupling, singlet tunneling
$\Phi_1\Phi_1'^\dagger+e^{-\ii\pi/3}\Phi_2\Phi_2'^\dagger+e^{\ii\pi/3}\Phi_3\Phi_3'^\dagger$, with primes for layer $l+1$ and phases
that depend on the convention for that layer, generated at order $J_\perp^2$.
The locked group adds two spin-2 triplet couplings, with $m=\pm2$ and $\mp1$,
generated at first order by the traceless part of the interlayer exchange
tensor, which has a nonzero component at the K point. In P6$_3$/mmc the layers are
stacked directly [Fig.~\ref{fig:structure}(c)], and already the Heisenberg group allows triplet tunneling
of spin 0 at first order in $J_\perp$, besides singlet tunneling, and the locked
group adds a spin-2 and a spin-1 coupling. Neither nearest interlayer bond
admits Dzyaloshinskii--Moriya exchange. The couplings generated at first and second order in
$J_\perp$ lie in these invariant spaces. The tunneling operators have dimension $2\Delta_\Phi$, below 3 at
large $N$, so a stack of Heisenberg DSLs is unstable below a scale set by
$J_\perp$. At the DSL--120$^\circ$ critical point the spin--orbit triplet channel of
R$\bar3$m, of dimension $2\times0.78$ at leading order, is more relevant than
the singlet channel, $2\times1.06$. In the rhombohedral compounds spin--orbit
coupling thus opens an interlayer channel, at lower order in $J_\perp$, that
the Heisenberg model lacks.

The size of this channel is not known for the superexchange,
but the magnetic dipole interaction gives a lower estimate. Its isotropic
part also cancels at the K point in R$\bar3$m, and summed over the adjacent layer its
traceless part at the K point is $2.2\times10^{-4}$, $3.4\times10^{-4}$ and $1.1\times10^{-3}$\,meV in KYbSe$_2$,
NaYbSe$_2$ and NaYbO$_2$, or $(5$--$21)\times10^{-4}J$. Further layers contribute less
than $10^{-6}$\,meV. With the eigenvalue $3-2\Delta_\Phi$ the crossover
$T_\perp\sim J(|J_\perp(\mathrm K)|/J)^{1/(3-2\Delta_\Phi)}$ is 26, 37 and 82\,mK at the critical point
($\Delta_\Phi=0.78$) and 2--10\,mK in the DSL with the next-to-leading-order large-$N$
value $\Delta_\Phi=1.02$, a scaling estimate with unknown prefactors. Because it couples the
triplet monopoles of adjacent layers, the spin--orbit channel favors
three-dimensional 120$^\circ$ order over valence-bond order.

Interlayer exchange also couples fermion bilinears of adjacent layers at first order. The
spin at the M points couples to the $\mathcal T$-odd spin--valley masses, and in R$\bar3$m the
form factor $\sum_je^{\ii\mathrm M\cdot\bm r_j}$ of the three neighbors in the next layer has unit modulus at
all three M points, so already Heisenberg $J_\perp$ generates
$J_\perp(\mathrm M)\,\bar\psi\sigma^a\tau^b\psi^{(l)}\,\bar\psi\sigma^a\tau^b\psi^{(l+1)}$. Its dimension $2\Delta_{\rm adj}$ is 2.92 at large $N$,
and 2.8 for the lattice value $\Delta_{\rm adj}\simeq1.4$ quoted in Ref.~\cite{He2022}, so the coupling is nearly
marginal, with eigenvalue 0.08--0.2, and its crossover scale
$J(|J_\perp(\mathrm M)|/J)^{1/(3-2\Delta_{\rm adj})}$ is negligible for any $|J_\perp(\mathrm M)|\ll J$. The monopole
channels, with eigenvalues near 1, dominate.

\begin{table*}[t]
\caption{Scaling dimensions used in this paper, at $N_f=4$ two-component flavors. ``LO'' and
``NLO'' denote leading and next-to-leading order in $1/N$, and $\epsilon$ denotes the expansion in
$d=4-2\epsilon$. The last column gives the quantity each dimension controls.}
\label{tab:dims}
\begin{ruledtabular}
\begin{tabular}{p{0.23\textwidth}p{0.21\textwidth}p{0.06\textwidth}p{0.15\textwidth}p{0.29\textwidth}}
operator & large-$N$ form & value & other estimate & controls\\
\hline
adjoint masses $\bar\psi\sigma^a\psi$, $\bar\psi\tau^b\psi$, $\bar\psi\sigma^a\tau^b\psi$ & $2-64/(3\pi^2N_f)$, LO~\cite{Hermele2005,Hermele2007} & 1.46 & 1.4(2) lattice~\cite{He2022} & M-point weight, $\alpha_{\rm M}=0.08$; random DM eigenvalue $4-2\Delta=1.08$\\
singlet mass $\bar\psi\psi$ & $2+128/(3\pi^2N_f)$, LO~\cite{Hermele2007,Chester2016} & 3.08 & 2.40--2.48, $\epsilon$~\cite{DiPietro2017} & field-induced chiral mass, $\kappa_{xy}$\\
SU(4) currents & exact & 2 & & Harris criterion for random currents\\
U(1) current $\bar\psi\gamma^\mu\psi$ & exact (gauge equation of motion) & 3 & & null quartics\\
tilt $\mathcal O_R$ & $3+64/(5\pi^2N_f)$, LO & 3.32 & $>3$ by unitarity~\cite{Minwalla1998} & spin--momentum locking, irrelevant\\
$\mathbf{20}'$ Lorentz scalar & $4-64/(\pi^2N_f)$, LO~\cite{Xu2008,Chester2016} & 2.38 & 3.17, $\epsilon$~\cite{Song2019} & XXZ channel; stability of the Heisenberg DSL\\
$\mathbf{20}'$ Lorentz vector & $4-128/(3\pi^2N_f)$, LO (this work) & 2.92 & 2.9--3.5, $\epsilon$ (this work) & Heisenberg-allowed; stability of the Heisenberg DSL\\
$\mathbf{20}'$ tensor, $\mathbf{45}\oplus\overline{\mathbf{45}}$ vectors & $4+O(1/N^2)$ & 4 & & six of the nine spin-1 quartics\\
$\mathbf{15}$ Lorentz vector & $4+128/(3\pi^2N_f)$, LO (this work) & 5.08 & & three of the nine spin-1 quartics\\
$\mathbf{84}$ Lorentz scalar & $4+64/(3\pi^2N_f)$, LO~\cite{Xu2008,Chester2016} & 4.54 & & $\Jpp$, $\Jzp$ at linear order\\
monopole $\Phi$ in the DSL & $0.265N_f-0.0383$, NLO~\cite{Pufu2014,Dyer2013} & 1.02 & $>1.046$ if stable~\cite{He2022}; 1.26(8) lattice & $\alpha_{\rm K}=3-2\Delta_\Phi$; interlayer tunneling $2\Delta_\Phi$; random valence-bond field $4-2\Delta_\Phi$\\
triplet $\Phi$ at the DSL--120$^\circ$ point & LO~\cite{Dupuis2019} & 0.78 & & $\alpha_{\rm K}=1.44$; $S^\pm$ below $T^*$ (easy plane)\\
$\Phi_z$ and singlets at the chiral-Ising point & QED$_3$ value, LO~\cite{Dupuis2019,Dupuis2021} & 1.06 & 1.09 with our NLO estimate & $S^{zz}$ below $T^*$; $R(\omega)\propto\omega^{-0.56}$ ($-0.62$)\\
spin-2 bilinear $T_{ab}$ of $\phi$ & $2\Delta_\phi=2$, LO & 2 ($y_T=1$) & $y_T=0.74$--$0.97$, $\epsilon$ & crossover scale $T^*$\\
\end{tabular}
\end{ruledtabular}
\end{table*}

\subsection{Scaling dimensions}
\label{sec:dims}
Table~\ref{tab:dims} collects the scaling dimensions used in this paper, with the order in
$1/N$ at which each is known and the quantity it controls. Only the conserved currents are
exact. The large-$N$ values at $N_f=4$ are not controlled, and where another estimate exists it
is listed. The stability of the Heisenberg DSL rests on $\Delta_{\mathbf{20}'}$, the
$\mathbf{20}'$ vector and $\Delta_\Phi$. For $\Delta_{\mathbf{20}'}$ and $\Delta_\Phi$ the large-$N$
values and the other estimates disagree on the side of 3 (or of the bound 1.046), and for the
vector the two-loop $\epsilon$-expansion estimates lie on both sides of 3 (Table~\ref{tab:vec}).

\section{The transition to \texorpdfstring{120$^\circ$}{120\textdegree} order}
\label{sec:critical}

\subsection{Critical theory and symmetry content}
Condensing the quantum-spin-Hall mass channel of the DSL and proliferating
the spin-triplet monopoles produces coplanar 120$^\circ$ order~\cite{Song2019,Song2020}. The
transition is described by the QED$_3$ chiral-Heisenberg
Gross--Neveu--Yukawa theory, in which a three-component scalar $\phi_a$
couples to $\bar\psi\sigma^a\psi$~\cite{Ghaemi2006,Dupuis2019}. The field $\phi_a$ carries
the quantum numbers of the quantum-spin-Hall masses (Table~\ref{tab:masses}):
spin 1, $C_6$ odd, $\mathcal T$ even, momentum zero. Its symmetric products
$\phi_a\phi_b$ contain spin 0 and spin 2, both lattice trivial. By the same
counting as in Sec.~\ref{sec:quartic}, the XXZ anisotropy couples linearly
to the traceless spin-2 bilinear $T_{ab}=\phi_a\phi_b-\tfrac13\delta_{ab}\phi^2$, while
$\Jpp$ and $\Jzp$ (lattice $E_2$) do not couple to it at linear order and enter
only through Eq.~\eqref{eq:deff}.

\subsection{Crossover exponent}
\begin{figure}[t]
\includegraphics[width=\columnwidth]{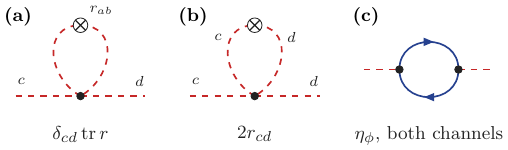}
\caption{One-loop graphs for the mass operator $r_{ab}\phi_a\phi_b$ (crossed
circle) at the chiral-Heisenberg point. Dashed lines are $\phi_a$ and solid
lines spinons. (a) Contraction with the loop index traced, $\delta_{cd}\operatorname{tr}r$,
present only for the singlet. (b) Contraction with open indices, $2r_{cd}$,
present in both channels. (c) Yukawa self-energy of $\phi_a$, which gives the
anomalous dimension $\eta_\phi$ common to both channels. Graph (a) gives
$y_T-y_S=4N_bu^*$.}
\label{fig:yT}
\end{figure}

Whether the XXZ anisotropy is relevant at the critical point depends on the
eigenvalue $y_T$ of the spin-2 bilinear. We use the one-loop renormalization-group equations of
Ref.~\cite{Dupuis2019} for four two-component flavors ($2N_f=4$ in their
notation) and $N_b=3$ components of $\phi_a$ (their Eqs.~(88)--(90), with
the fixed point of their Table~II), which reproduce
$\nu^{-1}=2-4.577\epsilon$ and $\nu=\tfrac12+1.144\epsilon$, with $\epsilon=4-D$ and $D$ the
spacetime dimension [their Eqs.~(122)--(125)]. A mass term $r_{ab}\phi_a\phi_b$ renormalizes through the quartic
vertex $u(\phi^2)^2$ with the tensor structure $\delta_{cd}\operatorname{tr}r+2r_{cd}$
[Fig.~\ref{fig:yT}(a,b)]. The traced contraction exists only for the singlet,
so the singlet receives $4(N_b+2)u^*$ and the traceless part $8u^*$. The
Yukawa self-energy of $\phi_a$ [Fig.~\ref{fig:yT}(c)], and through it the gauge
field, is the same in both channels at this order. The eigenvalues are
therefore related by
\begin{equation}
y_T=y_S+4N_bu^*=2-3.391\epsilon .
\end{equation}
Inverting the series as done for $\nu$ in Ref.~\cite{Dupuis2019} gives
$y_T\simeq0.74$ at $\epsilon=1$, and at large $N$ the spin-2 bilinear has
$\Delta_T=2$, $y_T=1$. Evaluated naively at $\epsilon=1$ the series gives
$y_T=-1.39$, so the value depends on the resummation. At one loop,
however, $y_T-y_S=4N_bu^*>0$ and the tuning eigenvalue $y_S=1/\nu$ is
positive, and these signs do not depend on the resummation. The XXZ
anisotropy is therefore relevant at the critical point. The crossover exponent
$\phi_T=\nu y_T=1+0.593\epsilon$, the combination usually resummed in O($N$) models, gives
$y_T=\phi_T/\nu\simeq0.97$ at $\epsilon=1$ with the resummed $\nu=1.64$, so the three estimates span
$y_T\simeq0.74$--1.

The field $\phi_a$ is the direction of the quantum-spin-Hall mass. A mass
$\bar\psi\sigma^z\psi$ makes the low-energy monopoles $\Phi_4\pm\ii\Phi_5$ (with
$\Phi_{4,5,6}\equiv\Phi_{x,y,z}$) carry $S^z=\pm1$~\cite{Song2020}, so their condensation
produces 120$^\circ$ order in the plane perpendicular to $\phi$, and $\phi$ is the
plane normal of the ordered state. Easy-plane anisotropy ($\Delta<1$) favors
spins in the $ab$ plane, $\bm m\parallel c$ (Sec.~\ref{sec:order}), and selects the
single component $\phi_z$, so the system crosses over from chiral-Heisenberg
to QED$_3$ chiral-Ising Gross--Neveu--Yukawa criticality, the theory with the traceless mass
$\bar\psi\sigma^z\psi$ studied in the $\epsilon$ expansion in Ref.~\cite{Zerf2018} (the large-$N$ theory of
Ref.~\cite{Boyack2019} couples the scalar to the singlet mass). Easy-axis
anisotropy ($\Delta>1$) selects the in-plane pair $(\phi_x,\phi_y)$ and leads to
QED$_3$ chiral-XY criticality, the QED$_3$--Gross--Neveu--XY class~\cite{Janssen2020}. Willsher
and Knolle assign the easy-plane case to the in-plane channels
$(\bar\psi\sigma^{x}\psi)^2+(\bar\psi\sigma^{y}\psi)^2$, that is, to the chiral-XY class, taking the
in-plane magnetic order as the order parameter~\cite{WillsherKnolle2025}.
Because $\phi$ is the plane normal, condensing $\phi_z$ keeps the U(1) spin
rotation about $c$ of the XXZ model and gives in-plane 120$^\circ$ order through
the monopoles $\Phi_4\pm\ii\Phi_5$, whereas condensing $\phi_x$ or $\phi_y$ puts the
ordered plane through the $c$ axis, against the easy-plane energy, and
would require a reorientation inside the ordered phase.

This selection can be tested in the spin model itself. The simplest spin
operator with the quantum numbers of $\phi_a$ (spin 1, $\mathcal T$ even, $C_6$ odd,
translation invariant) is the staggered vector chirality
\begin{equation}
K^a=\sum_\triangle\kappa^a_\triangle-\sum_\triangledown\kappa^a_\triangledown=2\sum_{\langle ij\rangle_\circlearrowleft}(\bm S_i\times\bm S_j)^a ,
\label{eq:K}
\end{equation}
with $\kappa_\triangle$ the counterclockwise vector chirality of a triangle and the last
sum over nearest-neighbor bonds oriented counterclockwise around the up
triangles. Its Fock decoupling on the flux state,
$\tfrac12(\ii\chi_{ij}f_j^\dagger\sigma^af_i+\mathrm{h.c.})$ on the same oriented bonds, projects
purely on the quantum-spin-Hall mass $\beta\sigma^a$, and a
classical 120$^\circ$ state with plane normal $\bm n$ has
$\langle\bm K\rangle/N_s=\pm(3\sqrt3/4)\bm n$. Table~\ref{tab:K} gives the ground-state ratio
$\langle K^zK^z\rangle/\langle K^xK^x\rangle$ of the $J_1$--$J_2$ XXZ model on clusters of 9 and 12 sites
(full diagonalization) and 21 sites (Lanczos). On 9 and 12 sites several
second-neighbor vectors join the same pair of sites through the periodic
boundaries. We add $J_2$ once per pair, so as not to multiply the coupling between
the same two spins, although this undercounts $J_2$ relative to the infinite
lattice. With this convention the ratio has the same sign at all $J_2$ studied and
at $T\le0.2J$, and on 9 sites it does not depend on an isotropic $J_2$, which there
reduces to a function of the sublattice total spins. On 21 sites each
second-neighbor vector joins a distinct pair, so no convention is needed, and the
ratio exceeds 1 for $\Delta=0.5$ and 0.9 and is below 1 for $\Delta=1.1$ at every $J_2$
studied, for isotropic and for XXZ-anisotropic $J_2$. Easy-plane anisotropy thus
enhances the $z$ correlations on every cluster and selects $\phi_z$, the reverse of
the selection of Ref.~\cite{WillsherKnolle2025} for both signs of $\Delta-1$. Adding
$J_2$ once per vector instead reverses the 12-site sign for an XXZ-anisotropic
$J_2\ge0.1J$ (1.0005 at $\Delta=1.1$, $J_2=0.1J$, and 0.93 at $\Delta=0.5$, $J_2=0.125J$).

\begin{table}[t]
\caption{Ground-state ratio $\langle K^zK^z\rangle/\langle K^xK^x\rangle$ of the staggered vector
chirality, Eq.~\eqref{eq:K}, for the $J_1$--$J_2$ XXZ model with isotropic $J_2$ added
once per pair of sites. The ratio is 1 at $\Delta=1$.}
\label{tab:K}
\begin{ruledtabular}
\begin{tabular}{ccccc}
sites & $J_2/J$ & $\Delta=0.5$ & $\Delta=0.9$ & $\Delta=1.1$\\
\hline
9 & any & 1.468 & 1.067 & 0.943\\
12 & 0 & 1.304 & 1.045 & 0.960\\
12 & 0.1 & 1.213 & 1.034 & 0.969\\
12 & 0.125 & 1.184 & 1.031 & 0.971\\
21 & 0 & 2.533 & 1.185 & 0.860\\
21 & 0.1 & 1.730 & 1.114 & 0.902\\
21 & 0.125 & 1.541 & 1.095 & 0.914\\
\end{tabular}
\end{ruledtabular}
\end{table}

The spinon estimate of Sec.~\ref{sec:seeds} gives $c_1>0$, so at
that level easy-plane anisotropy enhances the in-plane components of the
quantum-spin-Hall response, as in the mean-field theory of
Ref.~\cite{WillsherKnolle2025}. In the free-spinon theory, however, the Fock
interaction in the quantum-spin-Hall channel is repulsive, so this
response describes no ordering tendency and does not fix the sign of the
coupling of $\phi_a$ to the anisotropy at the critical point. In the spin model the
equal-time correlations of $K^a$ on 9, 12 and 21 sites and its static susceptibility on 12
sites (Table~\ref{tab:deffED}) give $c_1<0$, the sign of the correlations in Table~\ref{tab:K}. The two
selections predict opposite spin-resolved spectra at the K point
(Sec.~\ref{sec:predictions}).

The crossover occurs at the scale
\begin{equation}
T^*\sim J\,|\delta_{\rm eff}|^{1/y_T},\qquad y_T\simeq0.74\text{--}1 ,
\label{eq:Tstar}
\end{equation}
with $\delta_{\rm eff}$ from Eq.~\eqref{eq:deff} and $|c_1|$ of order one (0.3--1.6 on the clusters), so $T^*$ carries an
order-of-magnitude uncertainty from $c_1$ besides the range of $y_T$. The formula
assumes $|\delta_{\rm eff}|\ll1$. For the easy-axis case the exponents of the chiral-XY class cannot be
obtained reliably from the four-loop $\epsilon$ expansion. Reference~\cite{Zerf2020}
counts four-component flavors, so the triangular DSL corresponds to their
$N_f=2$. The Pad\'e approximants of their series reproduce their
tabulated values at their $N_f=4$ to all quoted digits but scatter widely at $N_f=2$
($1/\nu$ between $-1.0$ and $0.75$). We state the universality
classes without exponents.

\subsection{Monopole dimension and the measured exponents}
For an operator of dimension $\Delta_{\rm K}$ at the K point the dynamic
structure factor scales as $S(\mathrm K,\omega)\propto\omega^{2\Delta_{\rm K}-3}$, so the
measured exponent $\alpha$ of $S\propto\omega^{-\alpha}$ gives
$\Delta_{\rm K}=(3-\alpha)/2$. The K-point operators of the DSL are the triplet
monopoles. At next-to-leading order in $1/N_f$ their dimension is $0.265N_f-0.0383\simeq1.02$ in
the DSL~\cite{Pufu2014,Dyer2013,Song2019,He2022}, and at leading order it is $0.78$ at the
chiral-Heisenberg critical point~\cite{Dupuis2019}. Variational studies of the
$J_1$--$J_2$ model find the charge-1 monopoles gapless, with the predicted
momenta~\cite{Budaraju2025}. The conformal-bootstrap bound for a
stable DSL is $\Delta_\Phi>1.046$~\cite{He2022}, and the conformal bootstrap of Ref.~\cite{Albayrak2022}
excludes $\Delta_\Phi\simeq1.02$ if the lowest scalar of the $\mathbf{20}'$ sector is irrelevant.

\emph{Spin-resolved exponents below $T^*$.}---Below $T^*$ the pinned spin-Hall mass
splits the triplet monopoles, so the in-plane and out-of-plane spectra at the K point
acquire different exponents. At the chiral-Heisenberg point the
monopoles are organized by their total spin $s$. At leading order in $1/N$
the triplet ($s=1$) has $\Delta=0.78$ and the singlets ($s=0$), whose saddle has
no spin-Hall mass, have the QED$_3$ value $1.06$~\cite{Dupuis2021}. The three
components of the triplet are degenerate by spin-rotation symmetry, and the
$m_s=0$ member is a rotation of the maximally polarized state, whose
spin-Hall mass is antialigned with its spin. Below $T^*$ on the easy-plane side
only $\phi_z$ is critical, the spin-Hall mass is pinned along $z$, and this
rotation is no longer available. In the monopole background with mass $\pm M$
along $z$, the spin-up zero modes have energy $+M$ and the spin-down zero
modes $-M$, and two of the four are filled. Filling both spin-down or both
spin-up zero modes gives $\Phi_x\mp\ii\Phi_y$ ($S^z=\mp1$) with the saddle of
Ref.~\cite{Dupuis2019}, $\Delta=0.78156$ [their Eq.~(44)]. Filling one of each gives $\Phi_z$ and the
three singlets, a sector whose free energy is even in $M$, with positive
curvature and a global minimum at $M=0$, so its saddle has $M=0$ and
$\Delta=4\times0.26510=1.06038$, the QED$_3$ value~\cite{Dupuis2019,Dupuis2021}. The assignment of $\Phi_z$ to this
sector is ours. At leading order the components carrying $S^\pm(\mathrm K,\omega)$ and
$S^{zz}(\mathrm K,\omega)$ therefore have $\alpha_\pm=1.437$ and $\alpha_{zz}=0.879$, a
difference of 0.558. Applying the next-to-leading-order kernels of
Ref.~\cite{Dupuis2022} with one Gross--Neveu channel,
$\Delta^{(1)}=\Delta^{(1)}_{\rm QED_3}+\Delta^{(1)}_{\rm GN}$ with $\Delta^{(1)}_{\rm QED_3}=-0.038$ and the pure-scalar part
$\Delta^{(1)}_{\rm GN}=0.070$, half the difference between their QED$_3$--$\mathbb Z_2$ and QED$_3$ values (the
scalar--gauge mixing term vanishes for a traceless vertex, and the scalar kernel is the same for a
singlet and a $\sigma^z$ vertex at the $M=0$ saddle), raises $\Delta(\Phi_z)$ to about 1.09
($\alpha_{zz}\simeq0.82$) in our estimate. The correction to the polarized
states is not known and could be comparable to the predicted difference of exponents, so the
sign of $d\ln R/d\ln\omega$ is the controlled statement. On the easy-axis
side, with $(\phi_x,\phi_y)$ critical, the leading order leaves the triplet
degenerate. Quantizing the in-plane orientation of the spin-Hall mass as a
rotor with moment of inertia $N_f\chi_m/2$, where $\chi_m=0.990$ is the spin
susceptibility per valley in the monopole background and $N_f=4$ counts the two-component
flavors of the two valleys, places
$\Phi_x\pm\ii\Phi_y$ above $\Phi_z$ by $1/(N_f\chi_m)\simeq0.25$, so that $S^{zz}$ becomes the more
singular spectrum, a semiclassical estimate of a formally
subleading effect. These numbers, which reproduce Table~I of Ref.~\cite{Dupuis2019},
make the spin-resolved test of the easy-plane selection quantitative.
They predict a difference $\alpha_\pm-\alpha_{zz}\approx0.6$ in NaYbSe$_2$ if $\Delta\simeq0.5$, and the
opposite ordering for the chiral-XY selection.

The difference is measured most directly by longitudinal polarization
analysis at the K point, with the neutron polarization $\bm P\parallel c$ and $\bm Q$ in the $ab$
plane. The non-spin-flip channel then sees $g_c^2S^{zz}(\mathrm K,\omega)$ and the
spin-flip channel $g_{ab}^2S^{tt}(\mathrm K,\omega)$, with $t$ the in-plane direction normal
to $\bm Q$, and $S^{tt}=S^{xx}$ because the in-plane tensor at the K point is isotropic under $C_3$.
The $g$-normalized ratio
\begin{equation}
R(\omega)=\frac{g_c^2}{g_{ab}^2}\,\frac{I_{\rm SF}}{I_{\rm NSF}}=\frac{S^{xx}(\mathrm K,\omega)}{S^{zz}(\mathrm K,\omega)}\propto\omega^{-(\alpha_\pm-\alpha_{zz})}
\label{eq:Rw}
\end{equation}
grows toward low energy as $\omega^{-0.56}$ (as $\omega^{-0.62}$ with our $1/N$ estimate)
for $k_BT\lesssim\hbar\omega\lesssim k_BT^*$, independently of the nonuniversal amplitudes
of the two components. The exponent is a leading-order large-$N$ value, and for
$\Delta\simeq0.5$, where Eq.~\eqref{eq:Tstar} does not strictly apply, $k_BT^*$ is of order $J$. The
crossover formula gives 0.14--0.23\,meV for NaYbSe$_2$ with the $J$ of the same fits. Independently of these numbers, $S^{zz}(\mathrm K,\omega)$ stays
gapless, and $R$ grows toward low energy for the chiral-Ising selection and falls for the chiral-XY one.
In NaYbSe$_2$ the incoherent elastic line hides the spectrum below about
50\,$\mu$eV at an incident energy $E_i=1$\,meV, and exchange randomness of about 35\,$\mu$eV has been
inferred from the 40\,mK susceptibility peak~\cite{Scheie2024arXiv}. At 0.1\,K the
window 0.06--0.15\,meV then gives a rise of $R$ by a factor 1.7--1.8 at this order, with an
energy resolution of about 20--30\,$\mu$eV. With $J=0.551$\,meV~\cite{Scheie2024} the window is
0.11--0.27$J$, and with the $J=0.39$--0.46\,meV of the fits that give $\Delta\simeq0.5$ it is
0.13--0.38$J$. The window lies within a factor of
a few of $k_BT^*$, so the exponents are not fully asymptotic there. With
$g_{ab}/g_c=3.13/1.01$~\cite{Ranjith2019} the non-spin-flip intensity is about a tenth of
the spin-flip intensity, and a fifth with the Curie--Weiss values 2.81/1.27~\cite{Ranjith2019}, so with a
flipping ratio $F$, the ratio of correctly to wrongly analyzed neutrons, the leakage of the spin-flip signal into the non-spin-flip channel
is about $10/F$ (or $5/F$) of the latter and must be corrected, so keeping it below a third of the
signal requires $F\gtrsim15$--30. The counting time is set by the non-spin-flip channel, whose
rate is a few $10^{-3}$ of the unpolarized rate once polarizer, analyzer, the finer resolution and
the factor $g_c^2/g_{ab}^2$ are included, so the measurement needs samples of order a gram and
counting times of order weeks on a polarized time-of-flight spectrometer.

\begin{figure*}[t]
\includegraphics[width=0.96\textwidth]{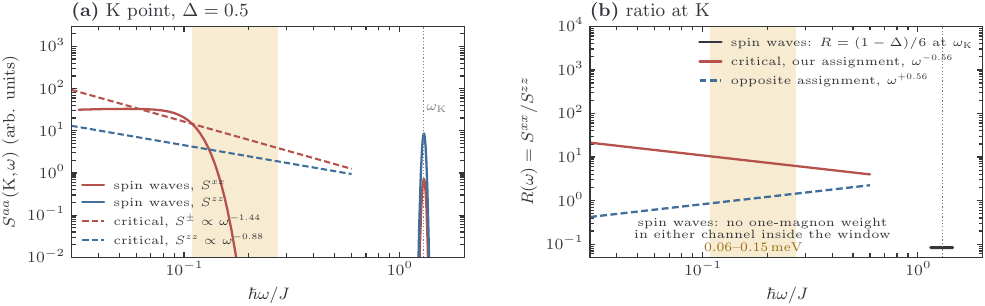}
\caption{The spin-resolved test at the K point. (a) $S^{xx}(\mathrm K,\omega)$ and $S^{zz}(\mathrm K,\omega)$ for
$\Delta=0.5$ from linear spin waves about the 120$^\circ$ state, integrated over a momentum window
$|\bm p|<0.1/a$ around the K point with a Gaussian resolution of $0.02J$ (solid), and the critical power laws
$\omega^{-1.44}$ and $\omega^{-0.88}$ of the chiral-Ising regime (dashed, common
amplitude). The spin waves put all $S^{zz}$ weight at the out-of-plane gap $\omega_{\rm K}=1.30J$,
and the in-plane branch within the window reaches only $\hbar\omega\simeq0.1J$. (b) The ratio $R=S^{xx}/S^{zz}$
of Eq.~\eqref{eq:Rw}, which grows toward low energy for the chiral-Ising selection and falls for the chiral-XY one,
whereas the ordered state gives no one-magnon weight in either channel inside the proposed
window (shaded, $0.06$--$0.15$\,meV for $J=0.551$\,meV) and $R=(1-\Delta)/6$ at the gap.}
\label{fig:Rw}
\end{figure*}

A conventional ordered state gives a different $R(\omega)$. In linear spin-wave
theory about the coplanar 120$^\circ$ state, $S^{zz}(\mathrm K,\omega)$ has weight only at the
out-of-plane mode, gapped at $\omega_{\rm K}=S\sqrt{J_{\rm in}(\mathrm K)J_{\rm out}(\mathrm K)}=\tfrac12\sqrt{13.5(1-\Delta)}\,J$ for $J_2=0$
($1.30J$ for $\Delta=0.5$), with weight $(S/2)\sqrt{J_{\rm in}(\mathrm K)/J_{\rm out}(\mathrm K)}=0.43$ there, while $S^{xx}(\mathrm K,\omega)$ contains the
in-plane Goldstone mode. Exactly at the K point both components vanish between zero energy
and $\omega_{\rm K}$. Over a finite window of $\bm Q$ around the K point, $S^{xx}$ collects the dispersing in-plane
mode while $S^{zz}$ has no one-magnon weight below $\omega_{\rm K}$, so $R$ is large below $\omega_{\rm K}$
and steps down to $(1-\Delta)/6$ at the gap, with no power law. Here
$J_{\rm in}(\bm k)=[J_{xy}(\bm k+\bm Q)+J_{xy}(\bm k-\bm Q)]/2-J_{xy}(\bm Q)$ and $J_{\rm out}(\bm k)=J_z(\bm k)-J_{xy}(\bm Q)$
are the in-plane and out-of-plane stiffnesses in the local frame, with $J_{xy}(\bm k)$ and $J_z(\bm k)$
the Fourier transforms of the in-plane and $zz$ exchange and $\bm Q$ the ordering vector K.
For $\Delta<0.978$ the gap lies above the window $0.11$--$0.27J$ ($J=0.551$\,meV) of the proposed
measurement and $S^{zz}$ has no one-magnon weight in it, whereas for $0.978<\Delta<0.996$
the gap falls inside the window. Finite resolution turns the step into an
apparent slope that depends on the resolution, between $-1.3$ and $-4.4$ for
typical settings. Exact diagonalization of 12 sites cannot address the
window, because its lowest K excitations lie at $0.56J$ ($S^{xx}$) and $0.97J$
($S^{zz}$) for $\Delta=0.5$, and its static ratios are $1.2$ (equal time) and $2.1$
(susceptibility) at $T=0.1J$. A gapless $S^{zz}\propto\omega^{-0.88}$ with
$R\propto\omega^{-0.56}$ independent of resolution separates the critical
spin liquid from a conventional ordered state. Magnon damping and
two-magnon continua add weight to $S^{xx}$ but cannot create $S^{zz}$ weight
below the out-of-plane gap. Short-range easy-plane correlations without
long-range order also leave the out-of-plane fluctuations at the K point gapped, on a
scale of order $J\sqrt{1-\Delta}$, so they too make $R$ grow toward low energy. The test is then a gapless $S^{zz}(\mathrm K,\omega)$ well below this
scale, with a power law that does not depend on resolution, and the sign of
$d\ln R/d\ln\omega$, which separates the chiral-Ising from the chiral-XY selection of the easy-plane case.
Figure~\ref{fig:Rw} shows the two cases.

For
$\Delta\simeq0.97$, $k_BT^*\simeq5$--$16\,\mu$eV for $y_T\le1$ ($T^*=0.06$--$0.19$\,K) and $R$ stays constant over the same
window, although $T^*$ is then comparable to the measurement temperature, so the
constancy is a statement about the crossover regime. The chiral-XY selection makes $R$
fall toward low energy. Yb ions
on Na sites, 4.8(10)\% in crystals grown from Te flux~\cite{Dai2021}, add quasi-elastic
paramagnetic scattering, so crystals with less site mixing are preferable.

\section{Selection of the \texorpdfstring{120$^\circ$}{120\textdegree} order}
\label{sec:order}

\subsection{Landau theory}
We describe the ordered state by the condensate $z_t=\langle\Phi_t\rangle$ of the
triplet monopoles $\Phi_{4,5,6}\equiv\Phi_{x,y,z}$, $t=x,y,z$, which transforms according to Table~III of
Ref.~\cite{Song2020}. The 120$^\circ$ manifold is
$z=e^{\ii\theta}(\bm n_1+\ii\bm n_2)/\sqrt2$ with orthonormal $\bm n_{1,2}$ and plane
normal $\bm m=\bm n_1\times\bm n_2$, and translations shift $\theta$ by $\pm2\pi/3$. We
found all invariant polynomials of degree 2, 4 and 6 in
$(\mathrm{Re}\,z,\mathrm{Im}\,z)$ under $G_H$ and under $G_L$, for both senses of
the locked rotation. At degree 2, $G_H$ allows only $|z|^2$, and $G_L$ adds
one invariant,
\begin{equation}
F_2=|z_z|^2-\tfrac12|z_\perp|^2,
\end{equation}
which on the manifold depends only on $m_z^2$ and selects the orientation
of the spin plane. Through degree 4 the in-plane angle $\theta$ of a state
with $\bm m\parallel c$ remains free. A sixfold harmonic $\cos6\theta$ first
appears at degree 6, from charge-6 monopole terms.

\subsection{Couplings that select the plane}
The coefficient of $F_2$ follows from $G_H$-equivariant polynomials
$f(z,\lambda)$, quadratic in $z$ and of degree $b$ in the couplings. At $b=1$
only $\Delta-1$ contributes. At $b=2$ the contributing products are
$J(\Delta-1)$, $(\Delta-1)^2$ and the combination $\Jzp^2-2\Jpp^2$, with $\Jpp\Jzp$
and the products of $J$ or $\Delta-1$ with $\Jpp$ or $\Jzp$ absent. These products feed the
same channel as Eq.~\eqref{eq:deff}.

The classical energy per site of any
120$^\circ$ state with plane normal $\bm m$ is
\begin{equation}
E_{\rm cl}=-\tfrac32J-\tfrac34J(\Delta-1)(1-m_z^2),
\end{equation}
for unit spin length. The bond-dependent couplings give exactly zero,
because $\sum_\varphi A_\varphi=0$ and the two neighbors of a site along each bond
orientation sum to minus its own spin. Classical relaxation
produces no $\Jpp$ or $\Jzp$ dependence at any order. For $\Delta<1$ the spins
lie in the $ab$ plane. In linear spin-wave theory, the $l=2$ part of the
zero-point energy as a function of $\bm m$ has a $\Jzp^2$ to $\Jpp^2$ ratio of
$-0.50000$ and no $\Jpp\Jzp$ part, in agreement with the equivariant maps.
In spin-wave theory $\Jzp^2$ favors $\bm m\parallel c$ and $\Jpp^2$ favors $\bm m$ in the
plane, but this sign need not carry over to the spin-liquid regime.

Beyond $l=2$, the product $\Jpp\Jzp$ generates at second order a term
$\propto m_z\rho^3\sin3\vartheta$, with $\rho$ and $\vartheta$ the modulus and azimuth of
the in-plane part of $\bm m$. It requires quartic terms in $z$, tilts planes
that contain the $c$ axis, and leaves $\bm m\parallel c$ stationary.

\subsection{In-plane angle}
For $\bm m\parallel c$ the anisotropy of the in-plane angle is a harmonic
$\cos n\theta$. Translations allow only $n\equiv0\bmod3$ and time reversal,
$\theta\to\theta+\pi$, requires $n$ even, so $n\equiv0\bmod6$. Under the U(1) spin
rotations about $z$, which are symmetries of the XXZ part, $\cos6\theta$
carries charge 6, $\Jpp$ charge 2 and $\Jzp$ charge 1. The lowest products of
the couplings with total charge 6 are $\Jpp^3$, then $\Jpp^2\Jzp^2$, $\Jpp\Jzp^4$
and $\Jzp^6$. In linear spin-wave theory the harmonics
$n=1$--$5$ vanish and the $\cos6\theta$ amplitude follows these orders, with coefficient
$b_6\simeq0.023J(\Jpp/J)^3$ for $\Jpp$ alone. For KYbSe$_2$ ($\Jpp/J\approx0.004$)
this is $1.5\times10^{-9}J$. The in-plane mode of the 120$^\circ$ state is a
pseudo-Goldstone mode whose gap, of order $J(b_6/J)^{1/2}\sim4\times10^{-5}J$, is
far below experimental resolution.

\section{Magnetic field}
\label{sec:field}

\subsection{Selection rules}
A field lowers the symmetry to the elements of $G_L$ that fix $\bm h$ and
the products of $\mathcal T$ with those that reverse it. All statements refer
to the Zeeman field $\bm h$, whose direction differs from that of $\bm B$ when $g_{ab}\neq g_c$. Table~\ref{tab:field}
lists the zeroth-order bilinears that become allowed. The chiral
mass $\bar\psi\psi$, which carries the scalar spin chirality, is activated for a
$c$-axis component and for in-plane fields away from the bond directions.
In the Heisenberg model with Zeeman coupling it is forbidden for every
field direction, because the Zeeman field is a spin triplet and
$\bar\psi\psi$ a singlet.

The selection rules above concern the Zeeman coupling. A $c$-axis field also threads the
flux $\Phi_\triangle=B_z\sqrt3a^2/4$ through each triangle. In a half-filled one-band Hubbard
description with hopping $t$ and repulsion $U$, the third-order ring exchange couples
it to the scalar chirality~\cite{SenChitra1995,Motrunich2006,Katsura2010},
\begin{equation}
H_\chi=J_\chi\sum_\triangle\bm S_i\cdot(\bm S_j\times\bm S_k),\qquad
J_\chi=\frac{24t^3}{U^2}\sin\frac{2\pi\Phi_\triangle}{\Phi_0},
\label{eq:orbital}
\end{equation}
with $\Phi_0=h/e$ and the sense of the sum fixed by the field. The chirality
is the flux of the emergent gauge field~\cite{Motrunich2006}, so this term reaches $\bar\psi\psi$
already in the Heisenberg model. With $J=4t^2/U$,
$J_\chi/J=6(t/U)\,2\pi\Phi_\triangle/\Phi_0\simeq7\times10^{-4}(t/U)$ per tesla in KYbSe$_2$
($a=4.115$\,\AA~\cite{Xing2021}), whereas the spin--orbit source is
$h_z|\Jzp^2-2\Jpp^2|/J^3\lesssim5\times10^{-5}$ per tesla. For the 4$f$ exchange, which runs through the
ligands, $t/U$ is not well defined, but for $t/U$ of a few percent the two are
comparable, so a chirality or thermal Hall signal in a $c$-axis field does not
isolate spin--orbit coupling. An in-plane field threads no flux through the
triangles of a layer, and the in-plane selection rules are unaffected.

\begin{table}[t]
\caption{Zeroth-order bilinears allowed in a field with Zeeman coupling (the U(1) gauge current
is omitted). The orbital coupling of a $c$-axis field, Eq.~\eqref{eq:orbital}~\cite{SenChitra1995,Motrunich2006}, adds $\bar\psi\psi$ also in the
Heisenberg model.}
\label{tab:field}
\begin{ruledtabular}
\begin{tabular}{lll}
field & locked & Heisenberg\\
\hline
$\parallel c$ & $\bar\psi\psi$, $\psi^\dagger\sigma^z\psi$ & $\psi^\dagger\sigma^z\psi$\\
$\parallel$ bond & $\psi^\dagger\sigma^x\psi$ & $\psi^\dagger\sigma^x\psi$\\
in plane, $\perp$ bond & $\bar\psi\psi$, $\psi^\dagger\sigma^{y,z}\psi$ & $\psi^\dagger\sigma^y\psi$\\
generic & $\bar\psi\psi$, $\psi^\dagger\sigma^{x,y,z}\psi$ & $\psi^\dagger(\hat{\bm h}\cdot\bm\sigma)\psi$\\
\end{tabular}
\end{ruledtabular}
\end{table}

\subsection{Order in the couplings}
Equivariant maps in the field $\bm h$ and the couplings $\lambda$ give the
leading coefficient of the chiral mass. For $\bm h\parallel c$ the term is
unique at order $h\lambda^2$,
\begin{equation}
m_\chi\propto h_z(\Jzp^2-2\Jpp^2),
\end{equation}
with the ratio of the two squares fixed by symmetry, because $\Jpp$ and
$\Jzp$ are two components of one spin-2~$\otimes E_2$ multiplet. For in-plane
fields a linear term is excluded by the locked threefold axis, and the
first term is
\begin{equation}
m_\chi\propto\Jpp\Jzp\,h_y(h_y^2-3h_x^2)=-\Jpp\Jzp|h_\parallel|^3\sin3\varphi ,
\end{equation}
with $\varphi$ measured from a Yb--Yb bond. It is odd in $\Jzp$ and vanishes for
fields along the bonds at every order. For KYbSe$_2$ the coefficients are
$|\Jzp^2-2\Jpp^2|/J^2\lesssim2\times10^{-4}$ and $|\Jpp\Jzp|/J^2\approx(0.2$--$1.3)\times10^{-4}$, the range spanned by the
fitted and the density-functional values with and without the dipolar term.

The out-of-plane magnetization $M_z$ has the same quantum numbers under
$G_L$ as the chiral mass. The corresponding maps give no $M_z$ for in-plane
fields at orders $(h,\lambda)$, $(h,\lambda^2)$ and $(h^3,\lambda)$, and at order
$(h^3,\lambda^2)$ only
\begin{equation}
M_z\propto\Jpp\Jzp|h_\parallel|^3\sin3\varphi ,
\end{equation}
followed by $\Jzp^3$ terms at the next order. It is uniform in the
R$\bar3$m compounds and alternates in sign between the layers of
CsYbSe$_2$. Within a layer no staggered magnetization is induced at any
order, since all Yb sites of a layer are related by translations.

\subsection{Spinon diagrams}
\label{sec:spinondiag}
\begin{figure*}[t]
\includegraphics[width=0.96\textwidth]{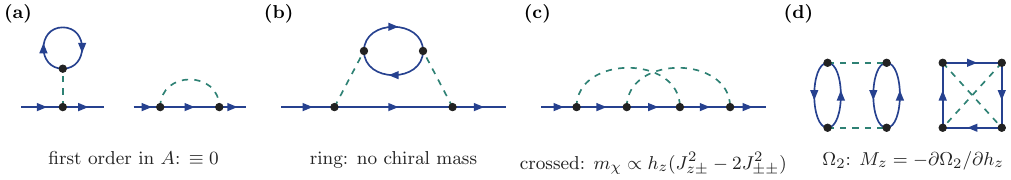}
\caption{Spinon diagrams in a Zeeman field. Solid lines are spinon propagators
of the staggered-flux state with the Zeeman splitting included exactly, and
dashed lines the anisotropic exchange $A_{ij}$. (a) Hartree and Fock
graphs, which vanish identically. (b) Ring graph, which gives no chiral
mass. (c) Crossed graph, the source of the chiral mass,
Eqs.~\eqref{eq:Mcz} and~\eqref{eq:Mcpar}. (d) Second-order vacuum graphs,
which give $M_z$.}
\label{fig:spinondiag}
\end{figure*}

The coefficients follow from the spinon perturbation theory in the
anisotropic part of the exchange, $A_{ij}=J_{ij}-J\mathds 1$ at $\Delta=1$,
around the staggered-flux mean-field state with hopping $-\chi_0t_{ij}$,
$\chi_0=0.148J$, with the Zeeman field kept exactly in the spinon
propagators (Fig.~\ref{fig:spinondiag}).

At first order (Hartree--Fock) $A$ drops out for every field direction.
On a spin-singlet bond the Fock term of a symmetric traceless tensor is
$\sum_{ac}A^{ac}\sigma^c\sigma^a=\operatorname{tr}A=0$. The
nearest-neighbor correlator of the staggered-flux band is even in the
chemical potential, because inversion combined with a sign gauge
transformation maps the band onto its negative, so the opposite Zeeman
shifts of the two spin species induce no bond triplet. The Hartree term of
the uniform magnetization cancels in the threefold sum over bonds. The
self-consistent mean field is the flux state with a rigid Zeeman
splitting, and chiral mass, $M_z$ and spinon thermal Hall response vanish at
this order, also for fields of generic direction.

At second order there are two self-energy graphs, the ring and the
crossed graph of Fig.~\ref{fig:spinondiag}(b,c), and the insertion graphs
vanish by the first-order result. We evaluated them in imaginary time on
tori up to $120\times120$. On a single flux triangle of unconstrained fermions with a
generic field and $O(1)$ anisotropy, the same graphs give the exact second-order
self-energy and grand potential. The Hermitian part of $\Sigma(i\pi T)$, projected on the
Dirac zero modes, gives the chiral mass $m_\chi$ as the energy of the zero modes,
calibrated with an explicit staggered flux that projects purely on the mass.
Only the crossed graph contributes. For a $c$-axis field
\begin{equation}
m_\chi=a_z\,h_z\,\frac{\Jzp^2-2\Jpp^2}{J^2},
\label{eq:Mcz}
\end{equation}
with $h_z$ the Zeeman energy $g_c\mu_BB_z$ and $a_z=-0.25$, $-0.29$, $-0.31$ and $-0.32$ at $T/J=0.04$, $0.02$, $0.01$ and
$0.005$, approaching $a_z\simeq-0.33$ for $T\to0$. Evaluating the self-energy in momentum space at the Dirac momenta gives
the same coefficients to within a few percent. The two squares enter in the ratio $-2$ and the cross term $\Jpp\Jzp$ is absent, as required by
symmetry. For in-plane fields
\begin{equation}
m_\chi=a_\parallel\,\frac{\Jpp\Jzp}{J^2}\,\frac{|\bm h_\parallel|^3}{J^2}\sin3\varphi,\qquad a_\parallel\simeq-0.27,
\label{eq:Mcpar}
\end{equation}
at $T=0.02J$ in the limit of small fields ($|\bm h_\parallel|\le0.08J$, while at $|\bm h_\parallel|=0.3J$ the coefficient is
$-0.15$), zero along the bonds and for $\Jpp$ or $\Jzp$ alone. The
magnetization follows from $M_z=-\partial\Omega_2/\partial h_z$ with the
second-order grand potential [Fig.~\ref{fig:spinondiag}(d)],
$M_z\simeq-0.4\,\Jpp\Jzp|\bm h_\parallel|^3\sin3\varphi/J^5$ per site at
$T=0.02J$ for small fields ($-0.26$ at $|\bm h_\parallel|=0.3J$) and $-0.10$ in the same units at $T=0.2J$. Gauge fluctuations and the
occupancy constraint are absent at this order. Exact diagonalization
(Sec.~\ref{sec:ed}) gives a 9-site $M_z$ at $T=0.2J$ of opposite sign and 11--18
times smaller for $|\bm h|=0.15$--$0.6J$, and chirality coefficients of opposite sign on 9 and 12 sites.
Exact diagonalization confirms the selection rules, while the coefficients are
order-of-magnitude estimates.

\begin{figure}[t]
\includegraphics[width=\linewidth]{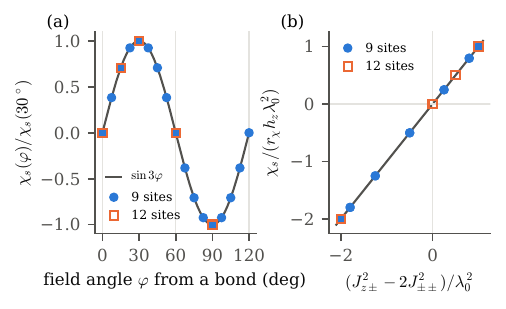}
\caption{Exact diagonalization of the spin model. (a) Scalar chirality for
in-plane fields at angle $\varphi$ from a bond, normalized to its value at
$30^\circ$, on 9- and 12-site clusters, compared with $\sin3\varphi$. (b)
Chirality for $c$-axis fields against $\Jzp^2-2\Jpp^2$ at coupling scale
$\lambda_0=0.03J$, each cluster scaled by its coefficient $r_\chi$.}
\label{fig:field}
\end{figure}

\subsection{Exact diagonalization}
\label{sec:ed}
The selection rules concern the spin model and can be tested without the
parton construction. We diagonalized Eq.~\eqref{eq:Hmaster} on periodic
clusters of 9 and 12 sites that keep $C_6$ and the bond mirrors, with
$J_2=0.1J$ (added once per pair of sites) and a Zeeman field, at $T=0.2J$ (Fig.~\ref{fig:field}). The
uniform scalar chirality per site,
$\chi_s=N_s^{-1}\sum_\triangle\bm S_i\cdot(\bm S_j\times\bm S_k)$ with the sum over
all up and down triangles of a cluster of $N_s$ sites, vanishes in the
Heisenberg model for every field. For in-plane fields ($|\bm h|=0.6J$, $\Delta=0.9$,
$\Jpp=0.08J$, $\Jzp=0.12J$) it follows $\sin3\varphi$ on 9 sites, and on 12 sites up to the
symmetry-allowed $\sin9\varphi$ harmonic, of relative size $10^{-4}$, and it reverses with $\Jzp$. For $c$-axis
fields ($\Delta=1$, $h_z=0.1J$) the coefficient
$r_\chi=\chi_s/[h_z(\Jzp^2-2\Jpp^2)]$ is $-0.092$ on 9 sites and $0.592$ on
12 sites, constant across $\Jpp/\Jzp$ to 0.8\% and 0.3\%. It depends on the
cluster, but the ratio $\Jpp^2:\Jzp^2=-2:1$ does not. The chirality vanishes at
$\Jzp^2=2\Jpp^2$ up to terms of order $h^3$. The
magnetization $M_z$ on 9 sites vanishes for the XXZ model and for $\Jpp$
alone, follows $\sin3\varphi$ exactly, reverses with $\Jzp$, and scales as
$\Jpp\Jzp|h|^3$. The coupling $\Jzp$ alone gives about 1.4\% of the mixed value, through the $\Jzp^3$ term.

\subsection{Single-ion contribution}
\label{sec:singleion}
The trigonal crystal-field terms $B_4^3O_4^3$ and $B_6^3O_6^3$ lock the Yb moment to
the lattice without any exchange, so an isolated ion also acquires an
out-of-plane moment in an in-plane field. We diagonalized the $J=7/2$
crystal-field Hamiltonian with the Zeeman term for the reported
parameters of KYbSe$_2$~\cite{Scheie2023}, NaYbSe$_2$~\cite{Zhang2021},
CsYbSe$_2$~\cite{Pocs2021} and YbZn$_2$GaO$_5$~\cite{Bag2024}. The KYbSe$_2$ parameters reproduce the
reported levels and $g$ factors within their errors. The NaYbSe$_2$ parameters, the 5\,K set
of Table~IV in the Supplemental Material of Ref.~\cite{Zhang2021}, diagonalized in the Stevens-operator
normalization defined in the same Supplemental Material, place the first excited doublet at 17.5\,meV, where 15.8\,meV is measured,
and give $g_c=1.21$, where about 1 is measured~\cite{Zhang2021,Ranjith2019}. Table~IV of Ref.~\cite{Zhang2021} lists
15.79\,meV and $g_c=1.05$ as calculated from the same parameters, a difference that no rescaling of
the individual operators removes. A sixth-order term
$B_6^0\simeq-1.1\times10^{-4}$\,meV restores the level and changes $c_{\rm SI}$ by 10\% and
$v'$ by 18\%. The CsYbSe$_2$ parameters were not fitted to neutron
spectroscopy. They give $g_c=1.44$, and because they are reported in rounded form,
$g_c$ can lie anywhere between 1.29 and 1.60, a range that contains the value 1.33 that Ref.~\cite{Pocs2021} obtains
from them, against 0.3 from electron spin
resonance~\cite{Xie2023}, so the CsYbSe$_2$ row of Table~\ref{tab:csi} rests on poorly constrained parameters. For in-plane fields the free-ion $M_z$ follows $\sin3\varphi$ and vanishes
for fields along the bonds, the twofold axes of the site. In KYbSe$_2$ it is
$5.3\times10^{-5}\mu_B$ per Yb at 1\,T and 1\,K. The single-ion response thus has the
angular dependence of the exchange response, and the field angle cannot
separate the two.

Projecting the field-dependent ground doublet onto the zero-field doublet
gives a single-ion pseudospin Hamiltonian $e_0(\bm B)+\bm h(\bm B)\cdot\bm S$ that is exact in the
field. Besides the Zeeman term it contains three pieces with the symmetry of
$M_z$, all proportional to $\sin3\varphi$ with a sign set by the orientation of
the ligand octahedra: a $c$-axis field $h_z=\kappa B^3\sin3\varphi$, and a
field-induced off-diagonal $g$ factor $\partial h_\parallel/\partial B_z=\lambda_t B^2\sin3\varphi$, which
tilts the moment operator and drops out of the Hamiltonian at $B_z=0$, and
a shift of the doublet center, $-\partial e_0/\partial B_z=v\mu_BB^3\sin3\varphi$, the Van Vleck
moment from the excited doublets, which does not depend on temperature. By
the Hellmann--Feynman theorem the tilt contributes
$\lambda_t B^2\langle S_\parallel\rangle/\mu_B$ to $M_z$ and the field $h_z$ contributes through the
transverse susceptibility. For an SU(2)-symmetric pseudospin exchange the
transverse susceptibility equals $\langle S_\parallel\rangle/h_\parallel$, because a transverse field
only rotates the magnetization, so
\begin{equation}
\begin{gathered}
M_z^{\rm SI}=\bigl(c_{\rm SI}M_{ab}+v'B\bigr)B^2\sin3\varphi,\\
c_{\rm SI}=\frac{\lambda_t}{g_{ab}\mu_B}+\frac{g_c\,\kappa}{g_{ab}^2\,\mu_B},
\end{gathered}
\label{eq:csi}
\end{equation}
with $M_{ab}$ the measured in-plane magnetization per Yb and
$v'=v-c_{\rm SI}\chi_{\rm VV}$, where $\chi_{\rm VV}B$ is the in-plane Van Vleck part of $M_{ab}$.
The pseudospin moment tilts rigidly out of the plane by the angle
$c_{\rm SI}B^2\sin3\varphi$, and at fixed field $M_z^{\rm SI}$ is linear in $M_{ab}$ with
coefficients that depend neither on temperature nor on the correlations.
An XXZ anisotropy corrects only the second term of $c_{\rm SI}$, at order
$|\Delta-1|$. Table~\ref{tab:csi} lists the coefficients. The tilt term dominates
$c_{\rm SI}$, which is about $5\times10^{-5}$\,T$^{-2}$ in KYbSe$_2$ and NaYbSe$_2$, a tilt of
0.3$^\circ$ at 10\,T. In CsYbSe$_2$ and YbZn$_2$GaO$_5$ the coefficient alternates between the
layers and the uniform $M_z$ cancels, as for the exchange response.

\begin{table}[t]
\caption{Single-ion coefficients for in-plane fields, namely the $c$-axis field
$\kappa$ and moment tilt $\lambda_t$ of the projected doublet, and the rigid-tilt
coefficient $c_{\rm SI}$ and Van Vleck coefficient $v'$ of Eq.~\eqref{eq:csi}, in units of
$10^{-6}$\,meV/T$^3$ ($\kappa$, $\lambda_t$), $10^{-5}$\,T$^{-2}$ ($c_{\rm SI}$) and $10^{-6}\,\mu_B$/T$^3$ ($v'$). The NaYbSe$_2$ value in parentheses includes
$B_6^0=-1.1\times10^{-4}$\,meV. Signs are relative to the in-plane
Zeeman term. In CsYbSe$_2$ and YbZn$_2$GaO$_5$ the values are per layer.}
\label{tab:csi}
\begin{ruledtabular}
\begin{tabular}{lcccc}
 & $\kappa$ & $\lambda_t$ & $c_{\rm SI}$ & $v'$\\
\hline
KYbSe$_2$ & 1.89 & 6.92 & 4.72 & $-1.27$\\
NaYbSe$_2$ & 1.36 & 7.02 & 4.57 (5.03) & $-1.42$ ($-1.68$)\\
CsYbSe$_2$ & 0.46 & $-14.1$ & $-6.83$ & 1.48\\
YbZn$_2$GaO$_5$ & 0.14 & 1.88 & 1.13 & $-0.09$\\
\end{tabular}
\end{ruledtabular}
\end{table}

To compare with the exchange response we added $\bm h(\bm B)\cdot\bm S$ to the KYbSe$_2$
model ($J=0.438$\,meV~\cite{Scheie2024}, $\Delta=0.98$ and $J_2=0.047J$~\cite{Scheie2023}) on the 9- and 12-site clusters,
for $B=0.5$--4\,T and $T=0.5$--2\,K. The single-ion $M_z$ obeys
Eq.~\eqref{eq:csi}, with the $h_z$ term within 1\% of its SU(2) value. The exchange
$M_z$ for $(\Jpp,\Jzp)=(0.004,0.015)J$ and $(0.0012,0.013)J$ is smaller by
factors of about 150 or more on 9 sites and of 0.4--300 on 12 sites, where its
sign is reversed, as for the chirality. The smallest factors occur at
$B\le1$\,T and $T=0.5$\,K, where the finite-size gap of the 12-site cluster
suppresses $M_{ab}$. For $B\ge2$\,T the 12-site factors are 5--300. The spinon value of
Sec.~\ref{sec:spinondiag} gives factors of 6 and 23 for the two sets at
$T=0.2J$, and smaller factors at lower temperature, because its coefficient
grows fourfold between $T=0.2J$ and $0.02J$ while $M_{ab}$ changes little. In
KYbSe$_2$ the single-ion moment is comparable to or larger than
the exchange moment, and torque magnetometry measures their sum.
Equation~\eqref{eq:csi} separates them. At fixed field, the
single-ion $M_z$ plotted against $M_{ab}$ for different temperatures is a straight
line, so the deviation from that line below $T\approx J$ comes from the exchange. Alternatively, a dilute
Lu-substituted crystal measures $c_{\rm SI}$ and $v'$ directly, since
Eq.~\eqref{eq:csi} holds for free pseudospins, up to the change of the crystal
field on dilution. The crystal-field fits alone
fix $c_{\rm SI}$ only to about 10\% and $v'$ to about 20\%, too coarsely to subtract
the single-ion part.

The tilt does not enter the Hamiltonian at $B_z=0$ and leaves the thermal
Hall response unchanged. The single-ion field reaches the chiral mass
through Eq.~\eqref{eq:Mcz}, $m_\chi=a_z\kappa B^3\sin3\varphi\,(\Jzp^2-2\Jpp^2)/J^2$, which is
$2.5\times10^{-4}$ and $8\times10^{-4}$ of Eq.~\eqref{eq:Mcpar} in KYbSe$_2$ for the two
parameter sets at $T=0.02J$. At this order the planar thermal Hall response is set
by the exchange, and $M_z$ largely by the crystal field. Field-induced
corrections to the exchange through the excited doublets, of relative
order $(\mu_BB/\Delta_{\rm CEF})^2\approx10^{-5}(B/{\rm T})^2$ with $\Delta_{\rm CEF}$ the crystal-field gap, are not included.

\section{Planar thermal Hall response}
\label{sec:hall}

The thermal Hall conductivity $\kappa_{xy}$ transforms as the $z$ component of
an axial vector and is odd in the field, $\kappa_{xy}(-\bm h)=-\kappa_{xy}(\bm h)$.
Expanding it in harmonics of the in-plane field angle $\varphi$ and imposing
the crystal point group gives the following through the twelfth harmonic:
\begin{itemize}
\item R$\bar3$m ($D_{3d}$; NaYbSe$_2$, KYbSe$_2$, NaYbO$_2$): $\sin3\varphi$ and $\sin9\varphi$
for in-plane fields. A single YbX$_6$ layer (layer group $p\bar3m1$) gives
the same harmonics.
\item P6$_3$/mmc ($D_{6h}$; CsYbSe$_2$): no harmonic. The twofold screw axis
$2_1=(6_3)^3$, or equivalently the horizontal mirror, maps $\bm h_\parallel$ to
$-\bm h_\parallel$ and leaves $\kappa_{xy}$ invariant, so $\kappa_{xy}\equiv0$ for in-plane
fields. The two layers of the cell carry opposite $\sin3\varphi$ and $\sin9\varphi$
coefficients, which cancel exactly in perfectly stacked crystals. Stacking
faults, surfaces or thin flakes restore a $\sin3\varphi$ response.
\item With a $c$-axis component, all four compounds allow $h_z$,
$h_z\cos6\varphi$ and $h_z\cos12\varphi$.
\end{itemize}
The harmonics hold for any heat carrier. In the Heisenberg pseudospin
model with Zeeman coupling a purely spatial mirror survives every field
direction, so $\kappa_{xy}$ vanishes for in-plane fields. A planar thermal Hall
signal in the R$\bar3$m compounds measures spin--orbit locking,
either of the exchange or of the single-ion response at higher order in
the field, through the excited crystal-field levels, which $D_{3d}$ also
allows. For the chiral mass of KYbSe$_2$ the one-ion part is below $10^{-3}$ of the
exchange part (Sec.~\ref{sec:singleion}). Obverse--reverse twins, domains with the two senses of rhombohedral stacking, reported in the KYbSe$_2$ crystal
arrays of Ref.~\cite{Scheie2023}, are related by a $60^\circ$ rotation that reverses
$\Jzp$ and the trigonal crystal-field terms, so they cancel part of every
$\sin3\varphi$ response.

In the DSL the spinon response follows from the chiral mass of
Sec.~\ref{sec:spinondiag}. A two-component Dirac cone with mass $m_\chi$ has the
Hall conductance $\mathrm{sgn}(m_\chi)/2$ inside the gap and $m_\chi/2|\epsilon|$ at
energy $|\epsilon|>|m_\chi|$. Integrating this Hall conductance with the thermal weight of
free fermions, each of the four cones contributes $\ln2\,k_Bm_\chi/(2\pi\hbar)$ for
$T\gg|m_\chi|$, so $\kappa_{xy}=2\ln2\,k_Bm_\chi/(\pi\hbar d_L)$ with $d_L$ the layer spacing,
if the cones sit at zero energy. With the
$T\to0$ coefficient, for KYbSe$_2$ [$J=0.438$\,meV~\cite{Scheie2024},
$|\Jzp^2-2\Jpp^2|/J^2\lesssim2\times10^{-4}$, $g_c=1.8(6)$~\cite{Scheie2023},
$d_L=c/3=7.56$\,\AA~\cite{Xing2021}] a $c$-axis field gives at most $|m_\chi|/B\approx8\times10^{-5}$\,K/T
and $|\kappa_{xy}|/B\approx9\times10^{-8}$\,W/(K\,m\,T), and in-plane fields of a few tesla
give values of the same order through $|\Jpp\Jzp|/J^2\simeq(0.2$--$1.3)\times10^{-4}$. These
are order-of-magnitude estimates for $T\ll J$, with the 30\% uncertainty of $g_c$. The cluster
coefficients of Sec.~\ref{sec:ed} differ in sign, and for $c$-axis fields the orbital term of
Eq.~\eqref{eq:orbital}, through which the spinons of a spin liquid acquire a thermal Hall
response~\cite{Katsura2010}, adds a contribution of similar size. The Zeeman field
shifts the spin-up and spin-down cones to $\pm g_c\mu_BB/2$, about $\pm0.6$\,K per
tesla, and including this shift in the thermal integral raises
$\kappa_{xy}$ by factors of 4.3, 2.2, 1.3 and 1.06 at 0.5, 1, 2 and 5\,K at 1\,T,
so that $|\kappa_{xy}|/B\approx2\times10^{-7}$\,W/(K\,m\,T) at most at 1\,K. With the first $1/N$
correction the chiral mass $\bar\psi\psi$ has $\Delta_{\bar\psi\psi}=2+128/(3\pi^2N_f)\simeq3.08$ at
$N_f=4$~\cite{Hermele2007,Chester2016}, slightly irrelevant, whereas Pad\'e approximants of the
$\epsilon$ expansion give 2.40--2.48~\cite{DiPietro2017}, slightly relevant, so gauge fluctuations may
reduce or enhance the response at low temperature. At $T=1$\,K and 1\,T the
free-spinon value is at most $3\times10^{-4}$ of the half-quantized $(\pi/12)k_B^2T/(\hbar d_L)$,
so a sizable planar thermal Hall signal would point to carriers other than
the spinons. Planar thermal Hall measurements resolve about $10^{-5}$--$10^{-4}$\,W/(K\,m) at
present, two orders above the spinon estimate at 10\,T, so the angular selection rules will
be tested on whatever carries the heat, in practice phonons, and the spinon amplitude is not
accessible.

\section{Quenched disorder}
\label{sec:disorder}

\subsection{Criterion}
Disorder destabilizes the DSL already without spin--orbit coupling, and spin--orbit
coupling adds one further channel. A static random coupling $\int d\tau\,d^2r\,V(\bm r)O(\bm r,\tau)$ with
$\overline{V(\bm r)V(\bm r')}=\mathcal D\,\delta(\bm r-\bm r')$ has replica scaling dimension
$[\mathcal D]=d+2z-2\Delta_O=4-2\Delta_O$ for $d=2$ and
$z=1$, the quantum form of the Harris criterion~\cite{Harris1974,Thomson2017,Dey2020},
and is relevant for $\Delta_O<2$. A defect breaks every
lattice symmetry locally, and $V(\bm r)$ multiplies $O$ by a random phase.
Lattice momentum and lattice irreducible representations give no
protection. Only time reversal, for non-magnetic disorder, and the spin
content of the random coupling constrain which operators are reached.
\begin{itemize}
\item Random isotropic exchange: spin 0, $\mathcal T$ even, present in any
material.
\item Random Dzyaloshinskii--Moriya exchange: spin 1, $\mathcal T$ even. It
requires spin--orbit coupling and the local loss of bond inversion, which
occurs near a Yb ion on an alkali site or a vacancy.
\item Random symmetric anisotropy: spin 2 (with a spin-0 trace part for
$\delta\Delta$), $\mathcal T$ even, requires spin--orbit coupling.
\item Frozen moments or random fields: spin 1, $\mathcal T$ odd.
\end{itemize}

\subsection{Classification}
Table~\ref{tab:disorder} classifies the 64 bilinears by time reversal,
spin, momentum and Lorentz content. Both Hermitian parts of the singlet
monopoles are $\mathcal T$ even and those of the triplets are $\mathcal T$ odd. The
$\mathcal T$-even assignments agree with Ref.~\cite{Dey2020}, in which random bond
disorder couples to the singlet masses and currents $M^{0b}$, $\mathcal J^{0b}_j$,
and random vector-chirality disorder, that is random Dzyaloshinskii--Moriya
exchange, to the triplet $M^{a0}$, $\mathcal J^{a0}_j$. Here
$M^{ab}=\bar\psi\sigma^a\tau^b\psi$ and $\mathcal J^{ab}_\mu=\bar\psi\gamma_\mu\sigma^a\tau^b\psi$, with
$\sigma^0=\tau^0=1$.

\begin{table}[t]
\caption{The 64 zeroth-order bilinears by time reversal, momentum and
spin, with their Lorentz content.}
\label{tab:disorder}
\begin{ruledtabular}
\begin{tabular}{llcl}
$\mathcal T$ & momentum & spin & content\\
\hline
even & 0 & 1 & 3 masses $M^{a0}$, 6 currents $\mathcal J^{a0}_j$\\
even & 0 & 0 & U(1) density\\
even & M & 1 & 9 densities $\mathcal J^{ab}_0$\\
even & M & 0 & 3 masses $M^{0b}$, 6 currents $\mathcal J^{0b}_j$\\
odd & 0 & 1 & 3 densities $\mathcal J^{a0}_0$\\
odd & 0 & 0 & mass $\bar\psi\psi$, U(1) currents\\
odd & M & 1 & 9 masses $M^{ab}$, 18 currents $\mathcal J^{ab}_j$\\
odd & M & 0 & 3 densities $\mathcal J^{0b}_0$\\
\end{tabular}
\end{ruledtabular}
\end{table}

\subsection{Consequences in the spin liquid}
The relevant dimensions are $\Delta_{\rm adj}=2-64/(3\pi^2N_f)\simeq1.46$
for the adjoint masses~\cite{Hermele2005}, $\Delta_\Phi\simeq1$ for the lowest
monopoles~\cite{Song2019,He2022} and 2 for the conserved SU(4) currents. The U(1) current to which the
gauge field couples has $\Delta=3$ in QED$_3$, so a random coupling to the
spinon density is irrelevant~\cite{Thomson2017}.
\begin{itemize}
\item Any non-magnetic defect produces random fugacities for the
single-flux singlet monopoles, which in the clean crystal are forbidden
only by lattice symmetry. Their replica eigenvalue is $4-2\Delta_\Phi\simeq2$, so with
the large-$N$ dimensions this is the most relevant disorder perturbation of
the DSL, and it acts as a
random field on the valence-bond order parameter. The conclusion holds for
any $\Delta_\Phi<2$. Reference~\cite{Dey2020} treats the charge-3 monopoles allowed in
the clean limit together with random currents, and the single-monopole fields
come in addition.
\item Random isotropic exchange also couples to the valence-bond masses
$M^{0b}$ and currents $\mathcal J^{0b}_j$. In the large-$N$ flow of
Ref.~\cite{Thomson2017}, disorder that preserves a non-Abelian flavor
subgroup only on average drives the random-current variance to strong
coupling. With the Pauli matrices acting
on the valley index this describes random bond disorder, so the DSL is
unstable to it even before monopoles are considered.
\item Random Dzyaloshinskii--Moriya exchange couples to the
quantum-spin-Hall masses, a channel closed for non-magnetic disorder in a
Heisenberg magnet, with tree-level replica eigenvalue $\simeq1.08$. In the
large-$N$ flow the same algebra applies with the Pauli matrices acting on
spin. Random spin-triplet couplings confined to one spin axis, random
$\bar\psi\sigma^z\psi$ and $\bar\psi\sigma^z\gamma^j\psi$, reach a fixed line with dynamical
exponent $z>1$. That line is unstable to the transverse
components~\cite{Thomson2017}, random currents along one axis destabilize
monopoles beyond a critical disorder strength~\cite{Dey2020}, and disorder in all
three components flows to strong disorder~\cite{Thomson2017}. On this lattice
the Dzyaloshinskii--Moriya vectors also feed the spin--valley densities, which
keeps even $c$-axis disorder off that line (Appendix~\ref{app:disrg}).
\item Random symmetric anisotropy opens no new channel. Its pure spin-2 part
reaches no bilinear and first enters through quartic operators with
$\Delta\gtrsim2.4$. Its trace part and its products with the uniform
anisotropy feed the channels above with amplitudes reduced by
$|\Delta-1|$, $\Jpp/J$ or $\Jzp/J$.
\end{itemize}

\subsection{One-loop flow}
Matching the microscopic random couplings onto the bilinears and integrating the
one-loop replicated flow, with the photon graphs specific to QED$_3$
(Appendix~\ref{app:disrg}), gives no disordered fixed point within weak disorder at
$N_f=4$. Bond disorder grows mainly into random valence-bond masses, and random
Dzyaloshinskii--Moriya exchange grows first into the quantum-spin-Hall masses and
then into the non-Abelian random spin currents. For $\delta J/J$ or $D/J$ of 0.1 the
disorder becomes strong at a scale $\ell\simeq5$--6.

\subsection{Consequences at the critical point}
At the chiral-Heisenberg critical point random Dzyaloshinskii--Moriya
exchange is a random field on $\phi_a$, with replica dimension $4-2\Delta_\phi$.
The dimension is $2$ at large $N$ ($\Delta_\phi=1$) and $0.4$ at one loop with
$\epsilon=1$ ($\Delta_\phi=1.8$ from $\eta_\phi=13/5$~\cite{Dupuis2019}), so the random field is relevant in both cases.
Random isotropic exchange couples to $\phi^2$ with replica dimension
$2/\nu-2$, which is negative when $d\nu>2$ with $d=2$, as in the classical
criterion~\cite{Harris1974}, and equals $-0.78$ for $\nu=1.644$ and $0$ for $\nu=1$. Random XXZ
anisotropy couples to $T_{ab}$ with replica dimension $2(y_T-1)$, between
$-0.52$ and 0 for $y_T=0.74$--1. The uniform
XXZ anisotropy is relevant, while its random part is irrelevant to marginal.

\subsection{Fate of the random valence-bond field}
In a two-dimensional valence-bond solid, weak bond randomness couples
linearly to the order parameter and destroys long-range order. The
topological defects that this nucleates carry spin-$\tfrac12$ moments and
form a sparse, strongly random spin network, and a valence-bond glass formed
by stronger pinning is unstable to the same spinful
defects~\cite{Kimchi2018}. The remnant moments can freeze at very low
temperature. In YbMgGaO$_4$ and YbZnGaO$_4$ the freezing lies about 20 times
below the specific-heat peak~\cite{Kimchi2018}. The random singlet-monopole
field is a random valence-bond field. We conjecture that the disordered DSL
reaches the same endpoint once this field grows. The thermodynamics and
heat transport of Lu-diluted NaYbSe$_2$ have been interpreted as a
valence-bond glass~\cite{Lyu2025} (Sec.~\ref{sec:predictions_conf}).

Such a network of moments has a characteristic field scaling. A broad
distribution of couplings $P(J)\sim J^{-\zeta}$ gives $C\sim T^{1-\zeta}$,
$\chi\sim T^{-\zeta}$ and the collapse
\begin{equation}
C[H,T]/T\sim H^{-\zeta}F_q(T/H),\qquad F_q(x)\sim x^q\ (x\ll1),
\end{equation}
where $q=0$ when all exchange is symmetric, $q=1$ with
Dzyaloshinskii--Moriya couplings in a layered magnet, and $q=2$ for
three-dimensional networks~\cite{Kimchi2018b}. In R$\bar3$m every Yb site is an
inversion center and all Yb are related by lattice translations, so the
midpoint of any Yb--Yb bond, within or between layers, is an inversion
center and all exchange between lattice Yb is symmetric. A network formed
by lattice moments far from defects is therefore expected to have $q=0$ if the effective
couplings between remnant moments inherit this inversion, which the random
valence-bond environment that generates them need not respect. Couplings involving Yb on Na sites, which lack bond inversion and
couple adjacent layers, give $q\ge1$. The exponent $q$ thus
identifies which moments carry the low-temperature entropy.

\subsection{Interlayer moments}
A Yb ion on an alkali site is a Kramers doublet coupled locally to the
layer by $V_{\rm imp}S^a_{\rm imp}O^a(0,\tau)$, with $O$ a $\mathcal T$-odd operator. The coupling
has dimension $1-\Delta_O$. For the triplet monopoles this is
$\approx-0.02$ in the DSL and $\approx+0.22$ at the critical point, $-0.46$ for the
spin--valley masses and $-1$ for the uniform spin density. The coupling is
marginal in the DSL and weakly relevant at the critical point, with a scale
$T_K\sim J(V_{\rm imp}/J)^{1/0.22}$ that is parametrically small. With these large-$N$
dimensions, such moments stay nearly free down to very low temperature.

\begin{table*}[tbp]
\caption{Observations relevant to the predictions.}
\label{tab:data}
\begin{ruledtabular}
\begin{tabular}{p{0.09\textwidth}p{0.19\textwidth}p{0.15\textwidth}p{0.12\textwidth}p{0.17\textwidth}p{0.17\textwidth}}
compound & zero-field order & spectrum near K & K-point exponent & low-$T$ specific heat & defects\\
\hline
NaYbSe$_2$ & none to 50\,mK~\cite{Ranjith2019,Dai2021}; ac-$\chi$ peak near 40\,mK~\cite{Scheie2024arXiv} & gapless continuum~\cite{Dai2021,Scheie2024arXiv} & $\alpha=1.74(6)$, one temperature~\cite{Scheie2024arXiv} & $\gamma\approx1.0(1)$ (NaCl flux)~\cite{Ranjith2019}, $\approx1.5$ (Te flux, from a figure)~\cite{Dai2021} J/mol\,K$^2$ & Yb on Na: 4.8(10)\% (X-ray), $\approx3$\% (ICP), Te flux~\cite{Dai2021}\\
KYbSe$_2$ & 120$^\circ$, $T_N=0.29$\,K~\cite{Scheie2023,Lee2024} & gapless ($\le40\,\mu$eV) at 0.3\,K~\cite{Scheie2023} & $\alpha=1.73(12)$, $\omega/T$ collapse~\cite{Scheie2023}; for $\hbar\omega\lesssim J$ above $T_N$ 1.4--1.5 (collapse), 0.6--0.95 (joint fit) (Sec.~\ref{sec:kybse2}) & $C/T$ linear in $T$ above $T_N$~\cite{Lee2024} & $(0.2\pm0.3)\%$ site mixing~\cite{Scheie2023}\\
CsYbSe$_2$ & short range below 0.4\,K, $\approx0.1\,\mu_B$~\cite{Xie2023} & broad continuum around K~\cite{Xie2023} & --- & broad hump~\cite{Xing2019} & ---\\
NaYbO$_2$ & none to 50\,mK~\cite{Bordelon2019} and 70\,mK~\cite{Ding2019} & continuum (powder)~\cite{Bordelon2019,Ding2019} & --- & $T^{2.04(2)}$~\cite{Bordelon2019} & 7--14\% free spins~\cite{Bordelon2019}\\
\end{tabular}
\end{ruledtabular}
\end{table*}

\section{Comparison with experiment}
\label{sec:data}

\subsection{Regime}
\begin{figure}[tbp]
\includegraphics[width=\linewidth]{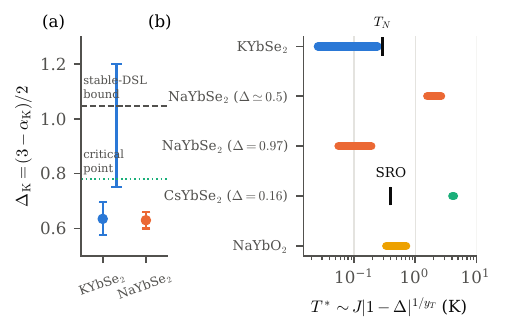}
\caption{(a) $\Delta_{\rm K}=(3-\alpha_{\rm K})/2$ from the reported K-point exponents of
KYbSe$_2$~\cite{Scheie2023} and NaYbSe$_2$~\cite{Scheie2024arXiv} (filled) and from the
range allowed by the KYbSe$_2$ data at $\hbar\omega\lesssim J$ above $T_N$ (bar,
Sec.~\ref{sec:kybse2}), with the
stable-DSL bound~\cite{He2022} and the large-$N$ monopole dimension at the
critical point~\cite{Dupuis2019}. (b) XXZ crossover scale
Eq.~\eqref{eq:Tstar} for each compound and parameter set, with $J=0.438$\,meV~\cite{Scheie2024} for
the $\Delta=0.98(6)$ estimate of KYbSe$_2$~\cite{Scheie2023} and otherwise the $J$ of the same source, and $c_1=1$. Bars span $y_T=0.74$--$1$. Bars with
$|1-\Delta|\gtrsim0.3$ lie outside the validity of Eq.~\eqref{eq:Tstar}, and CsYbSe$_2$ appears as a point. Markers show
$T_N$ of KYbSe$_2$ and the onset of short-range order (SRO) in CsYbSe$_2$.}
\label{fig:regimes}
\end{figure}

Taken at face value, the reported K-point exponents place the delafossites outside a
stable DSL, but the data below $J$ do not fix them. Table~\ref{tab:data} collects the
observations relevant here. The K-point
exponents $\alpha=1.73(12)$ in KYbSe$_2$, from an $\omega/T$ collapse near and above
$T_N$~\cite{Scheie2023}, and $\alpha=1.74(6)$ in NaYbSe$_2$ at 0.1\,K~\cite{Scheie2024arXiv}
give $\Delta_{\rm K}\simeq0.63$--$0.64$. The NaYbSe$_2$ exponent comes from a single
temperature and high energy transfer, deviates at low energy and depends
on the fit window~\cite{Scheie2024arXiv}. Willsher and Knolle found the KYbSe$_2$
exponent compatible with QED$_3$ Gross--Neveu criticality and a monopole
dimension $0.64$, with an uncertainty of $0.12$ that linear propagation
from $\alpha$ reduces to $0.06$~\cite{WillsherKnolle2025}. Scheie \emph{et al.}\ interpret the
KYbSe$_2$ scaling through a gapped $\mathbb Z_2$ liquid with an O(4) critical
point~\cite{Scheie2023} and place NaYbSe$_2$ inside the Heisenberg spin-liquid
phase~\cite{Scheie2024arXiv}. The value lies below the stable-DSL bound
$1.046$. Provided the measured exponent belongs to the lowest K-point
primary in its asymptotic regime, the window is therefore not described by a
stable DSL. Within the
Dirac description it points to the critical regime of the transition to
120$^\circ$ order, whose large-$N$ monopole dimension is $0.78$ (Fig.~\ref{fig:regimes}).
The existing KYbSe$_2$ fit, however, extends to $\hbar\omega\simeq3J$, and restricted
to $\hbar\omega\lesssim J$ the data do not fix the exponent (Sec.~\ref{sec:kybse2}).

The K-point weight itself is compatible with the DSL. Scheie
\emph{et al.}\ argue that a $\pi$-flux state, gapless at M, cannot produce
it~\cite{Scheie2023}. In the DSL the spin at the K point couples to the
triplet monopoles, which are critical, while the spinon bilinears carry the
weight at M. A QED$_3$--Higgs theory has been proposed for the transition from the
DSL to a gapped $\mathbb Z_2$ liquid, although its fixed point is asymptotically
unstable~\cite{Feuerpfeil2026PRB}. In a real-space analysis of the same KYbSe$_2$ data, correlations
spread sub-ballistically at 0.3\,K, with a dynamical exponent $1.4(2)$ that none of
the spin-wave, Schwinger-boson, Landau--Lifshitz or free-spinon descriptions reproduce,
a behavior attributed to quantum-critical hydrodynamics~\cite{Scheie2026}.

\subsection{Momentum-resolved scaling in \texorpdfstring{KYbSe$_2$}{KYbSe2}}
\label{sec:kybse2}
\begin{figure}[tbp]
\includegraphics[width=\linewidth]{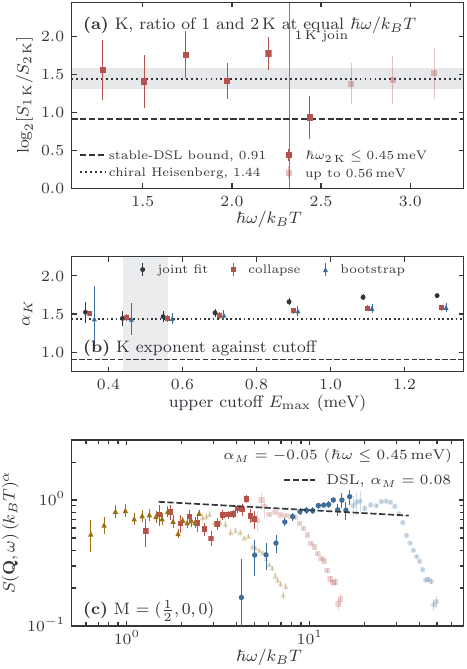}
\caption{K and M spectra of KYbSe$_2$ from the source data of Ref.~\cite{Scheie2023}.
(a) $\log_2$ of the ratio of the 1 and 2\,K spectra at the K point at equal $\hbar\omega/k_BT$, which equals
$\alpha_{\rm K}$ for $\omega/T$ scaling, with the stable-DSL bound and the large-$N$ chiral-Heisenberg value.
The band is the shift from a 10\% change of the relative normalization, the line marks
0.2\,meV, where the 1\,K cut switches between two incident energies, and faded points lie above 0.45\,meV at 2\,K.
(b) K exponent against the upper cutoff $E_{\max}$ for the joint fit of
Ref.~\cite{Scheie2023} (all temperatures) and the collapse of the 1 and
2\,K data, alone and averaged over random resamplings of the energy points, with the band marking the values of $J$. (c) M point at 0.3\,K (circles), 1\,K (squares) and 2\,K (triangles), scaled with
$\alpha_{\rm M}=-0.05$ below 0.45\,meV (faded points above, and 0.3\,K at or below $T_N$), with the DSL slope.}
\label{fig:kybse2}
\end{figure}

The DSL predicts different exponents at the K and M points. At the K point the spin couples to
the triplet monopoles, $\alpha_{\rm K}=3-2\Delta_\Phi$, with $\alpha_{\rm K}<0.91$ in a stable
DSL by the conformal-bootstrap bound~\cite{He2022} ($0.96$ from the next-to-leading-order large-$N$ monopole
dimension 1.02 at $N_f=4$, which violates the bound) and $1.44$ at the
chiral-Heisenberg point. At M it couples to the
$\mathcal T$-odd spin--valley masses $M^{ab}$, $\alpha_{\rm M}=3-2\Delta_{\rm adj}\simeq0.08$. We
reanalyzed the source data of Ref.~\cite{Scheie2023}: the K cut at
0.3, 1 and 2\,K and the path through M, both K points and two zone centers
from the same temperatures, which the authors subtracted with a 12\,K
background. Their definition, a power law in $\omega$ at $\hbar\omega/k_BT\ge1$
fitted jointly to all temperatures where their scaled data follow a power law
($\hbar\omega/k_BT\gtrsim1$ in their Fig.~6), reproduces their exponent,
$\alpha_{\rm K}=1.74\pm0.03$, and moving the lower cutoff spans 1.61--1.94, comparable
to their quoted uncertainty of $\pm0.12$. Their window extends from 0.11 to 1.3\,meV,
about $3J$, with $\chi^2$ per degree of freedom of 13.6, to which the quoted error is
scaled, and they note themselves that the scaling spans a single decade of $\hbar\omega/k_BT$. In their estimator the exponent stays at 1.45--1.52, set largely by the
0.3\,K data, for upper cutoffs up to 0.7\,meV and rises to 1.74 at 1.3\,meV. Their
base temperature, with the sample thermometer at 270\,mK~\cite{Scheie2023}, lies
at or below $T_N$, where the magnetic peaks already have a correlation length of
$47\pm10$\,\AA~\cite{Scheie2023}, so only the 1 and 2\,K data probe the regime above $T_N$.
Restricted to $\hbar\omega\le0.45$ or $0.56$\,meV, near the values of $J$ in
Refs.~\cite{Scheie2024,Scheie2023}, a collapse of these two
temperatures with the exponent left free, alone and averaged over random resamplings of the
energy points, gives $\alpha_{\rm K}=1.40$--$1.46$.
The point-by-point ratio of the two spectra at equal $\hbar\omega/k_BT$, and a fit with a
scaling function whose logarithm is quadratic in $\ln(\hbar\omega/k_BT)$, give 1.3--1.5, and cuts
through the K point along the path give values down to about 1.2. All these estimators
rest on the ratio of the two spectra at equal $\hbar\omega/k_BT$, which equals $2^{\alpha_{\rm K}}$ and
is available only over $\hbar\omega/k_BT\simeq1.3$--$2.6$ [Fig.~\ref{fig:kybse2}(a)], so a $\pm10\%$ change of
their relative normalization moves the exponent by $\pm0.14$. With the normalization
free the exponent is undetermined.

A fit with a smooth but otherwise free scaling function (Appendix~\ref{app:refit})
gives $\alpha_{\rm K}=1.43$--$1.47$ for $E_{\max}=0.45$--$0.56$\,meV, with 95\% intervals
$[1.08,1.80]$. Against the stable-DSL bound it favors $\alpha_{\rm K}=1.44$ by
$\Delta\chi^2\simeq6$--9, after scaling by the $\chi^2$ per degree of freedom, for a 10\% prior on the
relative normalization (3--4 with the reported errors), by 2--3 for a 20\% prior and by
nothing for a free normalization. The preference also rests on the 1\,K points below
0.2\,meV, where the reported cut joins two data sets, and starting the window at 0.2\,meV
reduces it to 0.2--4.

Within this window the two spectra have
different log-slopes, 0.9--1.3 at 1\,K, depending on the error model and weighting, and 0.2--0.5 at 2\,K, and the
estimator of Ref.~\cite{Scheie2023}, one power law fitted jointly to both
temperatures, gives 0.6--0.95 with $\chi^2$ per degree of freedom of 4--15.
Below $J$ the data therefore do not fix $\alpha_{\rm K}$ and allow both the
stable-DSL range and the chiral-Heisenberg value. A temperature-independent
continuum does not describe them.

At M the spectrum below $J$ is nearly flat, $\alpha_{\rm M}=-0.1\pm0.15$, as for the DSL
value 0.08 but also for a temperature-independent continuum (Appendix~\ref{app:refit}).
The analysis cannot fix the low-energy K exponent better than these ranges, because
only a factor of two in temperature lies above $T_N$, and it cannot separate spin
components.

\subsection{Predictions and existing data}
\label{sec:predictions_conf}
\emph{Gap.}---No compound shows a spin gap in zero field, consistent with
the absence of any induced mass.

\emph{Linear specific heat.}---The absence of induced densities and
currents excludes spinon Fermi pockets generated by spin--orbit coupling.
The sample dependence of $\gamma$ in NaYbSe$_2$, about 1.4--1.5\,J/mol\,K$^2$ in
crystals grown from Te flux, read from Fig.~1(e) of Ref.~\cite{Dai2021}, with 4.8(10)\% Yb on Na from x-ray refinement (about 3\%
from chemical analysis)~\cite{Dai2021} and $1.0(1)$ in crystals grown from NaCl
flux~\cite{Ranjith2019}, indicates an extrinsic origin, as expected
from the relevance of random singlet-monopole fields at every defect. The
frequency-dependent ac-susceptibility peak near 40\,mK~\cite{Scheie2024arXiv}
lies more than an order of magnitude below the specific-heat maxima near
1.1 and 2.75\,K~\cite{Ranjith2019}, as for the freezing of a defect network in
YbMgGaO$_4$~\cite{Kimchi2018}. Heat transport supports this. NaYbS$_2$ and
NaYbSe$_2$ show a nearly constant $C_{\rm mag}/T$ below 0.2\,K together with a
vanishing residual $\kappa_0/T$, which the authors attribute to spinons scattering
phonons~\cite{Li2024NaYbSe2kappa}. The low-energy entropy is thus not carried by mobile
fermions. A spinon Fermi surface would give a finite $\kappa_0/T$, whereas Dirac
spinons give none. In the dilution series
NaYb$_x$Lu$_{1-x}$Se$_2$ most of the entropy is released in a dimer peak near
2\,K, from pairs of Yb spins forming singlets, a broad shoulder near 1\,K grows with the connectivity of the Yb
lattice, and heat is carried by gapped itinerant excitations, with a gap of
about 270\,mK in NaYbSe$_2$, that are most abundant near the percolation threshold. Lyu
\emph{et al.}\ interpret these features as a valence-bond glass~\cite{Lyu2025}, the
endpoint expected for a random valence-bond field
(Sec.~\ref{sec:disorder}). We predict that $\gamma$ scales with the
site-mixing fraction and that the field-dependent specific heat collapses as
$C/T\sim H^{-\zeta}F_q(T/H)$, with $q=0$ for a network of lattice moments and
$q\ge1$ for one dominated by Yb on Na sites.

\emph{Spin anisotropy.}---In a stable Heisenberg DSL the spin-rotation-breaking
part of the correlations decays on cooling. In NaYbSe$_2$ the ratio of
in-plane to $c$-axis spin susceptibilities, after subtracting the Van
Vleck terms and normalizing by $g^2$, decreases by 25--30\% between 2 and
0.5\,K in the 1\,T data of Ref.~\cite{Ranjith2019}. Reference~\cite{Ranjith2019} states the in-plane Van Vleck
susceptibility as $7.8\times10^{-4}$\,cm$^3$/mol, but its quoted $0.0137\,\mu_B$/T and
the slope of its Fig.~3(b) give $7.65\times10^{-3}$\,cm$^3$/mol. A relevant XXZ anisotropy produces such a decrease, but a
$c$-axis Curie contribution from about 7\% quasi-free Yb would produce it
as well, and the data cannot exclude this.

\emph{Nuclear magnetic resonance (NMR).}---In NaYbSe$_2$, $1/T_1$ of $^{23}$Na rises on cooling at 0.9 and
1.5\,T in-plane and at 0.9 and 6\,T along $c$ down to 0.35\,K~\cite{Luther2024}, and
our reading of their data gives effective slopes between about $T^{-1}$
and $T^{-1.5}$. Luther \emph{et al.}\ attribute the rise to critical fluctuations
near an instability. In the critical regime the monopole contribution
$\propto T^{2\Delta_{\rm K}-1}$ (Appendix~\ref{app:nmr}) would decrease. For in-plane fields, order sets in
from about 2\,T~\cite{Ranjith2019}, and along $c$ an umbrella phase is reported from 6\,T at
0.4\,K~\cite{Luther2024}. The
in-plane data are therefore close to field-induced order, and the 6\,T
$c$-axis data are comparably close only in units of the effective Zeeman
field $g_c\mu_BB$. The rise more likely reflects this proximity.

\subsection{Crossover scales and predictions by compound}
\label{sec:scales}
Table~\ref{tab:scales} collects, for each compound and parameter set, the crossover scales
of this paper and the behavior of $R(\omega)$ they imply. The entries carry the uncertainties
discussed above: $c_1$ of order one (0.3--1.6 on the clusters, Table~\ref{tab:deffED}), the
dipolar shift of $\Delta$ (a factor 1.4--1.6 in $T^*$, Sec.~\ref{sec:seeds}), and the detuning from
the transition, which for KYbSe$_2$ gives $J|J_2-J_{2c}|^\nu/J^\nu=4$--30\,mK for $\nu=1.64$ and
70--220\,mK for $\nu=1$ with $J_{2c}/J=0.06$--0.09, so that the comparison of $T^*$ with $T_N$ is
one of orders of magnitude.

\begin{table*}[t]
\caption{Crossover scales by compound. $T^*$ is Eq.~\eqref{eq:Tstar} with $|c_1|=1$ and
$y_T=0.74$--1. The scale $T_{\rm SO}$ is the onset scale of the anisotropy at a multicritical Heisenberg
DSL (Sec.~\ref{sec:quartic}), and $T_\perp$ is the interlayer monopole crossover from the dipolar
coupling at the DSL--120$^\circ$ point (Sec.~\ref{sec:interlayer}). In rows marked $\dagger$,
$|1-\Delta|$ is too large for Eq.~\eqref{eq:Tstar}, which assumes $|\delta_{\rm eff}|\ll1$, and the entries
show only that the crossover scale is of order $J$.
The last column gives the behavior of $R(\omega)$ of Eq.~\eqref{eq:Rw} at the K point for the chiral-Ising selection
over the window $0.06$--0.15\,meV at 0.1\,K. The chiral-XY selection reverses the trend.}
\label{tab:scales}
\begin{ruledtabular}
\begin{tabular}{p{0.09\textwidth}p{0.2\textwidth}p{0.06\textwidth}p{0.14\textwidth}p{0.07\textwidth}p{0.06\textwidth}p{0.3\textwidth}}
compound & $\Delta$ (source) & $J$ (meV) & $T^*$ (K) & $T_{\rm SO}$ (mK) & $T_\perp$ (mK) & $R(\omega)$ at K\\
\hline
KYbSe$_2$ & 0.98(6), diffuse scattering~\cite{Scheie2023} & 0.438 & 0.03--0.10; 0.17--0.41 at $\Delta=0.92$ & 9 & 26 & one exponent above $T_N$ if $T^*<T_N$\\
 & 0.956, density functional~\cite{Villanova2023} & 0.474 & 0.08--0.24 & 36 & 26 & \\
NaYbSe$_2$ & 0.97, density functional~\cite{Villanova2023} & 0.551 & 0.06--0.19 & 22 & 37 & constant, within the crossover regime\\
 & 0.50--0.54, susceptibility and magnetization~\cite{Ranjith2019,Schmidt2021} & 0.39--0.46 & $\dagger$ of order $J$ (1.8--2.5) & $\dagger$ & 37 & rises toward low energy, $\omega^{-0.56}$ at leading order, if the compound lies near the transition\\
CsYbSe$_2$ & $\simeq1$, spin waves~\cite{Xie2023} & 0.395 & -- & -- & -- & $J_2/J\simeq0.03$, far from the transition; P6$_3$/mmc stacking\\
 & 0.16, crystal field~\cite{Pocs2021} & 0.44 & $\dagger$ of order $J$ (4.0--4.3) & $\dagger$ & & \\
NaYbO$_2$ & 0.88, representative point~\cite{Bordelon2019} & 0.51 & 0.34--0.71 & 190 & 82 & $J_2$ unknown\\
\end{tabular}
\end{ruledtabular}
\end{table*}

\emph{KYbSe$_2$.}---Its $T^*$ lies at or below $T_N$ for the best estimates of $\Delta$ and
reaches 0.4\,K at the lower edge of the diffuse-scattering range or for $c_1=1.6$. If $T^*<T_N$,
spin-resolved neutron scattering at the K point finds one exponent for $S^{zz}$ and $S^{\pm}$ above $T_N$.
If instead the Heisenberg DSL is multicritical, an anisotropy of this size acts at the DSL
itself only below $T_{\rm SO}$. In the ordered state the spins lie in the $ab$ plane, and the
in-plane mode is a pseudo-Goldstone mode with an unresolvable gap. The chiral-mass coefficients
are $|\Jzp^2-2\Jpp^2|/J^2\lesssim2\times10^{-4}$ and $|\Jpp\Jzp|/J^2\approx(0.2$--$1.3)\times10^{-4}$.

\emph{NaYbSe$_2$.}---The anisotropy estimates disagree. For the density-functional value
$|1-\Delta|$ is small and $T^*$ is comparable to the measurement temperature. For the
thermodynamic value $|1-\Delta|$ is not small, Eq.~\eqref{eq:Tstar} no longer applies, and the
anisotropy acts on the scale of $J$ itself. If NaYbSe$_2$ lies near the DSL--120$^\circ$
transition, which $J_2/J=0.071(15)$~\cite{Scheie2024} allows but does not establish, and which
Ref.~\cite{Scheie2024arXiv} disfavors by placing it inside the spin-liquid phase, the 0.1\,K
exponent would then belong to the chiral-Ising class, and its near-equality with the KYbSe$_2$
exponent would be a coincidence. Spin-resolved neutron scattering at the K point can distinguish the two:
for $\Delta\simeq0.5$ the in-plane monopoles $\Phi_4\pm\ii\Phi_5$, which order on the easy-plane
side in the chiral-Ising selection, make $S^{\pm}(\mathrm K,\omega)$ more singular than $S^{zz}(\mathrm K,\omega)$ for
$T\ll T^*$ (Sec.~\ref{sec:critical}), whereas the chiral-XY selection of Ref.~\cite{WillsherKnolle2025}
would make $R$ fall toward low energy. The specific-heat coefficient and the 40\,mK
ac-susceptibility peak should scale with site mixing.

\emph{CsYbSe$_2$.}---$J_2/J\approx0.03$~\cite{Xie2023} places it far from the spin-liquid
window, and with $\Delta=0.16$~\cite{Pocs2021} no Heisenberg critical regime is expected, although Xie
\emph{et al.}\ find $\Delta\simeq1$ within error from field-polarized spin waves~\cite{Xie2023}.
The planar thermal Hall response for in-plane fields vanishes identically in perfectly stacked
crystals, and $M_z$ for in-plane fields alternates between layers. Its planar response
must vanish by symmetry, which makes it the control for the R$\bar3$m compounds.

\emph{NaYbO$_2$.}---Its $g$ tensor is the best characterized of the four, so the collapse of
thermodynamic and NMR data against the Zeeman energy $|\bm h|$ for $c$-axis and in-plane fields
is a parameter-free test of the chiral-Heisenberg window. The 7--14\% free
spins~\cite{Bordelon2019} agree with the marginal coupling of local moments.

\emph{YbZn$_2$GaO$_5$.}---This $P6_3/mmc$ compound, with Yb--O layers 10.98\,\AA\ apart,
shows no order down to 0.06\,K, a magnetic specific heat $C\propto T^2$ below
0.1\,K with a contribution to $C/T$ linear in the field, and low-energy
neutron intensity at M and K with a gap near $\Gamma$~\cite{Bag2024}.
Matrix-product-state spectra of the $J_1$--$J_2$ XXZ model agree with its neutron
data for $\Delta\simeq1.35$ at a fixed $J_2/J_1=0.12$~\cite{Bag2024}, whereas its Curie--Weiss temperatures,
$-3.77$\,K for $\bm h\parallel c$ and $-5.22$\,K in the plane~\cite{Bag2024}, suggest easy-plane anisotropy. The Yb ions occupy the
Wyckoff site $2a$ of $P6_3/mmc$, with $D_{3d}$ symmetry~\cite{Bag2024,Zhao2025}. The
nearest-neighbor bond centers are $6g$ positions with symmetry $2/m$, hence
inversion centers, and the two Yb layers of the cell are related by the
$6_3$ screw, as in CsYbSe$_2$. The classification of
Secs.~\ref{sec:locked}--\ref{sec:monopoles}, which uses only the Yb site symmetry and the
locked group, carries over. With
$\Delta\simeq1.35$ the anisotropy is easy-axis, which in our analysis selects
$(\phi_x,\phi_y)$ and chiral-XY criticality near the transition. For $\Delta=1.3$ the
spin liquid gives way to stripe order at $J_2/J_1\simeq0.12$~\cite{Gallegos2025}, so the fitted point
lies at that boundary. As in CsYbSe$_2$, $\Jzp$ alternates between layers and the planar
thermal Hall response vanishes in perfectly stacked crystals. Neutron
powder diffraction on solid-state-synthesized samples, however, finds
nearly random Ga/Zn site mixing, 35\% of Ga on the Zn site and 60\% of Zn
on the Ga site, together with static displacements of Yb and O that
broaden the crystal-field levels to 7.5--10\,meV~\cite{Zhao2025}. The single-crystal
x-ray refinement of Ref.~\cite{Bag2024}, which barely distinguishes Zn from Ga,
and its neutron powder refinement found no mixing, whereas $^{71}$Ga NMR on crystals from the
same batch resolves two Ga sites with a broad distribution of quadrupolar frequencies, which
establishes Ga/Zn mixing and local crystal-field randomness, and $\mu$SR finds dynamic spins
without freezing down to 30\,mK~\cite{Zeroual2026}. Such mixing breaks the $D_{3d}$ site symmetry and the bond
inversion locally, so it produces the random Dzyaloshinskii--Moriya
exchange and random anisotropy of Sec.~\ref{sec:disorder}, which reach the
quantum-spin-Hall masses and, through the accompanying bond randomness,
the valence-bond masses and singlet monopoles. Between 0.1 and 0.3\,meV
the neutron intensity is larger at M than at K~\cite{Bag2024}, whereas the
large-$N$ dimensions make $S(\mathrm K,\omega)$ the more singular at the lowest
energies. That window lies at $\hbar\omega\simeq0.2$--$0.6J$ for $J_1\simeq0.5$\,meV, above
the asymptotic regime.

\subsection{Proposed experiments}
\label{sec:predictions}
\begin{enumerate}
\item Polarized neutron scattering at the K point in NaYbSe$_2$ and KYbSe$_2$, with
$\bm P\parallel c$. A gapless $S^{zz}(\mathrm K,\omega)$ well below the out-of-plane gap separates the
critical spin liquid from long- or short-range easy-plane order, and the sign of
$d\ln R/d\ln\omega$, with $R(\omega)$ of Eq.~\eqref{eq:Rw}, separates the chiral-Ising from the chiral-XY
selection of the easy-plane case. At leading order $R$ grows as $\omega^{-0.56}$
between 0.06 and 0.15\,meV at 0.1\,K in NaYbSe$_2$ if $\Delta\simeq0.5$ and the compound lies near
the transition, stays constant if $\Delta\simeq0.97$ and $y_T\le1$, and falls for the chiral-XY
selection. In KYbSe$_2$ above $T_N$ it should be constant if $T^*\lesssim T_N$.
\item Field-angle-resolved planar thermal Hall effect: $\sin3\varphi$ and
$\sin9\varphi$ with zeros along Yb--Yb bonds in the R$\bar3$m compounds,
identically zero in perfectly stacked CsYbSe$_2$.
\item Torque magnetometry with in-plane fields: $M_z\propto\sin3\varphi$ in the
R$\bar3$m compounds, canceling between layers in CsYbSe$_2$. The exchange part is
the deviation of $M_z$ from linearity in $M_{ab}$ at fixed field, or the
difference from a dilute Lu-substituted crystal, Eq.~\eqref{eq:csi}.
\item Specific heat and the ac-susceptibility peak of NaYbSe$_2$ against
controlled Yb/Na site mixing.
\item Collapse of low-temperature thermodynamics and NMR against the Zeeman energy $|\bm h|$
for $c$-axis and in-plane fields in NaYbO$_2$ and KYbSe$_2$, where the in-plane
field--temperature phase diagram, with its 120$^\circ$, up-up-down and oblique phases, is
known to 30\,mK~\cite{Lee2024}.
\end{enumerate}

\section{Conclusions}
\label{sec:conclusions}

A stable Heisenberg DSL remains stable, layer by layer, against the
spin--orbit-coupled exchange of the ytterbium delafossites. The locked symmetry
group alone forbids a Dirac mass and any displacement of the nodes at every order,
and, assuming adiabatic continuity with the Heisenberg spin liquid, it allows no new
monopole term through charge three. At linear order the exchange reaches spin-2
components of the quartic $\mathbf{20}'$ and $\mathbf{84}$, whose dimensions are those of
Heisenberg-allowed operators. At second order it produces a
spin--momentum-locked tilt of dimension above 3, an effective XXZ anisotropy
through $\Jzp^2-2\Jpp^2$, and nine spin-1 quartics without Heisenberg partners,
whose $O(1/N)$ dimensions make them irrelevant. The same computation shows that
the Heisenberg spin liquid allows a Lorentz-vector $\mathbf{20}'$ quartic of large-$N$
dimension 2.92, with two-loop $\epsilon$-expansion estimates of 2.9--3.5, so neither
expansion settles the stability of a single Heisenberg layer against its own
quartics. In the crystals, interlayer tunneling, which spin--orbit coupling opens
for the triplet monopoles at first order in the rhombohedral stacking, and the
spin-Peierls coupling to the lattice~\cite{Seifert2024} destabilize the DSL at low energy.

The anisotropy matters near the transition to 120$^\circ$ order, where it is
relevant. Its crossover scales may differ by an order of magnitude between
KYbSe$_2$ and NaYbSe$_2$, depending on which estimate of $\Delta$ holds, and at a
multicritical Heisenberg DSL an anisotropy of the size found in KYbSe$_2$ acts only
below tens of mK. Exact diagonalization of the staggered vector chirality shows
that easy-plane anisotropy selects its component along $c$. The easy-plane side is
then chiral-Ising-like, and the monopoles carrying $S^\pm$ and $S^{zz}$ split, with
the $g$-normalized ratio $R$ of Eq.~\eqref{eq:Rw} rising toward low energy as
$\omega^{-0.56}$ in NaYbSe$_2$ if $\Delta\simeq0.5$. If the measured K-point exponents
belong to the lowest monopole in its asymptotic regime, they are inconsistent with
a stable DSL. Restricted to $\hbar\omega\lesssim J$ and the two temperatures above
$T_N$, however, the KYbSe$_2$ data do not fix the exponent.

The bond-dependent couplings are invisible to the spinon mean field. Their
second-order effects, among them the field-induced chiral mass and the spinon
thermal Hall response, are of estimated relative size $10^{-4}$ in KYbSe$_2$,
comparable to the orbital chirality in a $c$-axis field. The out-of-plane moment in
an in-plane field is dominated by the single-ion crystal field, and CsYbSe$_2$ shows
no planar response in perfect stacking. The classification carries over to other
Yb triangular magnets with the same site symmetry, such as YbZn$_2$GaO$_5$.
Polarized neutron scattering at the K point gives the most direct test. A gapless
$S^{zz}$ and the sign of $d\ln R/d\ln\omega$ would show whether the delafossites lie
near the chiral-Ising critical point of the DSL.

\begin{acknowledgments}
The work of Y.I. was performed, in part, at the Aspen Center for Physics, which is supported by a grant from the Simons Foundation (1161654,
Troyer). This research was also supported in part by grant NSF PHY-2309135 to the Kavli Institute for
Theoretical Physics (KITP). Y.I. acknowledges support from the Abdus Salam International Centre for Theoretical Physics (ICTP) through the Associates Programme, from the Simons Foundation through Grant No. 284558FY19, from IIT Madras through the Institute
of Eminence (IoE) program for establishing QuCenDiEM (Project No. SP22231244CPETWOQCDHOC), and the International Centre for Theoretical Sciences (ICTS), Bengaluru for participation in the Discussion Meeting ``Fractionalized Quantum Matter'' (code: ICTS/DMFQM2025/07).
The author acknowledges the use of the AI tool Fable 5.1 (Anthropic) to verify the symmetry analysis, assist in writing parts of the code, perform data analysis, and prepare parts of the manuscript. The author reviewed all results and changes and takes full responsibility for the content.
\end{acknowledgments}

\section*{Data availability}
The data that support the findings of this article are available from the author upon reasonable request.

\appendix

\section{Methods}
\label{app:methods}

\emph{Dirac theory.}---The 16 real zero modes are obtained by diagonalizing the
Majorana form of the staggered-flux Hamiltonian on $L\times L$ tori with $L=8$ and
12. Each generator, with its sign gauge $\eta_i$, is applied to the lattice fermions
and projected on the zero modes, which gives the orthogonal $16\times16$ matrices $O$ of
Eq.~\eqref{eq:Mtrans}. Invariant operators are the common null space of $\{g-1\}$ over
the generators in the space of bilinears, gradient bilinears or quartics. The
SU(4) content of the quartics is obtained from the quadratic Casimir of the
flavor generators built from the zero-mode bilinears.

\emph{Equivariant maps.}---For an operator block $W$ and the coupling space $V$, the
maps from $\mathrm{Sym}^n(V)$ to $W$ that commute with $G_H$ are found as the null space of the
linear equivariance conditions, imposed with the finite lattice generators and
the exact spin-rotation generators, or by sampling invariant polynomials at
random points. Evaluated on the locked couplings, they give the allowed
coefficients. For the nine spin-1 quartics the orders follow from comparing, for each spin, how the lattice group acts
on the $n$-th powers of the coupling vector and on the operator blocks.

\emph{Monopole momenta.}---A uniform $2\pi$ flux is spread over the torus on top of the
staggered flux, and the monopole state fills the Dirac sea and the zero modes
of one spin species. Its translation eigenvalue is the expectation value of the
translation combined with a gauge transformation, divided by the same quantity
for the zero-flux ground state. The gauge transformation is fixed by angular
synchronization, the leading eigenvector of the matrix of bond-phase mismatches
between the translated and the original Hamiltonian, which spreads the Wilson-loop
mismatch of order $2\pi/L$ evenly, and it acts on the gauge charge with a background of
$-1$ per site.

\emph{Exact diagonalization.}---The clusters have periods $(3,0)$, $(0,3)$ (9 sites),
$(2,2)$, $(-2,4)$ (12 sites) and $(4,1)$, $(-1,5)$ (21 sites) in units of
$(\bm a_1,\bm a_2)$, and all keep $C_6$. On 9 and 12 sites several second-neighbor vectors
join the same pair, and $J_2$ is added once per pair unless stated. On 21 sites each
second-neighbor vector joins a distinct pair. The 9- and 12-site spectra are
obtained by full diagonalization, and the 21-site ground states by a matrix-free
Lanczos method in the full $2^{21}$-dimensional space.

\emph{Spinon saddle.}---The Fock decoupling of the isotropic exchange on the
staggered-flux state gives the hopping $-\chi_0t_{ij}$ with
$\chi_0=\tfrac34J|\langle f^\dagger_{i\sigma}f_{j\sigma}\rangle|=0.148J$, where the bond correlator per spin species is
0.1975. The second-order graphs are evaluated in imaginary time on tori up to
$120\times120$.

\section{Large-$N$ graphs}
\label{app:largeN}

\begin{figure*}[t]
\includegraphics[width=0.96\textwidth]{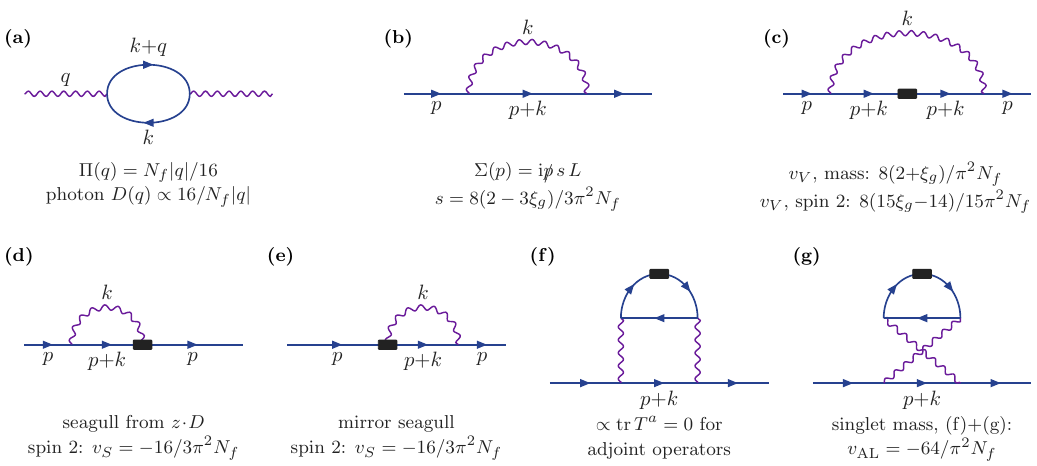}
\caption{Large-$N_f$ graphs for the dimensions of the adjoint masses and of
the spin-2 tilt $\mathcal O_R$, with the operator insertion drawn as a black box and the
photon as a wavy line. (a) Polarization, which fixes the photon propagator. (b)
Fermion self-energy. (c) Vertex correction across the insertion. (d,e)
Seagull graphs from the covariant derivative in the spin-2 operator. (f,g)
Closed-loop graphs, which vanish for flavor-adjoint operators and give
$-64/\pi^2N_f$ for the flavor-singlet mass~\cite{Hermele2007}.}
\label{fig:largeN}
\end{figure*}

We computed the large-$N$ dimensions of the adjoint masses and of $\mathcal O_R$
from the graphs of Fig.~\ref{fig:largeN}. The photon propagator, with gauge parameter $\xi_g$, is
$D_{\mu\nu}(q)=(16/N_f|q|)(\delta_{\mu\nu}-(1-\xi_g)q_\mu q_\nu/q^2)$, from the polarization
$\Pi(q)=N_f|q|/16$ [panel (a)]. In units of $1/\pi^2N_f$ the fermion self-energy
(b) contributes $8(2-3\xi_g)/3$, the vertex graph across the insertion (c)
$8(2+\xi_g)$ for a mass and $8(15\xi_g-14)/15$ for the spin-2 bilinear, and each
of the two seagull graphs (d,e), in which the photon attaches to the
covariant derivative inside the operator, $-16/3$. The closed-loop graphs
(f,g) are proportional to $\operatorname{tr}T^a=0$ for flavor-adjoint operators. The
gauge parameter cancels between (b) and (c), the conserved current has no
anomalous dimension, and
\begin{equation}
\Delta_{\rm adj}=2-\frac{64}{3\pi^2N_f},\qquad
\Delta_{\mathcal O_R}=3+\frac{64}{5\pi^2N_f},
\end{equation}
that is, $1.46$ and $3.32$ at $N_f=4$.
Both agree with Ref.~\cite{Hermele2005} [Eqs.~(17) and (21)] and with Eq.~(79)
of Ref.~\cite{Zhou2023} at spin 2, and the spin-2 value agrees with the large-$N_f$ dimension of
flavor non-singlet twist-2 operators of Ref.~\cite{Gracey1994} at $d=3$ for two-component flavors.

\section{NMR form factors}
\label{app:nmr}

The relaxation rate $1/T_1$ of a nucleus samples the spin fluctuations at
in-plane wave vector $\bm q$ with a weight $W_{\bm n}(\bm q)$ fixed by its
hyperfine couplings and the field direction $\bm n$. In NaYbSe$_2$ the
$^{23}$Na site ($3b$, site group of order 12) has six nearest Yb ions, three
in each adjacent layer, each set forming an elementary triangle, and the
$^{77}$Se site ($6c$, order 6) has three nearest Yb ions forming one
triangle. In both cases the Yb neighbors form a single orbit of the site
group, and the hyperfine tensor of each Yb--nucleus pair, constrained by the
mirror that fixes the pair, has five free parameters. With general tensors, the
$\Gamma$, K and M fluctuations all contribute for every field direction, so
symmetry imposes no field-direction selection rule on $1/T_1$.

An isotropic transferred hyperfine coupling gives $W(\mathrm K)=0$ exactly for
both nuclei, because each couples to complete elementary triangles and
$\sum_\triangle e^{\ii\bm K\cdot\bm r}=0$. The K-point fluctuations, carried by
the triplet monopoles, enter only through the anisotropic part of the
hyperfine tensor. We modeled each Yb--nucleus coupling as an isotropic part
plus a dipolar-shaped part and fixed the two parameters by the measured
uniform ($\bm q=0$) couplings of Ref.~\cite{Luther2024} ($^{23}$Na: $+11.0$ and
$-51.3$\,mT$/\mu_B$; $^{77}$Se: $-782$ and $-500$\,mT$/\mu_B$ for $\bm h\perp c$ and
$\bm h\parallel c$). With the two adjacent Yb layers fluctuating independently, as for a
two-dimensional spin liquid, this model gives $W(\mathrm K)/W(\mathrm M)\approx1.1$ for both nuclei, nearly
independent of the field direction. For $^{23}$Na, whose six neighbors lie three in each
layer, coherent K-point fluctuations of the two layers would raise the ratio to 2.1--2.3.
For $^{77}$Se the Yb--nucleus direction lies within about $1^\circ$ of the magic angle, so the
dipolar-shaped part is poorly constrained, but the ratio is insensitive to it. These ratios are outputs of the
hyperfine model, which is constrained only by the uniform couplings. In the
critical regime the K and M channels contribute
\begin{equation}
1/T_1\supset a_{\rm K}T^{2\Delta_{\rm K}-1}+a_{\rm M}T^{2\Delta_{\rm adj}-1},
\end{equation}
together with the contribution of the conserved spin density near $\Gamma$,
with $a_{\rm K}$ proportional to the square of the anisotropic part of the
hyperfine coupling. Within the model the monopole channel carries a weight
comparable to the M-point channel for both nuclei.

\section{One-loop flow of the matched disorder}
\label{app:disrg}
\begin{figure*}[t]
\includegraphics[width=0.96\textwidth]{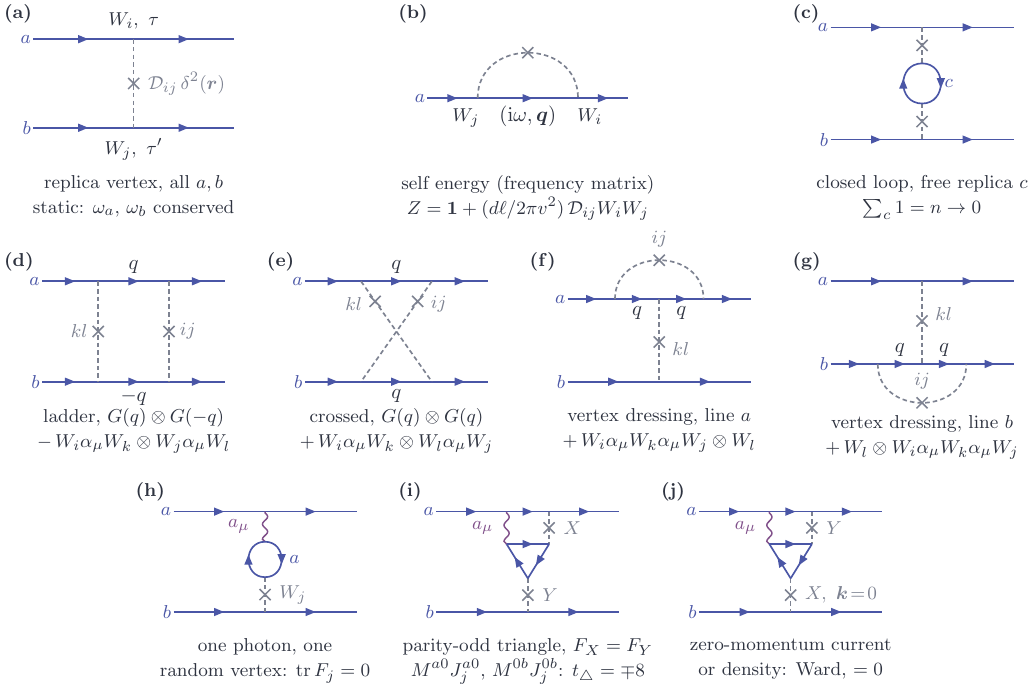}
\caption{One-loop graphs of the replicated disorder flow. (a) Replica vertex
with covariance $\mathcal D_{ij}$ between replicas $a$ and $b$. (b) Self-energy,
which renormalizes frequency. (c) Closed loops with a free replica sum,
which vanish as $n\to0$. (d) Ladder, (e) crossed and (f,g) vertex-dressing
graphs of the free fermions. (h) One photon and one random vertex on a loop,
zero for every adjoint vertex. (i) Parity-odd triangle, which couples a
random mass and a random current or density of the same flavor
($t_\triangle=\mp8$). (j) Zero-momentum current or density, zero by the Ward
identity.}
\label{fig:disorder}
\end{figure*}

To go beyond the tree-level eigenvalues we matched the microscopic random
couplings onto the 64 bilinears and integrated the one-loop replicated
flow, with the photon graphs that are specific to QED$_3$. Per replica the
action is $\int d\tau\,d^2r\,\psi^\dagger[\partial_\tau+v\bm\alpha\cdot(-i\nabla)+\sum_kV_k(\bm r)W_k]\psi$
with $\overline{V_k(\bm r)V_l(\bm r')}=\mathcal D_{kl}\delta(\bm r-\bm r')$, vertices normalized by
$\operatorname{tr}W_kW_l=8\delta_{kl}$, and $\mathcal D$ measured in units of $v^2$. The flow is
\begin{equation}
\frac{d\mathcal D_{ij}}{d\ell}=(4-\Delta_{O_i}-\Delta_{O_j})\mathcal D_{ij}+\frac1{2\pi}\big[B_{ij}(\mathcal D)+T_{ij}(\mathcal D)\big],
\label{eq:disflow}
\end{equation}
with $\Delta_{O_i}$ the clean dimension of the bilinear $O_i=\psi^\dagger W_i\psi$, $B$ the free-fermion ladder, crossed and
vertex graphs with the frequency renormalization
[Fig.~\ref{fig:disorder}(a)--(g)], and $T$ the photon graphs (h)--(j). With one
random vertex on a closed loop tied to the line by one photon (h), the
graph is proportional to the flavor trace of the vertex and vanishes for
every SU(4)-adjoint bilinear. Zero-momentum current and density insertions
cancel by the Ward identity (j). The parity-odd triangle (i), a random mass
and a random current or density of the same flavor on the loop, survives, and
it changes the mass vertex by $t_\triangle=-8$ per unit current variance and
$t_\triangle=+8$ per unit density variance. For our $\mathcal T$-even content it acts
on the quantum-spin-Hall mass with the spin currents and on the
valence-bond masses with the valley currents. The spin--valley densities
pair with $\mathcal T$-odd masses, which never carry a random field, and
receive no triangle. The free kernel reproduces $+2$, $-2$ and $0$ (in units
of $\mathcal D^2/2\pi$) for a random potential, a random mass and an Abelian
vector-potential pair, and with the triangle the flow reproduces
Eqs.~(50) and (51) of Ref.~\cite{Thomson2017} term by term, and their Eq.~(44)
up to two misprints in $\pi\beta_{t,a}$ ($g_{t,a}g_s\to g_s$ and $-2c\to-c$). Their
Eq.~(52), which should be the restriction of Eq.~(51) to one spin axis, has
$8g_{A,z}$ in $\pi\beta_{t,z}$ and $3g_{v,z}$ in $\pi\beta_s$, where Eq.~(51) gives $6g_{A,z}$ and $g_{v,z}$, and our flow
agrees with Eq.~(51). The fixed line is therefore $g_{t,z}=c-6g_{A,z}$, with
$z=1+(c-4g_{A,z})/2\pi$, $0\le g_{A,z}\le c/6$ and $c=\pi(4-2\Delta_{\rm adj})$, in place of their
Eqs.~(53) and (54). Their Eq.~(54) also lacks the factor $1/2\pi$ of their Eq.~(41).

\emph{Matching.}---Decoupling a bond exchange $S_i^aK^{ab}S_j^b$ on the
spin-singlet flux state gives the hopping
$-\tfrac14K^{ab}[\chi_{ij}f_j^\dagger\sigma^b\sigma^af_i+\mathrm{h.c.}]$. A random isotropic
exchange multiplies the mean-field bond term. A random
Dzyaloshinskii--Moriya vector $\bm D$ gives $(i/2)\chi_{ij}f_j^\dagger(\bm D\cdot\bm\sigma)f_i+\mathrm{h.c.}$
A symmetric traceless random anisotropy ($\delta\Jpp$, $\delta\Jzp$) gives exactly
zero, since $K^{ab}\sigma^b\sigma^a=\operatorname{tr}K$, and a random $\delta\Delta$ acts as
$\delta J=J\delta\Delta/3$. Projecting independent random bonds on the zero
modes gives the covariances. Bond disorder reaches only the valence-bond
masses $M^{0b}$ (one third of $\operatorname{tr}\mathcal D=2\sqrt3\,(\delta J/J)^2$) and the valley currents
$\mathcal J^{0b}_j$ (two thirds). In each valley the mass and the current along
one direction come from one random amplitude, with mass, parallel and
perpendicular variances in the ratio $2:1:3$ and a mass--current covariance
equal to the geometric mean. Each component of $\bm D$ reaches the
quantum-spin-Hall masses, the spin currents and the spin--valley densities
in the ratio $1:2:3$, with $\operatorname{tr}\mathcal D=(8/9)\sqrt3\,(D/J)^2$ and variance $0.257(D/J)^2$ in the
mass channel, and equal variances of the three components give a spin-rotation
invariant covariance.

\emph{Flows.}---For bond disorder the flow closes on the mass variance $m$,
the parallel and perpendicular current variances $p_\parallel$, $p_\perp$ and the
mass--current covariance $x$,
\begin{align}
\pi\,\dot m&=cm+m^2-3m(p_\parallel+p_\perp)-x^2,\nonumber\\
\pi\,\dot p_{\parallel,\perp}&=m^2+(p_\parallel+p_\perp)^2\mp2x^2,\label{eq:bondflow}\\
\pi\,\dot x&=x\big[\tfrac c2+\tfrac32m-\tfrac72(p_\parallel+p_\perp)\big],\nonumber
\end{align}
with the U(1) density projected out and each variance taken per valley component, so that
$m_0=0.385(\delta J/J)^2$. The current variance obeys
$\pi\,d(p_\parallel+p_\perp)/d\ell=2m^2+2(p_\parallel+p_\perp)^2$ and only grows, so the clean point is
the only fixed point. For $\delta J/J=0.05$, 0.1 and 0.2 the largest
covariance eigenvalue reaches 1 at $\ell=6.3$, 5.0 and 3.8, carried to
about 77\% by the valence-bond masses, and the scale is set by the tree-level
growth of the random masses. Random Dzyaloshinskii--Moriya exchange closes
on 13 couplings. The quantum-spin-Hall masses grow first. With any
in-plane component of $\bm D$ the weight then passes to the non-Abelian
random spin currents, and for $\bm D\parallel c$ the flow passes near the line
$g_{t,z}=c$ of Ref.~\cite{Thomson2017} and leaves it along the spin--valley
densities, relevant there with eigenvalue $2(4-2\Delta_{\rm adj})\simeq2.2$. At second
order random Dzyaloshinskii--Moriya exchange also generates random
valence-bond masses. For $D/J=0.1$ the largest eigenvalue reaches 1 at
$\ell\simeq5$--6 for every ratio of $c$-axis to in-plane components, and bond and
Dzyaloshinskii--Moriya disorder together give $\ell\simeq5$--5.5. No matched
ensemble has a disordered fixed point at one loop, and every fixed point
of the flow has relevant directions. The scheme is the large-$N_f$ flow
extrapolated to $N_f=4$, at one loop in the disorder, with Gaussian,
bond-uncorrelated disorder matched without gradients. The disordered fixed
points it contains, such as $g_{t,z}=c\simeq3.4$, lie outside weak disorder
at $N_f=4$.

\section{Large-$N$ dimensions of the quartic operators}
\label{app:quartic}

We write a derivative-free quartic as $\bar\psi_a\psi_b\bar\psi_c\psi_d\,C^{ab,cd}$, with $a$ combining
flavor and Dirac indices and $C$ antisymmetric under $a\leftrightarrow c$ and under $b\leftrightarrow d$. For
$N_f=4$ this space has dimension $28^2=784$. At $O(1/N)$ the dimension follows from the
logarithmic parts of the graphs with one photon between two legs of the operator at
zero momentum. With the photon propagator $D_{\mu\nu}(k)=16(\delta_{\mu\nu}-(1-\xi)k_\mu k_\nu/k^2)/(N_f|k|)$, a
photon between a $\bar\psi$ leg and a $\psi$ leg replaces $C$ by
$(8/\pi^2N_f)\langle(\gamma^\mu\,\hat k\cdot\gamma)\,C\,(\hat k\cdot\gamma\,\gamma^\nu)(\delta_{\mu\nu}-(1-\xi)\hat k_\mu\hat k_\nu)\rangle\ln\Lambda$, with the
matrices acting on the two legs and the angular average over $\hat k$. A photon between two
$\psi$ or two $\bar\psi$ legs carries the opposite sign. The legs contribute through the fermion
wave function, which we fix by requiring the conserved current to have no anomalous
dimension. For operators with a flavor-singlet scalar channel we add the closed-loop
graph with two photons, normalized by $\Delta_{\bar\psi\psi}=2+128/(3\pi^2N_f)$~\cite{Hermele2007}. The
eigenvalues of the resulting $784\times784$ operator include the known dimensions of the adjoint mass,
$2-64/(3\pi^2N_f)$, and of the Lorentz-scalar $\mathbf{20}'$ and $\mathbf{84}$, $4-64/(\pi^2N_f)$ and
$4+64/(3\pi^2N_f)$~\cite{Xu2008,Chester2016}.

Operators that contain the U(1) gauge current as a factor, in any Fierz channel, vanish by
the equation of motion of the gauge field. They span 189 dimensions, in the singlet and
the $\mathbf{15}$ only, and the physical dimensions there are those of the operator compressed on
their orthogonal complement. For the $\mathbf{15}$ we repeated the computation at general $N$
with one adjoint component and all Dirac structures. For the Lorentz vector, the product of
$\bar\psi\psi$ with the conserved adjoint current, the closed-loop graphs reduce to their value for
$\bar\psi\psi$ alone, because the insertions of a conserved current at zero momentum into the loop
sum to zero by the Ward identity, and the coefficient approaches $\lambda=16/3$ as $1/N$, that is
$\Delta_{\bar\psi\psi}+2$. For the Lorentz scalar, the product with the adjoint mass, the same
treatment gives $8/3$, $\Delta_{\bar\psi\psi}+\Delta_{\rm adj}$, but omits the two-photon graphs that connect the
loop to the legs of the adjoint mass. No locked invariant lies in that block, so this
value enters none of our results. The finite-$N$ approach depends on the choice of
representatives of the null operators, but the limit does not. The invariant counts of
Table~\ref{tab:quarticdims} use the projected generators of Sec.~\ref{sec:psg}. The results do not depend on the Dirac basis or on the angular quadrature.

\section{Refit of the KYbSe$_2$ data}
\label{app:refit}

Table~\ref{tab:refit} collects the estimators of Sec.~\ref{sec:kybse2} and this appendix.

\begin{table*}[t]
\caption{Exponent $\alpha_{\rm K}$ of the KYbSe$_2$ K-point spectrum from the source data of
Ref.~\cite{Scheie2023} by estimator, temperatures and energy window (the two values correspond
to $E_{\max}=0.45$ and 0.56\,meV where two are given). $\Delta\chi^2$ is the difference
$\Delta\chi^2(0.91)-\Delta\chi^2(1.44)$ scaled by the $\chi^2$ per degree of freedom, with errors
raised to the point-to-point scatter (reported errors in parentheses).}
\label{tab:refit}
\begin{ruledtabular}
\begin{tabular}{p{0.3\textwidth}p{0.08\textwidth}p{0.11\textwidth}p{0.16\textwidth}p{0.3\textwidth}}
estimator & $T$ (K) & $\hbar\omega$ (meV) & $\alpha_{\rm K}$ & remarks\\
\hline
one power law in $\omega$, $\hbar\omega/k_BT\ge1$ (Ref.~\cite{Scheie2023}) & 0.3, 1, 2 & 0.11--1.3 & $1.740\pm0.034$ & $\chi^2$ per degree of freedom 13.6; 1.61--1.94 with the lower cutoff\\
same, upper cutoff $\le0.7$ & 0.3, 1, 2 & 0.11--0.7 & 1.45--1.52 & set largely by the 0.3\,K data\\
collapse with free exponent, alone and averaged over resamplings of the energy points & 1, 2 & 0.11--0.45, 0.56 & 1.40--1.46 & rests on the ratio of the two spectra; $\pm0.14$ per 10\% of relative normalization\\
point-by-point ratio, and $\ln\mathcal F$ quadratic in $\ln(\hbar\omega/k_BT)$ & 1, 2 & same & 1.3--1.5 & \\
$\log_2$ of the 1 and 2\,K ratio at equal $\hbar\omega/k_BT$ & 1, 2 & same & 1.47--1.49 (1.41--1.42) & 1.28--1.29 from 0.21\,meV [Fig.~\ref{fig:kybse2}(a)]\\
free scaling function, 10\% prior on the normalization & 1, 2 & same & 1.43, 1.47; 95\% [1.08, 1.80] & $\Delta\chi^2=6.2$, 9.1 (2.6, 3.8)\\
same, 20\% prior & 1, 2 & same & & $\Delta\chi^2=2.0$--2.8 (0.7--1.0)\\
same, free normalization & 1, 2 & same & undetermined & $\Delta\chi^2=0$\\
same, 10\% prior, from 0.21\,meV & 1, 2 & 0.21--0.45, 0.56 & 1.19, 1.34 & $\Delta\chi^2=0.2$, 4; 95\% [0.74, 1.63] at 0.45\,meV\\
same, with the 0.3\,K data & 0.3, 1, 2 & 0.11--0.45, 0.56 & 1.5--1.55 & only if 0.3\,K lies in the scaling regime\\
one power law, 1 and 2\,K only & 1, 2 & 0.11--0.45, 0.56 & 0.6--0.95 & $\chi^2$ per degree of freedom 4--15\\
single-temperature log-slope & 1 & same & 0.9--1.3 & 0.2--0.5 at 2\,K; error-model dependent\\
collapse of the path cuts through K & 1, 2 & same & $\approx1.2$ & \\
three-temperature collapse, flat background subtracted & 0.3, 1, 2 & 0.11--1.3 & 1.8--2.2 & background bounded by the zone centers\\
M point, collapse & 1, 2 & $\le0.45$ & $\alpha_{\rm M}=-0.1\pm0.15$ & $-0.05$ in Fig.~\ref{fig:kybse2}(c); DSL value 0.08\\
\end{tabular}
\end{ruledtabular}
\end{table*}

To separate the exponent from the shape of $\mathcal F$ we fitted
$S=T^{-\alpha}\mathcal F(\hbar\omega/k_BT)$ with $\ln\mathcal F$ piecewise linear in $\ln(\hbar\omega/k_BT)$ and
a penalty on its curvature, which leaves pure power laws unpenalized. At each
$\alpha$ we minimized $\chi^2$ over all other parameters, with a Gaussian prior on the relative normalization $r$
of the 1 and 2\,K data. The fit gives $\alpha_{\rm K}=1.43$--$1.47$ for
$E_{\max}=0.45$--$0.56$\,meV, with 68\% intervals $[1.26,1.63]$ and 95\%
intervals $[1.08,1.80]$. The spacing of the nodes of $\ln\mathcal F$ and the error model move
$\hat\alpha$ by at most 0.05. Against the
stable-DSL bound the fit favors $\alpha_{\rm K}=1.44$ by
$\Delta\chi^2(0.91)-\Delta\chi^2(1.44)=9$--11 for a 10\% prior on $r$, 2.4--3.5 for a 20\%
prior and 0 for free $r$. These are raw values. Scaled by the $\chi^2$ per degree of
freedom, 1.26--1.43 with errors raised to the point-to-point scatter and 2.9--3.4 with
the reported errors, they become 6.2--9.1 and 2.6--3.8 at 10\%, and 2.0--2.8 and 0.7--1.0
at 20\%. The data fix only the combination
$\ln r-\alpha\ln2$, to about $\pm0.065$, so the preference is carried by the
assumed normalization, and varying the penalty strength over four decades
keeps the raw difference at 6--11. It also rests on the lowest-energy 1\,K points.
The reported error bars shrink by factors of 2.5--2.9 between 0.19 and
0.21\,meV at all three temperatures, the join of the $E_i=1.55$\,meV data below 0.2\,meV with the
$E_i=3.32$\,meV data above it~\cite{Scheie2026} (the switch quoted for Fig.~2 of Ref.~\cite{Scheie2023} lies at 0.5\,meV),
and across it the 1\,K intensity falls from 4.22 to 2.70, against about
12\% expected from the local log-slope of 1.1--1.3. Starting the window at 0.2\,meV
gives $\alpha_{\rm K}=1.19$ (95\% $[0.74,1.63]$) at $E_{\max}=0.45$\,meV and 1.34 at
0.56\,meV, with $\Delta\chi^2$ differences of 0.3 and 5 (0.2 and 4 scaled). A common additive
background, which the negative-energy data bound to about $\pm0.05$, changes
none of this, and including the 0.3\,K data would raise the exponent to
1.5--1.55 only if these data lay in the scaling regime, which their
temperature at or below $T_N$ does not allow us to assume. Test spectra
built with a known exponent, $\alpha=0.91$ or 1.44, the shape of the measured spectra and
Gaussian noise at the published error bars are recovered without bias by this fit, and its
95\% intervals contain the true exponent in 95--97\% of the test cases, whereas parametric crossover forms for $\mathcal F$ fitted to the
same test spectra return $\alpha\simeq1.3$--$1.55$ for a true $\alpha=0.91$, so a preference
derived from such forms is an artifact of the assumed shape.

At M the spectrum below $J$ is nearly flat, $\alpha_{\rm M}=-0.1\pm0.15$ from the
two temperatures above $T_N$, and it excludes $\alpha_{\rm M}\gtrsim0.3$. The value agrees with
the DSL value but cannot distinguish it from a temperature-independent
continuum, and the 2\,K spectrum lies 13--18\% below the 0.3\,K one. At
0.3\,K the M intensity is suppressed by a factor of about 2 below 0.18\,meV,
just under the roton-like minimum of the dispersion at 0.20\,meV~\cite{Scheie2023}, and above 0.5\,meV the M spectrum
falls toward the band top at every temperature, so $\omega/T$ scaling at M
holds only below $J$. Over 0.2--0.8\,meV the ratio $S(\mathrm K)/S(\mathrm M)$ falls as
$\omega^{-2.0}$ at 0.3 and 1\,K, so K is much more singular than M, as the
DSL requires, although this finite-window slope cannot be compared
directly with the difference of exponents. Over the full window, subtracting a flat
non-magnetic background, bounded by the zone centers, steepens the
three-temperature collapse at the K point to 1.8--2.2, and a $\pm10\%$ change of the relative
normalization of the 0.3\,K data shifts it by $\pm0.06$. Below $J$ a common additive
background is bounded by the negative-energy data and changes none of the results above.

\nocite{Maksimov2022}
\bibliography{arXiv_v1_prb_yb}
\end{document}